\documentclass[fleqn,usenatbib,useAMS]{mnras}

\usepackage{graphicx}	
\usepackage{amsmath}	
\usepackage{amssymb}	
\usepackage{multicol}        
\usepackage{bm}		
\usepackage{pdflscape}	

\usepackage{xcolor}
\usepackage{color}
\usepackage{graphicx}
\usepackage{dcolumn}
\usepackage{bm}
\usepackage{hyperref}
\usepackage{amsmath}
\usepackage{multirow}
\usepackage{cancel}

\newcommand{\Ms}{{\rm ~M}_\odot}

\newcommand{\bh}{{\rm BH}}

\newcommand{\gw}{{\rm GW}}

\newcommand{\bbh}{{\rm BBH}}
\newcommand{\ham}{{\rm HM}}

\newcommand{\nbsix}{\textsc{Nbody6++GPU~}}
\newcommand{\sse}{\textsc{SSE~}}
\newcommand{\bse}{\textsc{BSE~}}
\newcommand{\mcl}{\textsc{McLuster~}}

\newcommand{\dragonii}{\textsc{Dragon-II~}}

\usepackage[T1]{fontenc}
\usepackage{ae,aecompl}

\title[The \textsc{Dragon-II} simulations - I.]{The \textsc{Dragon-II} simulations - I. Evolution of single and binary compact objects in star clusters with up to 1 million stars}

\author[M. Arca Sedda et al]{Manuel Arca Sedda$^{1,2,3}$
\thanks{Contact e-mail:\href{mailto:m.arcasedda@gmail.com}{m.arcasedda@gmail.com}},
Albrecht W. H. Kamlah$^{4,3}$, Rainer Spurzem$^{3,5,6}$, 
\newauthor 
Mirek Giersz$^{7}$, Peter Berczik$^{3,8,9}$, Sara Rastello$^{2,10}$, Giuliano Iorio$^{2,10,11}$, 
\newauthor 
Michela Mapelli$^{2,10,11}$, Massimiliano Gatto$^{12}$, Eva K. Grebel$^{3}$
\\
$^{1}$ Gran Sasso Science Institute (GSSI), 67100 L’Aquila, Italy\\
$^{2}$ Physics and Astronomy Department Galileo Galilei, University of Padova, Vicolo dell'Osservatorio 3, I--35122, Padova, Italy\\
$^{3}$ Astronomisches Rechen-Institut, Zentrum f\"{u}r Astronomie der Universit\"{a}t  Heidelberg, M\"onchhofstr. 12-14, D-69120 Heidelberg, Germany\\
$^{4}$ Max-Planck-Institut f\"ur Astronomie, K\"onigstuhl 17, 69117 Heidelberg, Germany \\
$^{5}$ National Astronomical Observatories and Key Laboratory of Computational Astrophysics, Chinese Academy of Sciences, \\ 20A Datun Rd.,Chaoyang District, 100101, Beijing, China\\
$^{6}$ Kavli Institute for Astronomy and Astrophysics, Peking University, Yiheyuan Lu 5, Haidian Qu, 100871, Beijing, China\\
$^{7}$ Nicolaus Copernicus Astronomical Centre, Polish Academy of Sciences, ul. Bartycka 18, 00-716 Warsaw, Poland\\
$^{8}$ Konkoly Observatory, Research Centre for Astronomy and 
Earth Sciences, E\"otv\"os Lor\'and Research Network (ELKH), \\
MTA Centre of Excellence, Konkoly Thege Mikl\'os \'ut 15-17, 
1121 Budapest, Hungary \\
$^{9}$ Main Astronomical Observatory, National Academy of Sciences of Ukraine, 27 Akademika Zabolotnoho St., 03680, Kyiv, Ukraine \\
$^{10}$ INFN-Padova, Via Marzolo 8, I--35131 Padova, Italy\\
$^{11}$ INAF-Padova, Vicolo dell'Osservatorio 5, I--35122 Padova, Italy\\
$^{12}$ INAF – Osservatorio Astronomico di Capodimonte, Via Moiariello 16, 80131 Naples, Italy
}

\date{Last updated 2020 June 10; in original form 2013 September 5}

\pubyear{2020}

\begin{document}
\label{firstpage}
\pagerange{\pageref{firstpage}--\pageref{lastpage}}
\maketitle

\begin{abstract}
We present the first results of the \textsc{Dragon-II} simulations, a suite of 19 $N$-body simulations of star clusters with up to $10^6$ stars, with up to $33\%$ of them initially paired in binaries. In this work, we describe the main evolution of the clusters and their compact objects (COs). All \textsc{Dragon-II} clusters form in their centre a black hole (BH) subsystem with a density $10-100$ times larger than the stellar density, with the cluster core containing $50-80\%$ of the whole BH population. In all models, the BH average mass steeply decreases as a consequence of BH burning, reaching values $\langle m_{\rm BH}\rangle < 15$ M$_\odot$ within $10-30$ relaxation times. Generally, our clusters retain only BHs lighter than $30$ M$_\odot$ over $30$ relaxation times. Looser clusters retain a higher binary fraction, because in such environments binaries are less likely disrupted by dynamical encounters. We find that BH-main sequence star binaries have properties similar to recently observed systems. Double CO binaries (DCOBs) ejected from the cluster exhibit larger mass ratios and heavier primary masses than ejected binaries hosting a single CO (SCOBs). Ejected SCOBs have BH masses $m_{\rm BH} = 3-20$ M$_\odot$, definitely lower than those in DCOBs ($m_{\rm BH} = 10-100$ M$_\odot$). 
\end{abstract}


\section{Introduction}

Massive star clusters in the range $(10^4-10^6)\Ms$, like globular clusters or young massive clusters, represent galactic repositories of  stellar compact  objects, and are ideal laboratories to study the interplay of stellar evolution and dynamics.
Several hundreds of stellar black holes (BHs), neutron stars (NSs), and white dwarfs (WDs) are expected to form in a typical massive cluster.

In the last decade, it became clear that the fraction of BHs that massive clusters can retain is much larger than previously thought, as suggested by numerous theoretical and numerical works \citep[see e.g.][]{2013ApJ...763L..15M,2014ApJ...790..119W,2016MNRAS.458.1450W,2018MNRAS.478.1844A,2018A&A...617A..69P, 2018MNRAS.479.4652A,2016MNRAS.462.2333P,2021NatAs...5..957G, 2022MNRAS.516.3266K}, providing support to the crescent number of observations of stellar BH candidates in Galactic clusters \citep{2012Natur.490...71S,2015MNRAS.453.3918M,2018MNRAS.475L..15G,2013ApJ...777...69C,2017MNRAS.467.2199B,giesers19}.

The progress in stellar evolution of massive stars \citep{2007Natur.450..390W,2015MNRAS.450.4070W,2018MNRAS.480.2011G,2010ApJ...714.1217B,2016A&A...594A..97B,2019MNRAS.485..889S,2021MNRAS.504..146V}, partly triggered by the discovery of gravitational-wave (GW) emission by merging BH and NS binaries \citep{2021arXiv211103606T,2021arXiv211103634T}, has completely changed our understanding of BHs. Stellar models demonstrated that the evolution of single massive stars is significantly influenced by the possible development of so-called pair instability supernovae (PISN), which causes the complete disruption of stars that develop an He core with a mass of $M_{\rm He} = 64-135 \Ms$, and pulsational pair instability supernovae (PPISN), a mechanism that leads to an enhanced mass-loss in stars with a He core mass of $M_{\rm He} = 32-64\Ms$. This leads to a maximum stellar BH mass in the range $m_{\rm BH, max} = (40-60)\Ms$, depending on the theoretical model adopted and the stellar metallicity. Direct consequence of these two processes is the well known upper-mass gap of BHs, a region of the mass-spectrum where no remnants are expected \citep{2007Natur.450..390W}. The boundaries of the upper-mass gap are highly uncertain and depend on the adopted stellar evolution model and metallicity \citep{2007Natur.450..390W,2015MNRAS.450.4070W,2016A&A...594A..97B,2017MNRAS.470.4739S,2021MNRAS.504..146V,2021MNRAS.501.4514C,2022arXiv221111774I}. Only stars with a zero age main sequence mass beyond $M_{\rm ZAMS} > (200-250)\Ms$ can avoid PISN and, depending on their metallicity, directly collapse to an intermediate-mass BH with little mass loss in the process \citep[see e.g.][]{2017MNRAS.470.4739S}. Stellar collisions might lead to the formation of BHs in the upper-mass gap \citep[e.g.][]{2019MNRAS.485..889S}, thus suggesting that star clusters could be perfect laboratories to form mass-gap BHs  \citep[e.g.][]{2019MNRAS.487.2947D,2020ApJS..247...48K,2021MNRAS.501.5257R,2021MNRAS.507.3612R,2022MNRAS.512..884R,2022A&A...665A..20B}, but it is unclear how the stellar merger frequency depends on the cluster initial properties \citep{2022MNRAS.512..884R} or the stellar conditions at merger \citep{2022arXiv220403493B,2022MNRAS.516.1072C}. 

More in general, the formation of a population of compact objects can significantly affect star cluster dynamics. Massive stars and BHs rapidly sink into the cluster centre via mass-segregation, possibly forming a massive subsystem on a core-collapse timescale \citep[e.g.][]{1987degc.book.....S,2013MNRAS.432.2779B,2018A&A...617A..69P,2018MNRAS.479.4652A,2022arXiv220901564L} which can contract and determine the onset of runaway stellar collisions if the time does not exceed the stellar evolution timescale \citep[e.g.][]{1987degc.book.....S, 2002ApJ...576..899P,2004Natur.428..724P,2004MNRAS.352....1F,2013MNRAS.432.2779B,2016MNRAS.459.3432M, 2015MNRAS.454.3150G,2018MNRAS.479.4652A,2021A&A...649A.160V,2022arXiv220915066V,2022MNRAS.514.5879M}. The runaway growth of a massive star can be hampered by the formation of tight binaries that supply energy to the cluster core, cause BH ejection, deplete the cluster's BH reservoir, and eventually kick each other out via super-elastic encounters \citep{2013MNRAS.432.2779B}.

The competing effect of binary energy supply and stellar collisions likely depends on the cluster mass, density, metallicity, the fraction of primordial binaries, the initial mass function and its boundaries, the natal kicks of BHs and NSs, and the compact object mass spectrum. Typically, the exploration of a tiny part of such parameter space is performed with numerical models capable of simultaneously accounting for stellar dynamics and evolution, either via direct $N$-body \citep[e.g.][]{2015MNRAS.450.4070W, 2018MNRAS.473..909B, 2021MNRAS.500.3002B, 2019MNRAS.487.2947D,2021MNRAS.507.3612R,2021MNRAS.507.5132D, 2022MNRAS.513.4527C} or Monte Carlo techniques \citep[e.g.][]{2016PhRvD..93h4029R,2017MNRAS.464L..36A,2019PhRvD.100d3027R,2020ApJS..247...48K,2022MNRAS.514.5879M}. 

Direct $N$-body simulations offer most likely the highest level of accuracy in terms of stellar dynamics modelling, but their computational cost forced the vast majority of works in the literature to focus on star clusters with less than a few $\times~10^5$ stars and/or with a relatively small fraction of primordial binaries \citep{2018MNRAS.473..909B, 2021MNRAS.500.3002B, 2019MNRAS.487.2947D,2021MNRAS.507.3612R,2021MNRAS.507.5132D}, with a few notable exceptions. For example, several works have explored the impact of a large primordial binary fraction, up to $100\%$, on the dynamics of isotropic \citep[e.g.][]{2006MNRAS.368..677H,2007MNRAS.374..344T,2020A&A...638A.155P} and anisotropic \citep{2022MNRAS.509.3815P,2022MNRAS.515.1830P} low-mass star cluster models, i.e. with $N < 20,000$, with equal-mass stars, and recently in intermediate-mass GCs, i.e. $N\sim 10^5$ \citep{2022MNRAS.509.4713W}. With regards to simulations tailored to represent massive globular clusters, the DRAGON simulations remain the only one that exploited $10^6$ particles \citep{2016MNRAS.458.1450W}.

Since the development of such pioneering simulations, and especially after the discovery of GWs, numerical tools underwent major upgrades in terms of stellar evolution and treatment of relativistic binaries.

In this work, we present the \dragonii simulation database, a suite of 19 direct $N$-body simulations performed with the \nbsix code\footnote{\url{https://github.com/nbody6ppgpu/Nbody6PPGPU-beijing}} representing star clusters with $N=(0.12-1)\times 10^6$ stars, half-mass radius densities in the $\rho_h = 1.3\times 10^4 - 6.9 \times 10^6$ M$_\odot$ pc$^{-3}$ range, and a fraction $f_{\rm 2b} = 0.10-0.33$ of stars initially paired in primordial binaries.   
This work, which is the first one of a series, focuses on the evolution of single and binary BHs and compact objects in massive and dense star clusters, paying particular attention to the relation between the BH population (mass, average BH mass, density) and the cluster properties (mass, radius). Our \dragonii models explore a portion of the parameter space still uncharted by direct $N$-body simulations, thus complementing previous works that either rely on Monte Carlo simulations or exploit star cluster models with old stellar evolution recipes or a significantly smaller number of stars.

The paper is organised as follows: Section \ref{sec:methods} describes the main properties of the \dragonii clusters and the improvements integrated in the \nbsix code; Section \ref{sec:res} presents our main results in terms of overall star cluster evolution (Section \ref{sec:clu}), main properties of single and binary compact objects (Sections \ref{sec:cob} - \ref{sec:BHS}), and the possible implementation of $N$-body outputs into semi-analytic tools (Section \ref{sec:scal}); whilst Section \ref{sec:sum} is devoted to summarise the main outcomes of our work.

\section{Numerical Methods}
\label{sec:methods}

All the \dragonii models are carried out exploiting the \nbsix code \citep{2015MNRAS.450.4070W}, which represents the current state-of-the-art of direct $N$-body codes optimised to exploit
GPU-accelerated high-performance supercomputing \citep{1999JCoAM.109..407S,2012MNRAS.424..545N,2015MNRAS.450.4070W} altogether with several recently developed codes, like \textsc{Petar} \citep{2020MNRAS.497..536W} or \textsc{Bifrost} \citep{2022arXiv221002472R}. \nbsix belongs to a long-standing family of direct $N$-body integrators initiated by Sverre Aarseth and developed for almost 50 years \citep{1974A&A....37..183A,1999JCoAM.109..407S,1999PASP..111.1333A,2003gnbs.book.....A,2008LNP...760.....A,2012MNRAS.424..545N,2015MNRAS.450.4070W,2022MNRAS.511.4060K}. 

\nbsix implements a 4th-order Hermite integrator with individual block-time steps \citep{1986LNP...267..156M,1995ApJ...443L..93H} and sophisticated algorithms for close encounters and few-body dynamics, namely the Kustaanheimo-Stiefel (KS) regularisation \citep{Stiefel1965}, the Ahmad-Cohen (AC) scheme for neighbours \citep{1973JCoPh..12..389A}, and algorithmic chain regularisation \citep{1999MNRAS.310..745M,2008AJ....135.2398M}, which enables us to closely follow the evolution of binaries with periods $10^{-10}$ times smaller than the dynamical timescales of star clusters, which typically exceed O($10$) Myr.

In the last few years, the code underwent a series of major upgrades related to the treatment of relativistic compact objects \citep{2021MNRAS.501.5257R}, the implementation of flexible stellar evolution recipes \citep{2022MNRAS.511.4060K}, and the inclusion of a dedicated treatment for spins \citep[][this work]{2020A&A...639A..41B, 2022MNRAS.511.4060K}. Here, we expand the possible choices for BH natal spin distribution and implement relativistic recoil for post-merger remnants. 
In the following, we briefly summarize the features of the code that are most relevant for this work, and discuss the newest upgrades that we implemented into the code and use here for the first time. 

\subsection{Stellar evolution}

\nbsix implements stellar evolution for single and binary stars via the \sse and \bse routines \citep{2000MNRAS.315..543H,2002MNRAS.329..897H}, which we heavily updated to include up-to-date prescriptions for the evolution of massive stars. We refer the reader to \cite{2022MNRAS.511.4060K} for a comprehensive discussion about the updated stellar evolution encoded in \nbsix. 

In this work, we adopt the level-B of stellar evolution as defined in \citep[][, see their Table A1]{2022MNRAS.511.4060K}. This implies that our models take into account the formation of  electron-capture supernovae (ECSNe, following \citealt{2008ApJS..174..223B}), the delayed SN scheme \citep{2012ApJ...749...91F}, and the development of pair-instability (PISN) and pulsational pair instability supernovae (PPISN) \citep{2016A&A...594A..97B}. For the formation of compact objects, we adopt mass loss from \cite{2010ApJ...714.1217B} with additional metallicity-dependent correction factors taken from \cite{2001A&A...369..574V} and a dedicated treatment for mass loss of hot and massive H-rich O/B stars \citep{2001A&A...369..574V}. 
The adopted stellar evolution models imply that the maximum BH mass attainable by massive stars with zero-age main-sequence mass $<150\Ms$ is $m_{\rm BH, max} = 40.5\Ms$ \citep{2016A&A...594A..97B,2020A&A...639A..41B,2022MNRAS.511.4060K}. The BHs falling in the so-called upper mass-gap can still form via stellar collisions, accretion of stellar material onto stellar BHs, and BH-BH mergers, as we discuss in our companion papers.

Natal kicks for NSs forming via ECSNe, accretion induced collapse (AIC), and merger-induced collapse (MIC) are drawn from a Maxwellian distribution with dispersion $3$ km$/$s \citep[see][]{2018ApJ...865...61G,2022MNRAS.511.4060K}, whilst for all other NSs we adopt a Maxwellian distribution with dispersion $265$ km$/$s \citep{2005MNRAS.360..974H}. This latter value is adopted also for BHs, but the kick amplitude is reduced by a factor that accounts for the amount of fallback material \citep{2012ApJ...749...91F}.

For binary stars, we model common envelope evolution via the parametrised $\alpha_{\rm CE}-\lambda_{\rm CE}$ scheme, according to which it is possible to regulate the fraction of orbital energy injected into the envelope ($\alpha_{\rm CE}$) and to scale the binding energy of the envelope by a factor $\lambda_{\rm CE}$ in a way similar, but not equal, to the one followed by \cite{2014A&A...563A..83C} \citep[further details about these parameters are discussed in][]{2022MNRAS.511.4060K}. In this work, we adopt $\alpha_{\rm CE} =  3$ \citep{2018MNRAS.480.2011G, 2022MNRAS.511.4060K}.

\subsection{Dynamics of compact objects}

In particularly dense clusters, stellar interactions can trigger collisions among stars and/or compact objects. The aftermath of such collisions is still a poorly understood process that can crucially affect the formation and evolution of stellar BHs.
Whilst the outcome of stellar mergers is better understood, also thanks to recent detailed hydrodynamical simulations coupled with stellar evolution models \citep{2022arXiv220403493B,2022MNRAS.516.1072C}, it is still  unclear how much mass a massive star can accrete onto a stellar BH. Several works have shown that in the case of a star with a mass $\sim (1-10)\Ms$ merging with a stellar BH, there is little accretion as most of the energy is radiated away via jets, although the mechanism is highly uncertain and likely depends on the star structure and evolutionary stage \citep{2013ApJ...767...25G,2015ApJ...798L..19M, 2020ApJ...894..147C,2022ApJ...933..203K}. Hydrodynamical simulations of star--BH close encounters have shown that up to $70\%$ of the star mass remains bound to the BH, but energy arguments suggest that even a tiny amount of accreted matter, $O(10^{-3}-10^{-2}\Ms)$ would suffice to evaporate the accretion disk and halt the BH growth \citep{2022ApJ...933..203K}. Nonetheless, recent simulations modelling the common envelope phase of a tight star--BH binary have shown that the BH accretes the stellar core and expels the envelope, a process accompanied by a SN-like transient and spin-up of the BH to nearly extreme values regardless of the initial spin \citep{2020ApJ...892...13S}. In multiple main-sequence star collisions, the merger product is expected to have a compact core and a tenuous envelope with densities as low as $10^{-10}$ g cm$^{-3}$ \citep{2009A&A...497..255G}. Therefore, if: a) most of the merger product mass is in the core \citep{2009A&A...497..255G}, and b) the core can efficiently feed the BH \citep{2020ApJ...892...13S}, it is reasonable to assume that a BH would accrete a significant fraction of it. 

Given the aforementioned uncertainties, in \nbsix we parametrise the outcome of star-BH collisions via the fraction of star mass accreted onto the BH, $f_c$ \citep{2021MNRAS.500.3002B,2021MNRAS.501.5257R}. Throughout this paper we adopt $f_c = 0.5$.

Natal spins are another poorly known property of stellar BHs. \nbsix implements the so-called ``Geneva'', ``MESA'', and ``Fuller'' models \citep{2020A&A...636A.104B,2021MNRAS.500.3002B,2022MNRAS.511.4060K}, and four additional choices implemented in this work, namely: zero-spins, uniform spin distribution, Gaussian spin distribution with mean value $\chi = 0.5$ and dispersion $\sigma_\chi = 0.2$, and a Maxwellian distribution with dispersion $\sigma_\chi = 0.2$.

\nbsix also features a treatment for compact binary mergers based on an orbit-averaged formalism \citep{1963PhRv..131..435P,1964PhRv..136.1224P}, which enables us to follow the formation and evolution of in-cluster compact binary mergers, a feature implemented in a number of recent works modelling young star clusters \citep{2019MNRAS.487.2947D, 2020MNRAS.497.1043D,2020MNRAS.498..495D,2021MNRAS.507.5132D,2021MNRAS.501.5257R,2022MNRAS.512..884R,2021MNRAS.507.3612R}.

In this work, we present the implementation of three new features of the code: mass and spin of the merger remnant, calculated via numerical relativity fitting formulas  \citep{2017PhRvD..95f4024J,2020ApJ...894..133A}, and the recoil kick imparted by asymmetric GW emission promptly after merging events \citep{2007PhRvL..98w1102C,2008PhRvD..77d4028L,2012PhRvD..85h4015L}. We follow the implementation depicted in our previous works \citep[e.g.][]{2020ApJ...894..133A,2021ApJ...920..128A}.

\begin{eqnarray}
\vec{v}_\gw   =& v_m\hat{e}_{\bot,1} + v_\bot(\cos \xi \hat{e}_{\bot,1} + \sin \xi \hat{e}_{\bot,2}) + v_\parallel \hat{e}_\parallel, \label{eqKick1}\\
v_m         =& A\eta^2 \sqrt{1-4\eta} (1+B\eta), \\
v_\bot      =& \displaystyle{\frac{H\eta^2}{1+q_\bbh}}\left(S_{2,\parallel} - q_\bbh S_{1,\parallel} \right), \\
v_\parallel =& \displaystyle{\frac{16\eta^2}{1+q_\bbh}}\left[ V_{11} + V_A \Xi_\parallel + V_B \Xi_\parallel^2 + V_C \Xi_\parallel^3 \right] \times \nonumber \\
             & \times \left| \vec{S}_{2,\bot} - q_\bbh\vec{S}_{1,\bot} \right| \cos(\phi_\Delta - \phi_1).  \label{eqKick2}
\end{eqnarray}
Here $\eta \equiv q_\bbh/(1+q_\bbh)^2$ is the symmetric mass ratio, $\vec{\Xi} \equiv 2(\vec{S}_2 + q_\bbh^2 \vec{S}_1) / (1 + q_\bbh)^2$, and the subscripts $\bot$ and $\parallel$ mark the perpendicular and parallel directions of the BH spin vector ($\vec{S}$) with respect to the direction of the binary angular momentum. We adopt $A = 1.2 \times 10^4$ km s$^{-1}$, $B = -0.93$, $H = 6.9\times 10^3$ km s$^{-1}$, and $\xi = 145^\circ$ \citep{2007PhRvL..98i1101G,2008PhRvD..77d4028L}, $V_{11} = 3677.76$ km s$^{-1}$, and $V_{A,B,C} = (2.481, 1.793, 1.507)\times 10^3$ km s$^{-1}$. The quantity $\phi_\Delta$ represents the angle between the direction of the infall at merger (which we randomly draw in the binary orbital plane) and the in-plane component of the quantity $\vec{\Delta} \equiv (M_a+M_b)^2 (\vec{S}_b - q_\bbh \vec{S}_a)/(1+q_\bbh)$, while $\phi_1 = 0-2\pi$ is the phase of the binary, extracted randomly between the two limiting values. 

\subsection{Massive star cluster models with up to one million stars}
We generate the 19 \dragonii star clusters with the updated \mcl software \citep{2011MNRAS.417.2300K}, as described in \cite{2022MNRAS.511.4060K} and \cite{2022MNRAS.514.5739L}. 

All \dragonii star clusters are modelled via \cite{1966AJ.....71...64K} dynamical models with a central dimensionless  potential well $W_0 = 6$, and are characterised by three values of the half-mass radius, $R_{\ham} = 0.47,~0.80,~1.75$ pc, four values of the initial number of stars, $N = (1.2,~3,~6,~10)\times 10^5$, and two values of the primordial binary fraction, as described below. All clusters have the same metallicity $Z = 0.0005$, a value typical of several clusters proposed to host a dense subsystem of stellar BHs, like NGC3201 or a central intermediate-mass black hole (IMBH), like NGC6254 \citep[see e.g.][]{2018MNRAS.478.1844A,2018MNRAS.479.4652A,2020ApJ...898..162W}.  
 
All simulations were conducted on the Juwels BOOSTER supercomputer and the GRACE HPC workstation over a $\sim 2$ yr timespan. Eventually, the whole \dragonii database consists of almost 35 Tb of data.

Stellar masses are drawn from the \cite{2001MNRAS.322..231K} initial mass function limited between $m_* = 0.08-150\Ms$, which implies an initial average stellar mass is $\langle m_* \rangle \simeq 0.59\Ms$. The corresponding initial mass and density scale in \dragonii clusters are $M_c = (0.7-5.9)\times 10^5\Ms$ and densities $ \rho_c \simeq 1.3\times 10^4 - 6.9 \times 10^6 ~\Ms$ pc$^{-3}$, respectively.

All \dragonii clusters move on a circular orbit at a distance of $13.3$ kpc from the centre of a galaxy whose gravitational potential is modelled via a simple Keplerian potential assuming a total galaxy mass of $M_g = 1.78\times 10^{11}\Ms$. As a consequence, our clusters have initially a tidal radius in the range $R_{\rm tid} = 67-138$ pc and they can all be considered as underfilling systems, thus the gravitational field has a smaller impact on the cluster evolution with respect to internal dynamics, at least at the beginning. \dragonii clusters would underfill their Roche lobe even in the case of a rather extremely eccentric orbit, e.g. $e = 0.9$.

We assume that a fraction of the total number of stars is initially paired in a {\it primordial} binary system. Following in \mcl, we define the binary fraction as the ratio between the number of binaries and the sum of single stars and binaries, $f_b = n_b/(n_s+n_b)$. We set a $f_b = 0.05-0.2$ depending on the cluster model as summarized in Table \ref{tab:t1}. Our simulation grid contains two sets that differ only in $f_b$, thus their comparison could unveil some effects triggered by primordial binary dynamics. Note also our definition of $f_b$ implies that the number of stars in binaries over the total is $f_{\rm 2b} = 2f_b/(1+f_b)= 0.10-0.33$. 

Binaries are initialised assuming the same mass function of single stars and a uniform mass ratio distribution in the range $q=0.1-1$ for stars heavier than $m_*>5\Ms$ or random pairing for the lighter ones \citep{2012ApJ...751....4K,2012Sci...337..444S,2014ApJS..213...34K}. Following previous works on the same topics, we adopt a thermal distribution of the eccentricity and a semi-major axis distribution flat in logarithmic values, with an upper limit set to 50 AU and a lower limit set by the sum of the stars' radii \citep{2015MNRAS.450.4070W, 2022MNRAS.511.4060K}. 

In the majority of the cases, for each value of $R_{\ham}$ and $N$ we run two simulations with different random seeds to explore possible dependencies on the randomness of the star distribution. The only exception is the case $R_{\ham} = 0.47$ pc and $N = 300$k stars, which was limited to only one model because of the available computational time. 

The simulations are performed until either the average mass of stellar BHs falls below $\langle m_{\bh}\rangle \lesssim 15\Ms$, no BHs with a mass above $30\Ms$ are retained in the cluster, or the simulated time exceeds at least one relaxation time \citep{1987degc.book.....S, 2008gady.book.....B, 2021MNRAS.507.3312G}, which can be expressed in the form \citep{2021MNRAS.507.3312G}
\begin{equation}
    T_{\rm rlx} = 282 {\rm Myr} \frac{1}{m_* \ln{\gamma_n N}} \sqrt{\frac{M_c}{10^5\Ms}} \left(\frac{R_{\ham}}{1{\rm pc}}\right)^{3/2},
\end{equation}
where $\gamma_n = 0.11-0.4$ for a monochromatic mass spectrum \citep{1996MNRAS.279.1037G,2008gady.book.....B} but it can be as low as $\gamma_n=0.02$ for a multi-mass mass spectrum \citep{1996MNRAS.279.1037G}. These choices result in a physical simulated time ranging between $T_{\rm sim} \sim 0.1-2.3$ Gyr and lead to an optimal balance between the computational cost of the simulations and the portion of parameter space that can be explored. 
Table \ref{tab:t1} summarizes the main properties of \dragonii models. 

\begin{table*}
    \centering
    \begin{tabular}{ccccccc|cc|cc|cc|cc|cc|cc|cc}
    \hline
    \hline
        $N_*$ & $M_c$ & $R_h$ & $f_b$ &  $N_{\rm sim}$ & $T_{\rm rlx}$ &$T_{\rm seg}$ & \multicolumn{2}{|c}{$T_{\rm sim}$} & \multicolumn{2}{|c}{$N_{\rm GW, in}$}& \multicolumn{2}{|c}{$N_{\rm GW, out}$} & \multicolumn{2}{|c}{$M_{\rm max}$} & \multicolumn{2}{|c}{$M_{\rm max,fin}$} & \multicolumn{2}{|c}{$N_{>30}$} & \multicolumn{2}{|c}{$N_{>40}$} \\
        $10^3$ & $10^5\Ms$ & pc & &  & Myr & Myr & \multicolumn{2}{|c}{Myr} & \multicolumn{2}{|c}{}&\multicolumn{2}{|c}{}& \multicolumn{2}{|c}{$\Ms$} & \multicolumn{2}{|c}{$\Ms$}& \multicolumn{2}{|c}{}& \multicolumn{2}{|c}{}\\
    \hline
    120 & 0.7 & 1.75 & 0.05& 2& 99 & 2.1 & 2379 & 2326 & 0 & 2 & 2& 0&  64 &   76 & 25 &  34 &   0 &   2 &  0 &  0 \\
    300 & 1.8 & 1.75 & 0.05& 2& 142 & 2.7 & 1196 & 1422 & 0 & 2 & 2& 2&  69 &   77 & 40 &  40 &  13 &  13 &  5 &  1 \\
    1000& 5.9 & 1.75 & 0.05& 2& 233 & 3.4 &  207 &  194 & 1 & 1 & 4& 4&  81 &  146 & 52 &  70 & 149 & 169 & 72 & 85 \\
    120 & 0.7 & 1.75 & 0.2 & 2& 99 & 2.1 & 1710 & 1540 & 2 & 2 & 0& 2& 232 &   81 & 38 &  28 &   2 &   0 &  0 &  0 \\
    300 & 1.7 & 1.75 & 0.2 & 2& 142 & 2.7 &  519 &  793 & 1 & 0 & 7& 5&  92 &   77 & 65 &  47 &  26 &  26 &  8 & 14 \\
    600 & 3.5 & 1.75 & 0.2 & 2& 189 & 3.4 &  205 &  126 & 0 & 0 & 2& 5&  87 &  144 & 59 &  84 &  95 & 103 & 45 & 65 \\
    120 & 0.7 & 0.80 & 0.2 & 2& 30 & 0.7 & 1154 & 1201 & 4 & 3 & 4& 2& 120 &  132 & 21 &  27 &   0 &   0 &  0 &  0 \\
    300 & 1.7 & 0.80 & 0.2 & 2& 44 & 0.8 &  307 &  309 & 1 & 0 & 1& 0&  93 &  107 & 40 &  43 &  15 &  11 &  2 &  2 \\
    120 & 0.7 & 0.47 & 0.2 & 2& 14 & 0.3 & 1149 &  530 & 2 & 2 & 3& 1& 350 &   92 & 50 &  30 &   1 &   0 &  1 &  0 \\
    300 & 1.7 & 0.47 & 0.2 & 1& 30 & 0.4 &  148 &    - & 4 & - & 3& -&   245 &  - &  48 & - & 22 &   - &  9 &  - \\
    \hline
    \end{tabular}
    \caption{Col. 1-4: initial number of stars, cluster mass, half-mass radius, and primordial binary fraction. Col. 5: number of independent realisations. Col. 6-7: initial relaxation and segregation time. Col. 8: simulated time. Col. 9-10: number of mergers inside the cluster. Col. 11: maximum BH mass during the simulation. Col. 12: maximum BH mass at the end of the simulation. Col. 13-14: number of BHs with a mass $m_{\rm BH}>30\Ms$ or $>40\Ms$ at the last simulation snapshot.}
    \label{tab:t1}
\end{table*}

\begin{figure}
    \centering
    \includegraphics[width=\columnwidth]{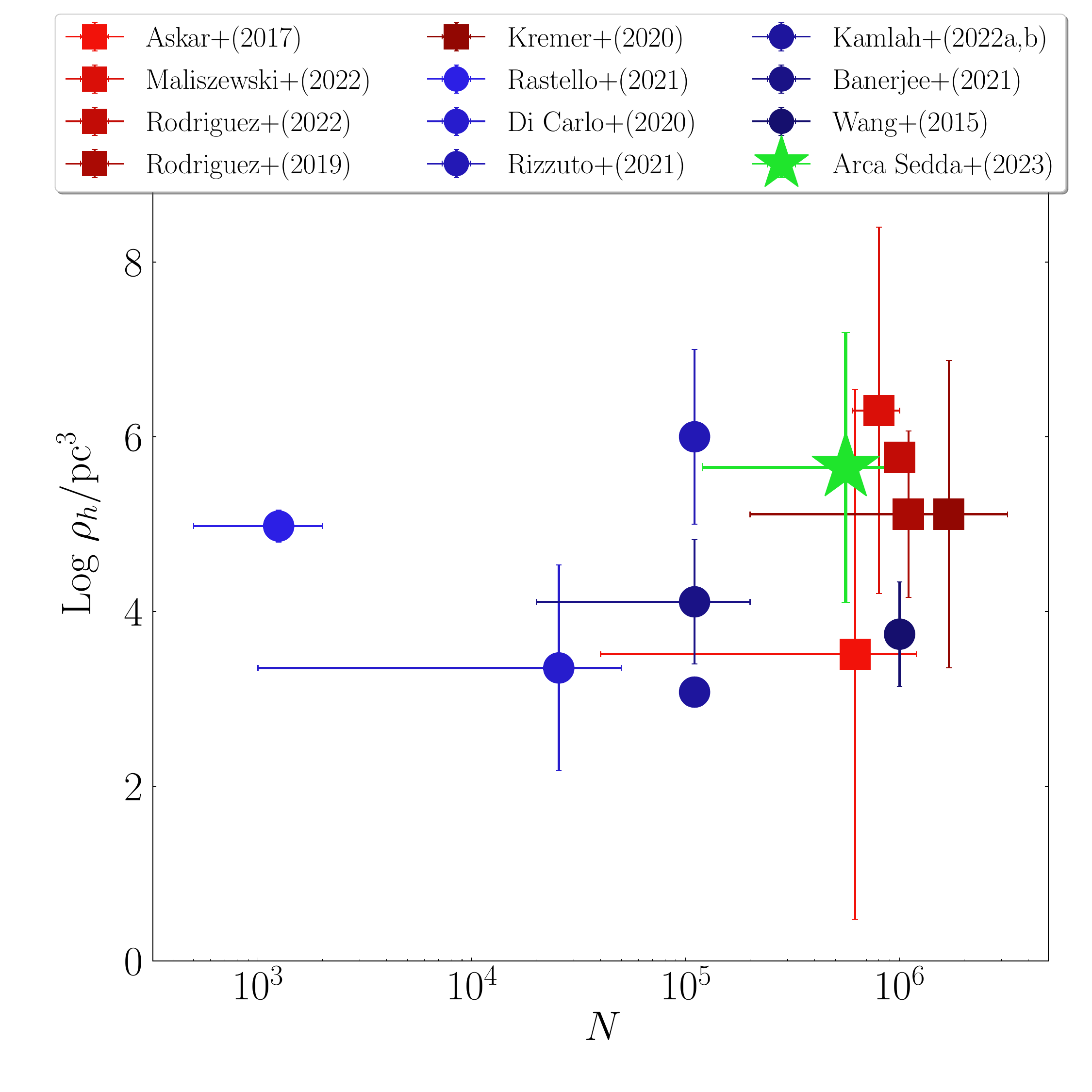}
    \caption{Initial density, calculated at the half-mass radius, as a function of number of stars for several grids of direct $N$-body (blue points) and Monte Carlo (red boxes) simulations. The \dragonii cluster database is represented by the green star.}
    \label{fig:f1}
\end{figure}

As sketched in Figure \ref{fig:f1}, in comparison to the most recent studies based on $N$-body \citep[e.g.][]{2015MNRAS.450.4070W, 2018MNRAS.473..909B, 2021MNRAS.500.3002B, 2019MNRAS.487.2947D,2021MNRAS.507.3612R,2021MNRAS.507.5132D} and Monte Carlo simulations \citep[e.g.][]{2016PhRvD..93h4029R,2017MNRAS.464L..36A,2019PhRvD.100d3027R,2020ApJS..247...48K,2022MNRAS.514.5879M}, the \dragonii clusters occupy a region of the $N$-$\rho_h$ plane mostly populated by Monte Carlo simulation grids. This, coupled with the fact that simulations with $N>10^5$ stars usually adopt a binary fraction $<20\%$, makes our \dragonii simulations an unprecedented grid of models that complements, and further expands, the phase space accessible with direct $N$-body models.

\section{Results}
\label{sec:res}

\subsection{Star cluster evolution}
\label{sec:clu}

\begin{figure*}
    \centering
    \includegraphics[width=0.8\textwidth]{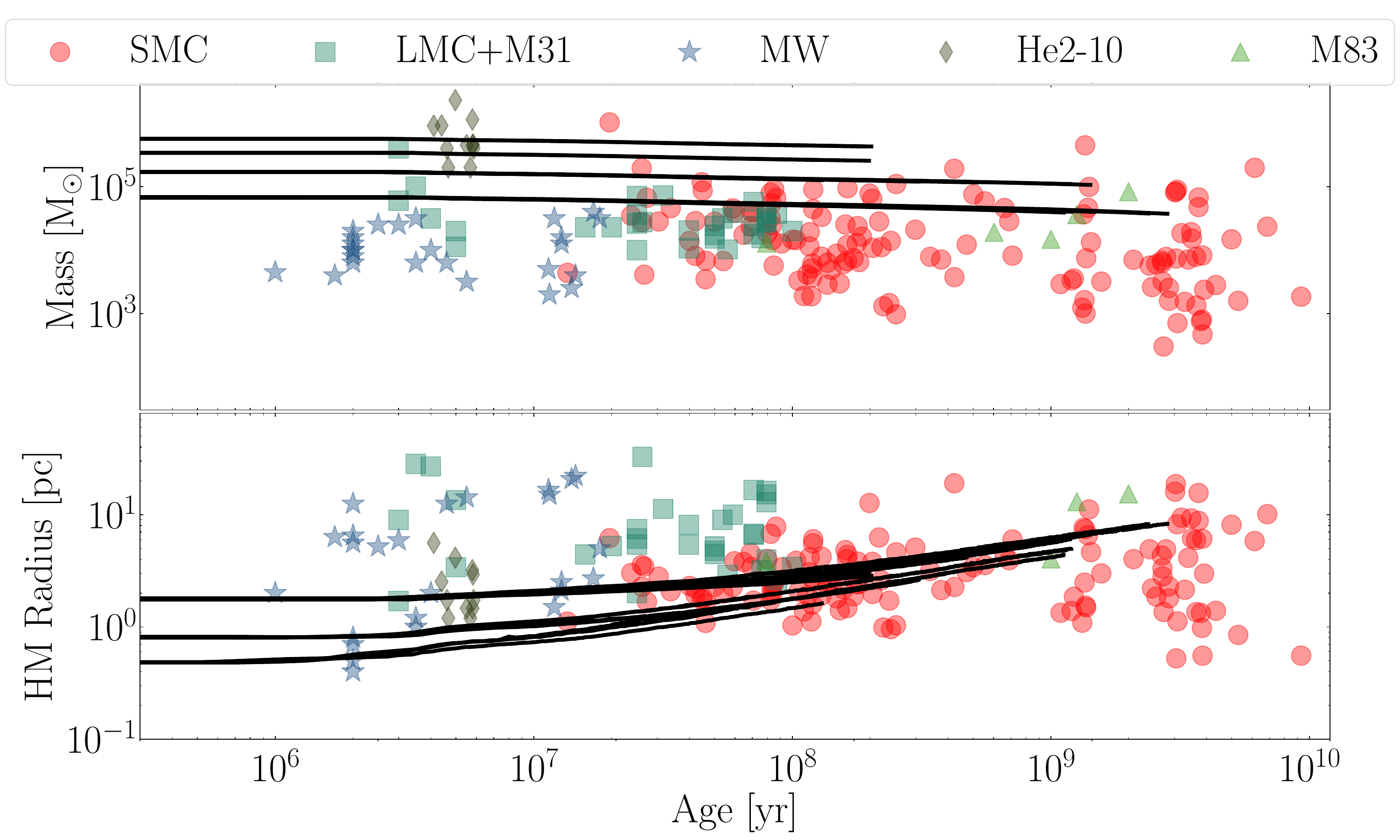}\\
    \caption{Time evolution of the mass (top panel) and half-mass radius (bottom) of \dragonii clusters (black lines) compared to observed massive clusters in the Milky Way (MW, blue stars), the two Magellanic Clouds (green squares and red points), the Andromeda galaxy (M31, green squares), the Henize 2-10 starburst galaxy (He2-10, grey diamonds), and the M83 galaxy (light green triangles).}
    \label{fig:f2}
\end{figure*}

The \dragonii clusters were originally devised to explore compact object dynamics, compact binary mergers, and intermediate-mass black hole build-up in dense star clusters, thus they are not meant to be representative of any observed cluster. 
Nonetheless, it is interesting to compare in Figure \ref{fig:f2} the time evolution of the modelled mass and half-mass radius with relatively young, i.e. typical ages $0.1-1$ Gyr, massive star clusters in the Milky Way (MW), the Small (SMC) and Large Magellanic Cloud (LMC), M31 \citep{2010ARA&A..48..431P,2021MNRAS.507.3312G}, the Henize 2-10 starburst dwarf galaxy \citep{2014ApJ...794...34N}, and the M83 galaxy \citep{2015MNRAS.452..525R}. Over the simulated time, our models overlap with observed clusters, thus indicating that the adopted initial conditions lead to numerical models that can represent one possible evolutionary pathway of some observed clusters.

We find that the mass and half-mass radius evolution is well described by the following relations:
\begin{eqnarray}
M_{\rm cl}(t) &=& M_{\rm cl,0}\left[1 + \alpha_M\left(\frac{t}{T_{\rm rlx}}\right)^{-\beta_M}\right],\label{eq:Mtime}\\
R_\ham(t) &=& R_{\ham,0}\left[1+\frac{t}{\alpha_R T_{\rm rlx}}\right]^{\beta_R}. \label{eq:rhalftime}
\end{eqnarray}

\begin{table}
    \centering
    \begin{tabular}{ccccc}
    \hline\hline
        $R_\ham$ [pc] & $\alpha_M$   & $\beta_M$  & $\alpha_R$ & $\beta_R$     \\
    \hline
        $1.75$     &  $ 0.29-0.33$& $0.28-0.31$& $0.65-0.81$& $0.45-0.59$   \\
        $0.80$     &  $ 0.18-0.19$& $0.34$     & $0.7$      & $0.57$        \\
        $0.47$     &  $ 0.15$     & $0.35-0.4$ & $0.78-1.1$ & $0.44-0.56 $  \\
    \hline
    \end{tabular}
    \caption{Fitting parameters in Equations \ref{eq:Mtime}-\ref{eq:rhalftime}.}
    \label{tab:fit}
\end{table}

The values of the fitting parameters, which are summarised in Table \ref{tab:fit}, are independent of the initial cluster mass, and weakly depend on the initial value of the half-mass radius. This owes to the fact that the mass-segregation time scales with $M_c^{1/2} R_\ham^{3/2}$, thus it is mostly affected by the choice of the half-mass radius.

\begin{figure}
    \centering
    \includegraphics[width=\columnwidth]{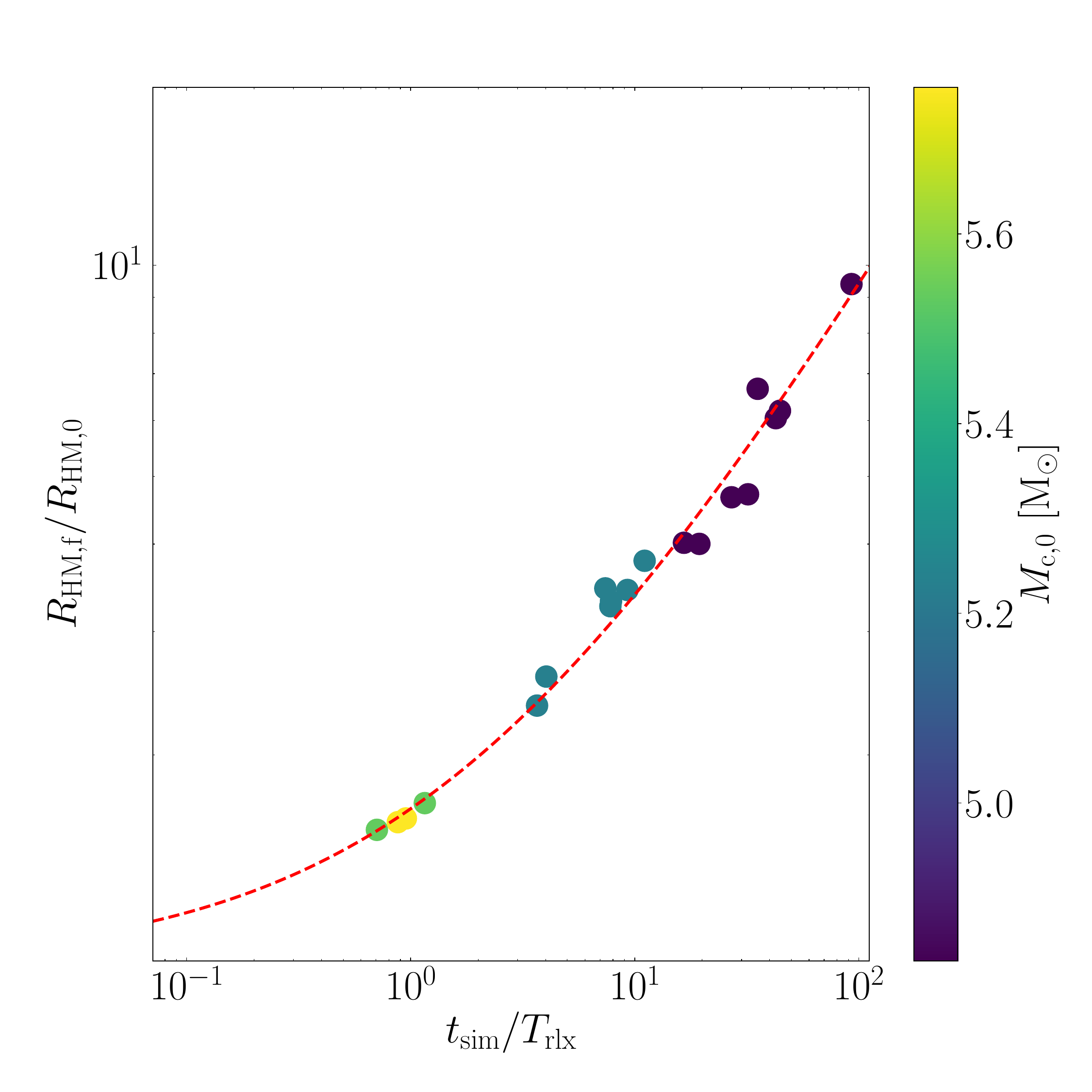}
    \caption{Half-mass radius calculated at the end of each simulation, normalised to the initial value, as a function of the total simulated time normalised to the cluster initial relaxation time. The dashed line is a power-law fit of the data, while each point represents one simulation, with the colour map representing the initial cluster mass.}
    \label{fig:rhrlx}
\end{figure}

Figure \ref{fig:rhrlx} shows the ratio between the final and initial values of $R_\ham$ as a function of the simulated time, normalised to the initial relaxation time. The plot clearly highlights how the cluster expansion depends only on the dynamical age of the cluster, regardless of the initial cluster mass.

By the end of the simulations, our clusters have typically lost $\sim 25-50\%$ of their initial mass and their radius has expanded by a factor of $1.5-10$, thus implying a reduction of the density at the half-mass radius by up to four orders of magnitude and a reduction of the velocity dispersion of around $1-1.5$ times. The drop in density and velocity dispersion crucially affects the rates at which dynamical interactions take place.

A thorough comparison among \dragonii simulations and the models discussed in the past literature is made hard by the many different assumptions of previous works, like the use of equal-mass stars to represent the cluster, the different binary fraction, the properties of the primordial binary population, the lack of a dedicated treatment to deal with compact binaries, and the use of outdated prescriptions for the evolution of massive stars ($m_{\rm ZAMS} > 50\Ms$). 

In order to test the new features of the \nbsix code, we have carried out an extensive comparison of the evolution of star clusters with 110,000 stars in $N$-body and Monte Carlo simulations in our companion paper \citep{2022MNRAS.511.4060K}, where we have shown, among other things, that $N$-body models of the same clusters seem to evolve toward sparser configurations compared to Monte Carlo models with large tidal radii simulated with the \textsc{MOCCA} code. This difference is likely due to the different criteria used to identify escapers in the two methods, which can lead to an early removal of escaping stars in \textsc{MOCCA} simulations compared to \nbsix.

\subsection{Stellar and compact object binaries}
\label{sec:cob}

\begin{figure}
    \centering
    \includegraphics[width=\columnwidth]{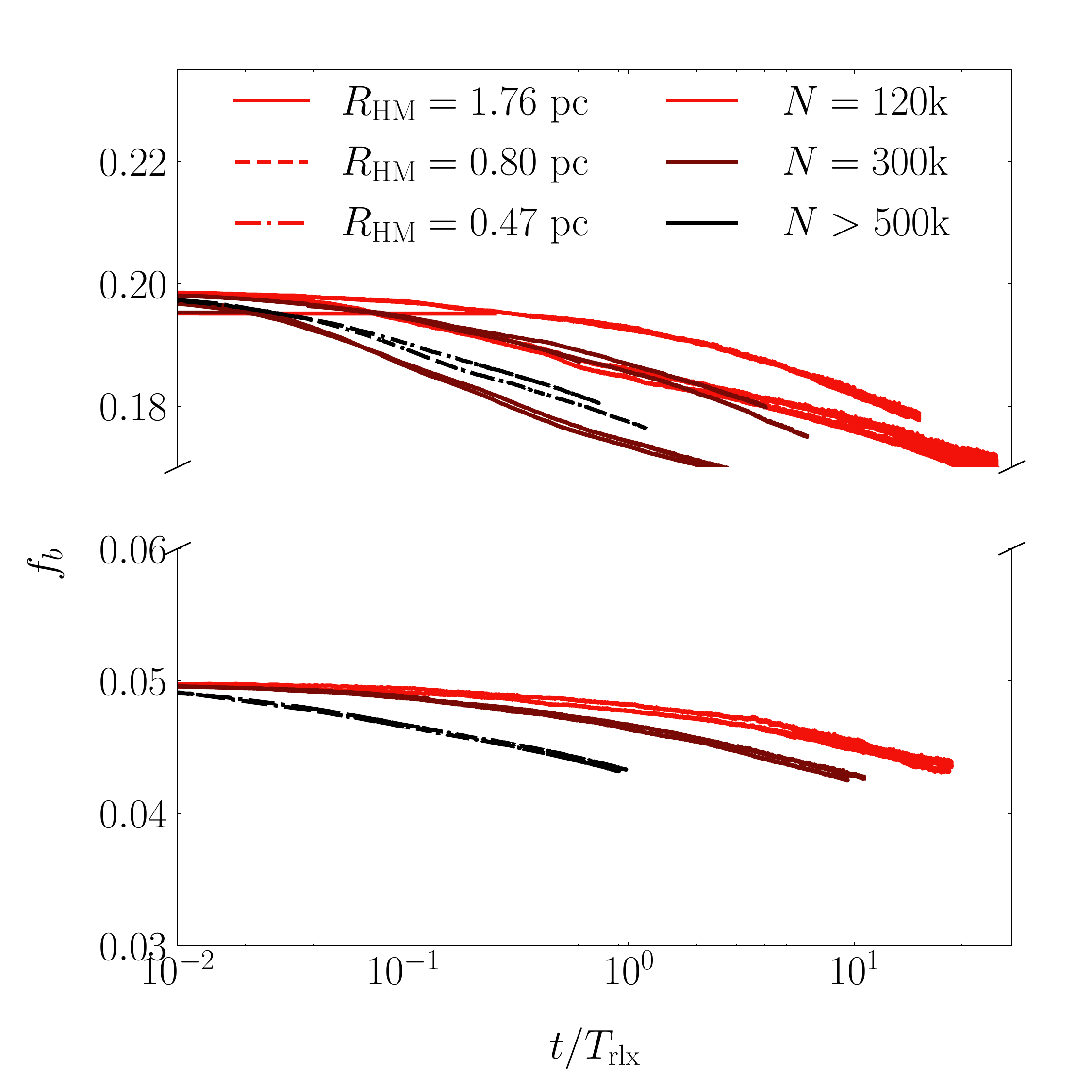}
    \caption{Binary fraction as a function of  time normalised to the relaxation time, taking into account all binaries.}
    \label{fig:bins}
\end{figure}

\begin{figure}
    \centering
    \includegraphics[width=0.9\columnwidth]{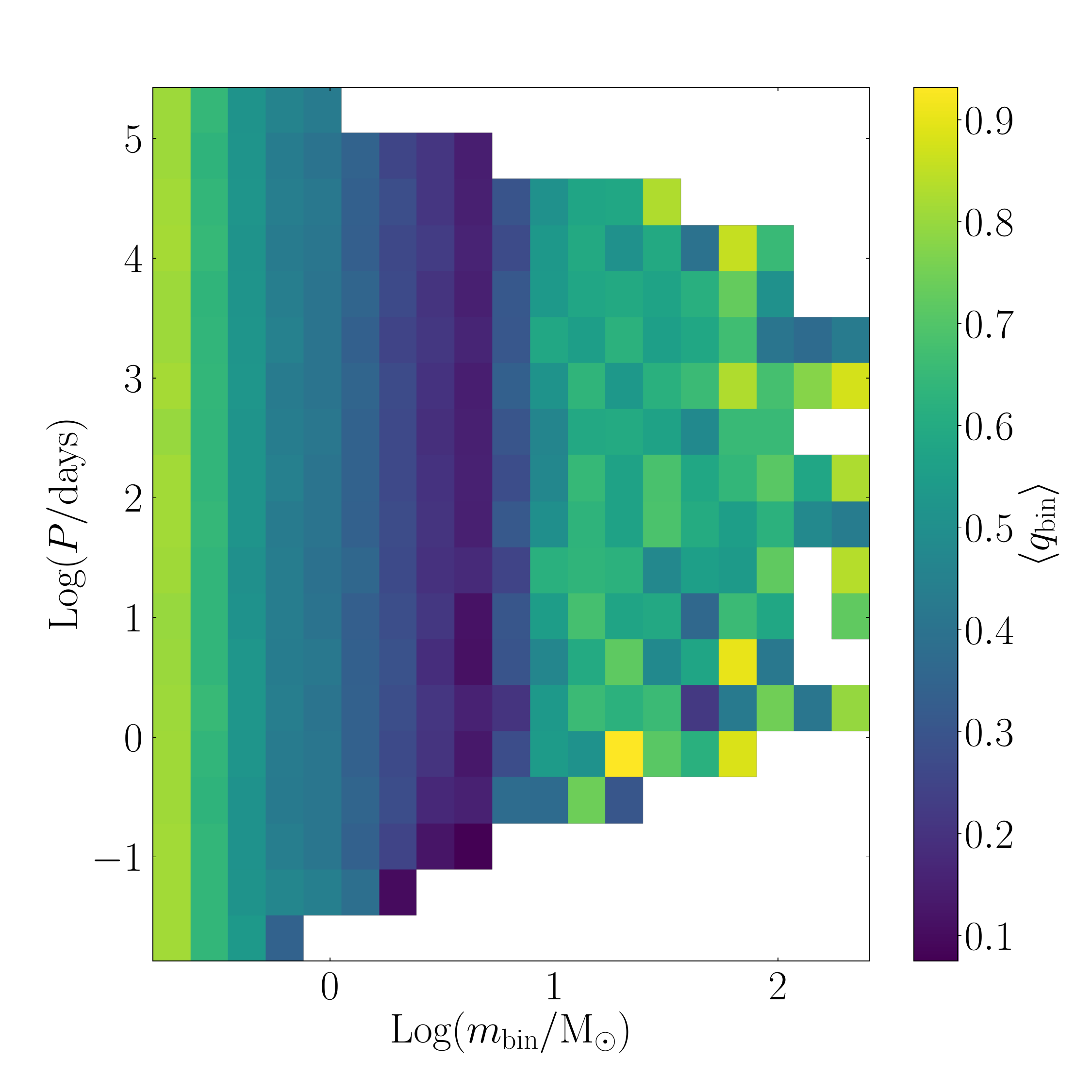}\\
    \caption{Initial values of total mass (x-axis) and period (y-axis) for all primordial binaries in one of the models with $R_\ham = 1.75$ pc, $N=120$k, and $f_b=0.2$. The colour map marks the mean value of the mass ratio inside each pixel.}
    \label{fig:smain}
\end{figure}

Mass-segregation of the most massive stars enhances strong dynamical interactions, which can trigger the ejection of the tightest binaries, the ionisation of the loosest ones, and the formation and hardening of new binaries. In the \dragonii clusters, the processes responsible for the formation and disruption of binaries counterbalance efficiently, determining a slow variation of the overall binary fraction. As shown in Figure \ref{fig:bins}, the binary fraction decreases by a small fraction, down to $f_{b,fin} \sim 0.16-0.18$ in models starting with $f_b=0.2$ and to $f_{b,fin}=0.04-0.05$ in models with $f_b = 0.05$. Interestingly, this variation in the binary fraction is similar, within the simulation time, to results obtained for lower-$N$ cluster simulations \citep[see e.g.][]{2006MNRAS.368..677H}.
The decrease of the binary fraction is mostly due to the disruption of the softest binaries in the cluster and, for a small fraction ($< 5\%$), to hard binaries that are ejected in strong dynamical interactions. These binaries have typical semi-major axes broadly distributed in the $10^{-2}-5\times 10^2$ AU. 
For the sake of comparison, Figure \ref{fig:smain} shows the initial period-mass distribution and mass-ratio of the population of primordial binaries in our models.

\begin{figure*}
\centering
\includegraphics[width=0.47\textwidth]{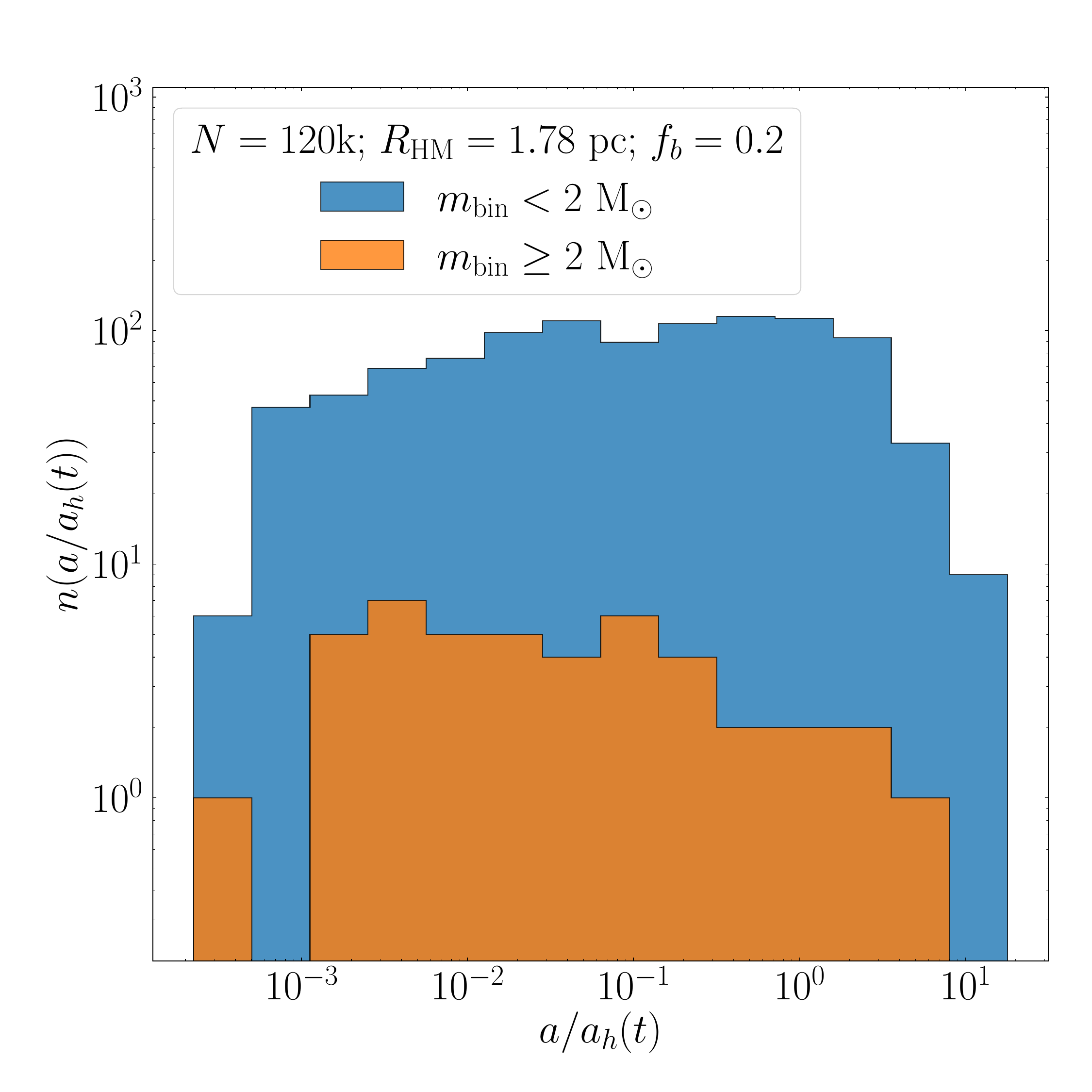}
\includegraphics[width=0.47\textwidth]{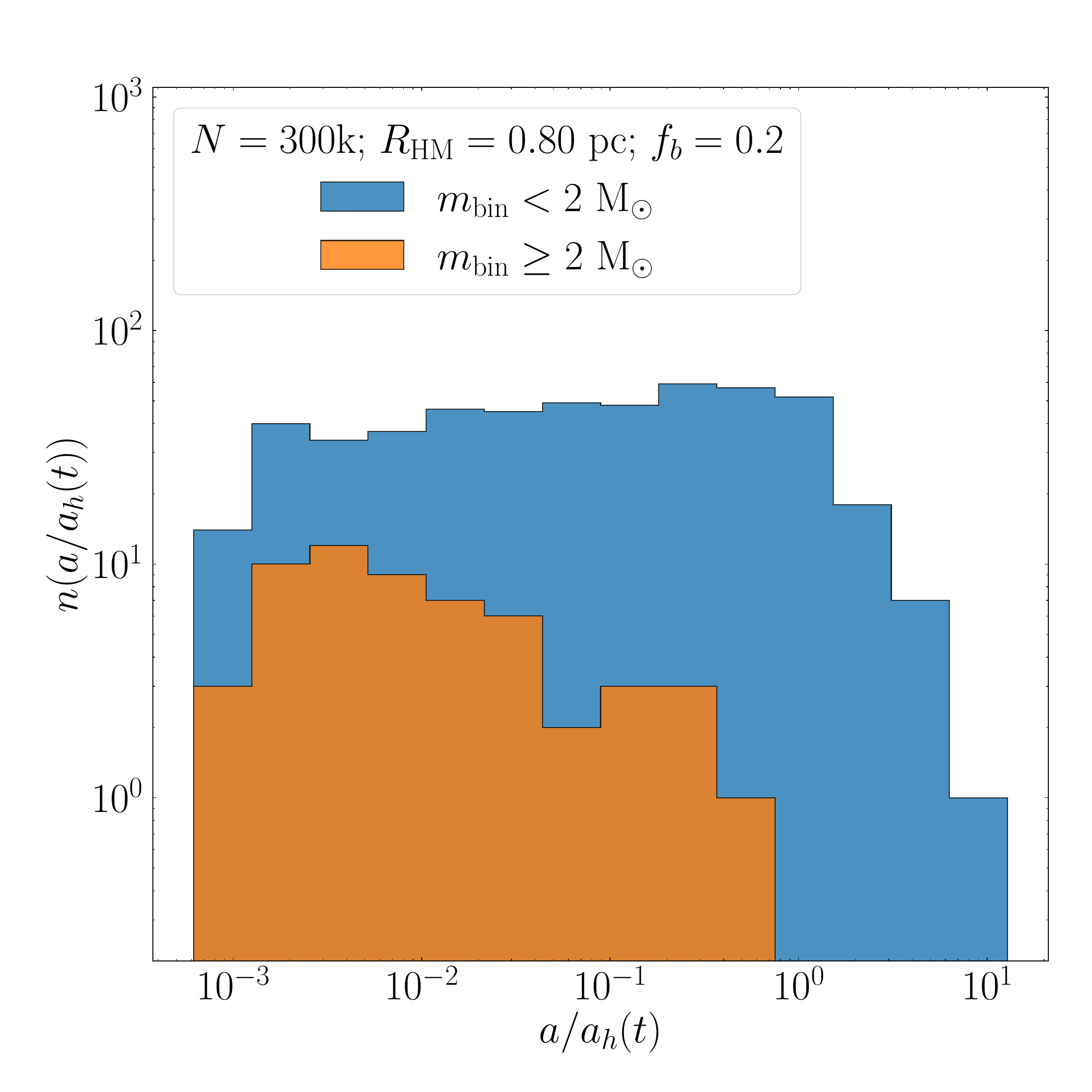}
\includegraphics[width=0.47\textwidth]{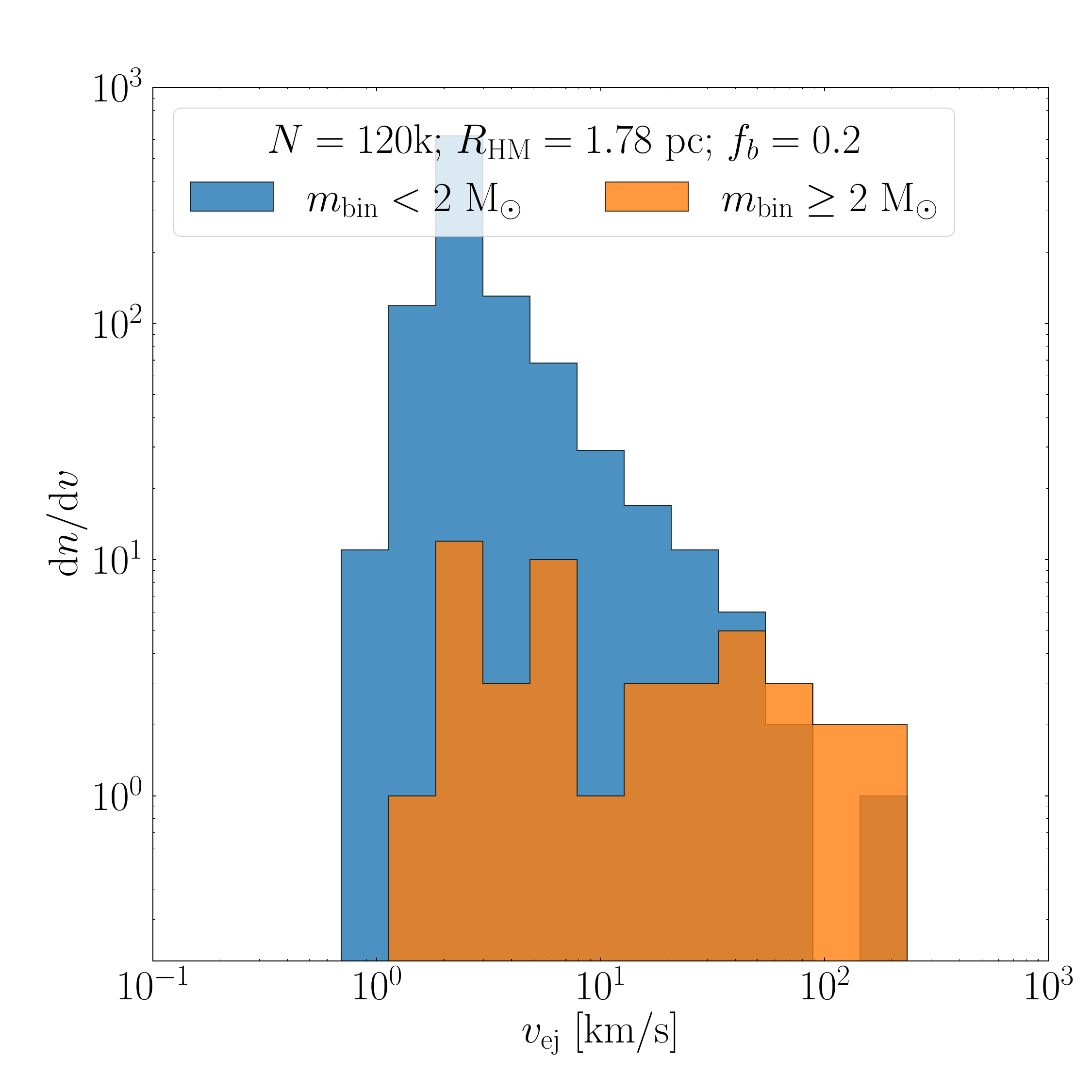}
\includegraphics[width=0.47\textwidth]{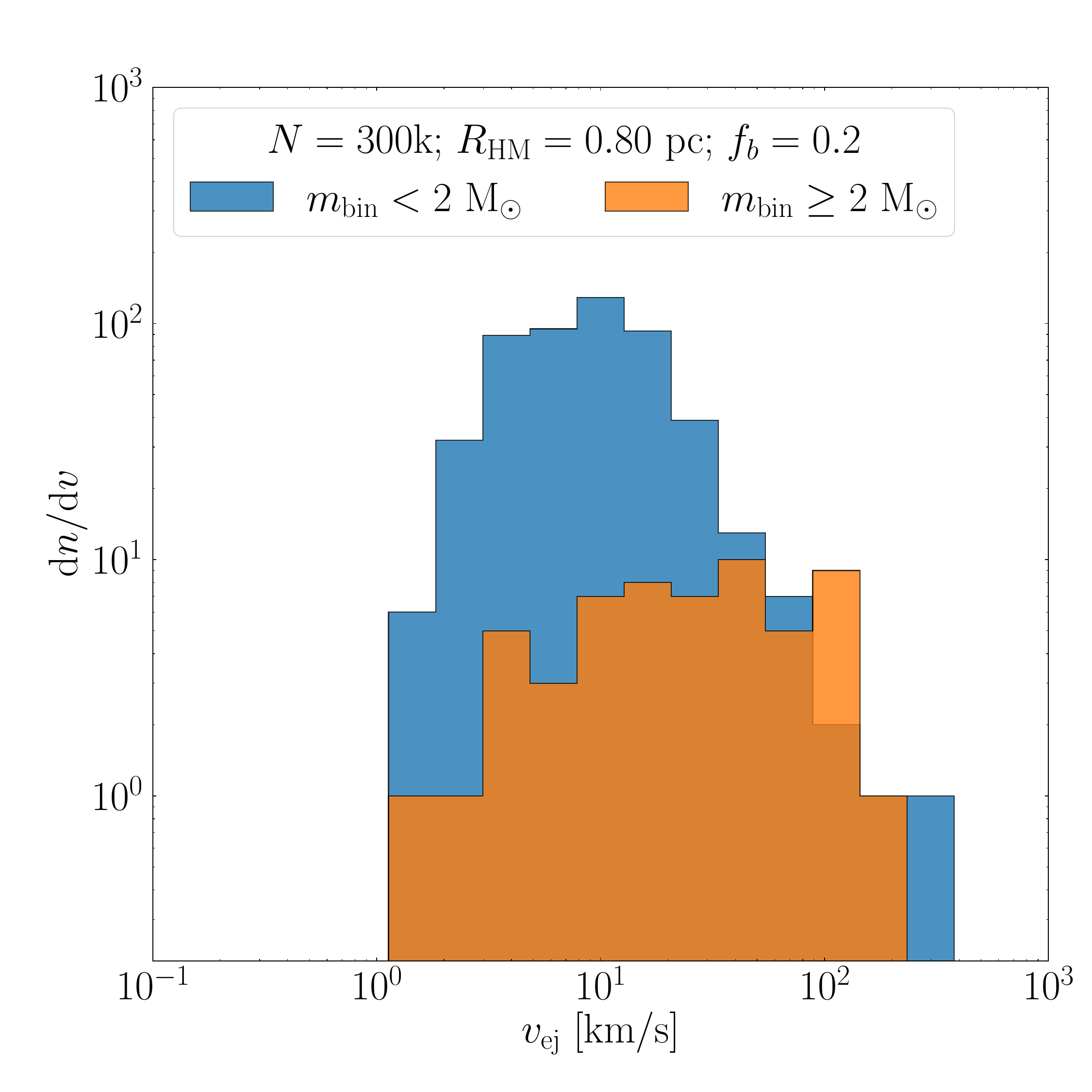}
\caption{Upper panels: Distribution of the semi-major axis of ejected binaries normalised to the hard-binary separation measured at the moment of ejection for binaries with a mass $m_{\rm bin} < 2\Ms$ (blue steps) or heavier (orange steps). Lower panels: ejection velocities for the same classes of binaries. The plots refer to two simulations with $(R_\ham,~ N) = (1.78{\rm ~pc},~ 120{\rm k})$ (left-hand panels) and ($0.8{\rm ~pc},~300{\rm k}$) (right-hand panels).}
\label{fig:binej}
\end{figure*}

Figure \ref{fig:binej} shows the distribution of the ratio between the semi-major axis of ejected binaries and the hard-binary separation, both measured at the moment of the ejection, and the ejection velocity distribution for two different simulations. The plot makes clear that the vast majority of ejected binaries are hard and that this population is dominated mostly by binaries with a mass $m_{\rm bin} < 2\Ms$. The velocities of the ejected binaries generally remain in the range of $1-100$ km s$^{-1}$, too small compared to the circular velocity of the Galaxy to permit the identification of these escapers as former cluster members.

\begin{figure}
\includegraphics[width=\columnwidth]{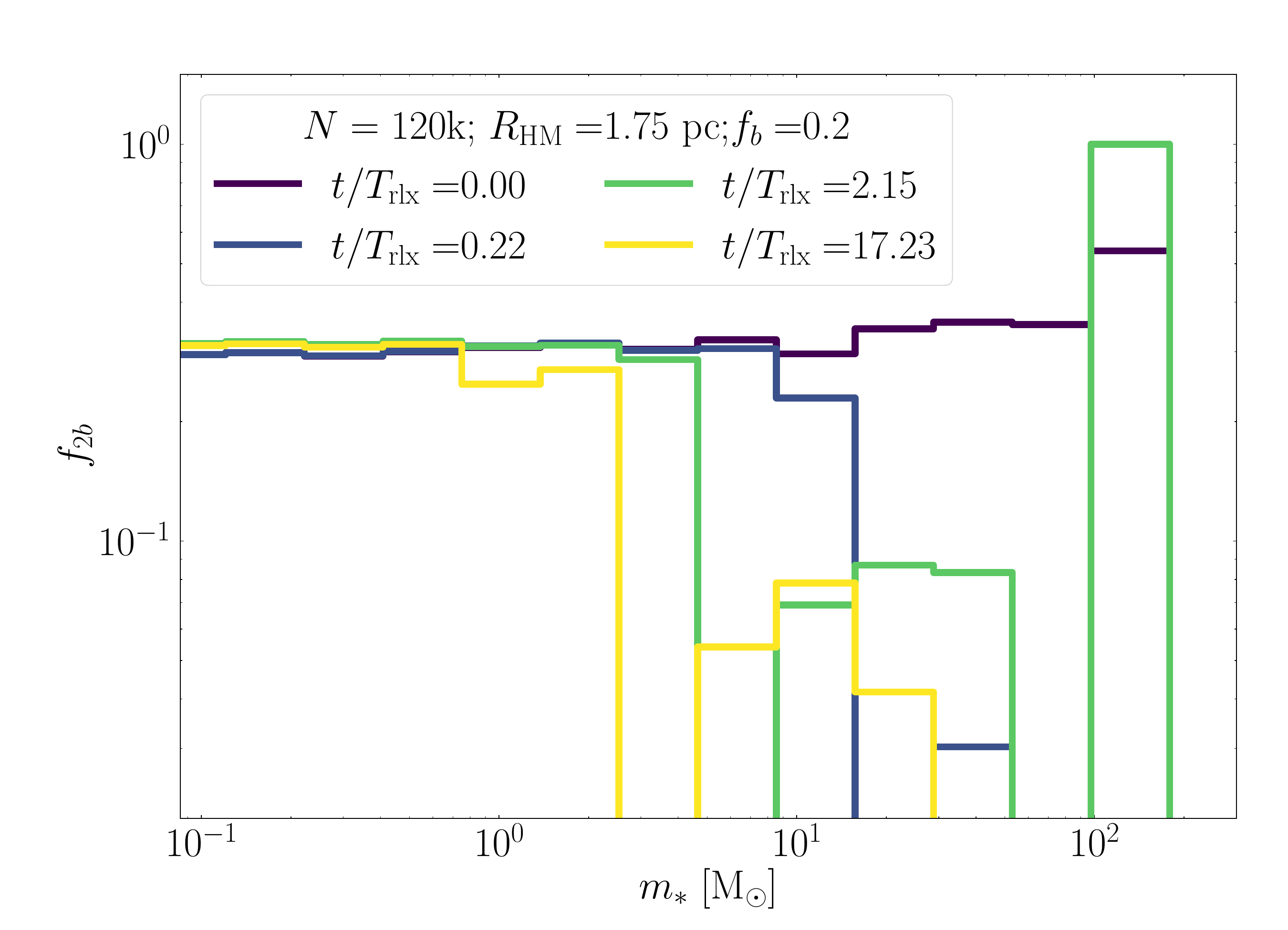}\\
\includegraphics[width=\columnwidth]{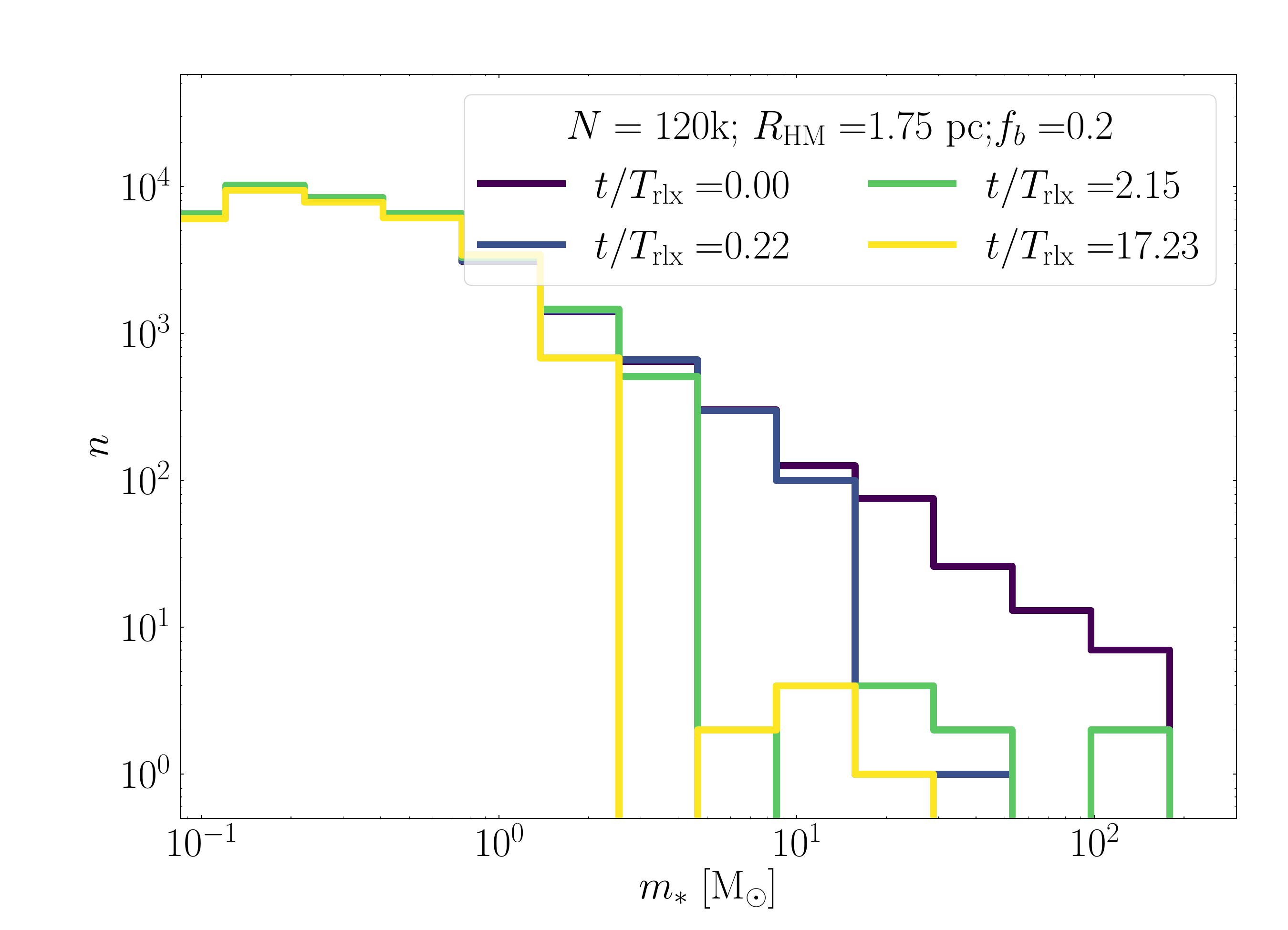}
\caption{Top: fraction of objects in binaries for different mass bins and different times. Bottom: number of binaries with mass in a given bin at different times. We show a simulation with $f_b=0.2$, $N=120$k stars, and $R_\ham=1.75$ pc.}
\label{fig:bmass}
\end{figure}

The upper panel of Figure \ref{fig:bmass} shows the variation of the fraction of binaries normalised to the total number of stars in a given mass bin and at a given time. Initially, around $35-50\%$ of all stars with a mass above $20\Ms$ are initially binary members, with the maximum percentage achieved for stars heavier than $100\Ms$. However, the population of heavy objects is rapidly depleted (note that $t/T_{\rm rlx} = 0.22$ corresponds in this case to $t = 18.8$ Myr) owing mostly to stellar/binary evolution, which causes a sharp drop in their number. The maximum stellar mass keeps decreasing over time, whilst a small population of binaries with components in the $5-100\Ms$ develops -- clearly owing to the formation of binaries with one or two BHs. The mass distribution of objects in binary systems, shown in the lower panel of Figure \ref{fig:bmass}, highlights that the number of binaries with at least one component heavier than $10\Ms$ is relatively small compared to the total number of objects in binaries. Assuming initially $N=120,000$ stars and $f_b=0.2$, we see that less than $1,000$ binaries contain a component with a mass $m_* > 10\Ms$, most of them being former components of a primordial binary.

The progenitors of compact objects, which are the most massive stars and stellar binaries in the cluster, have already sunk into the cluster centre when compact objects form. Therefore, to dissect the properties of compact binaries in \dragonii clusters, we focus on binaries forming within the cluster half-mass radius, calculated along the cluster evolution. 

\begin{figure*}
    \centering
    \includegraphics[width=\columnwidth]{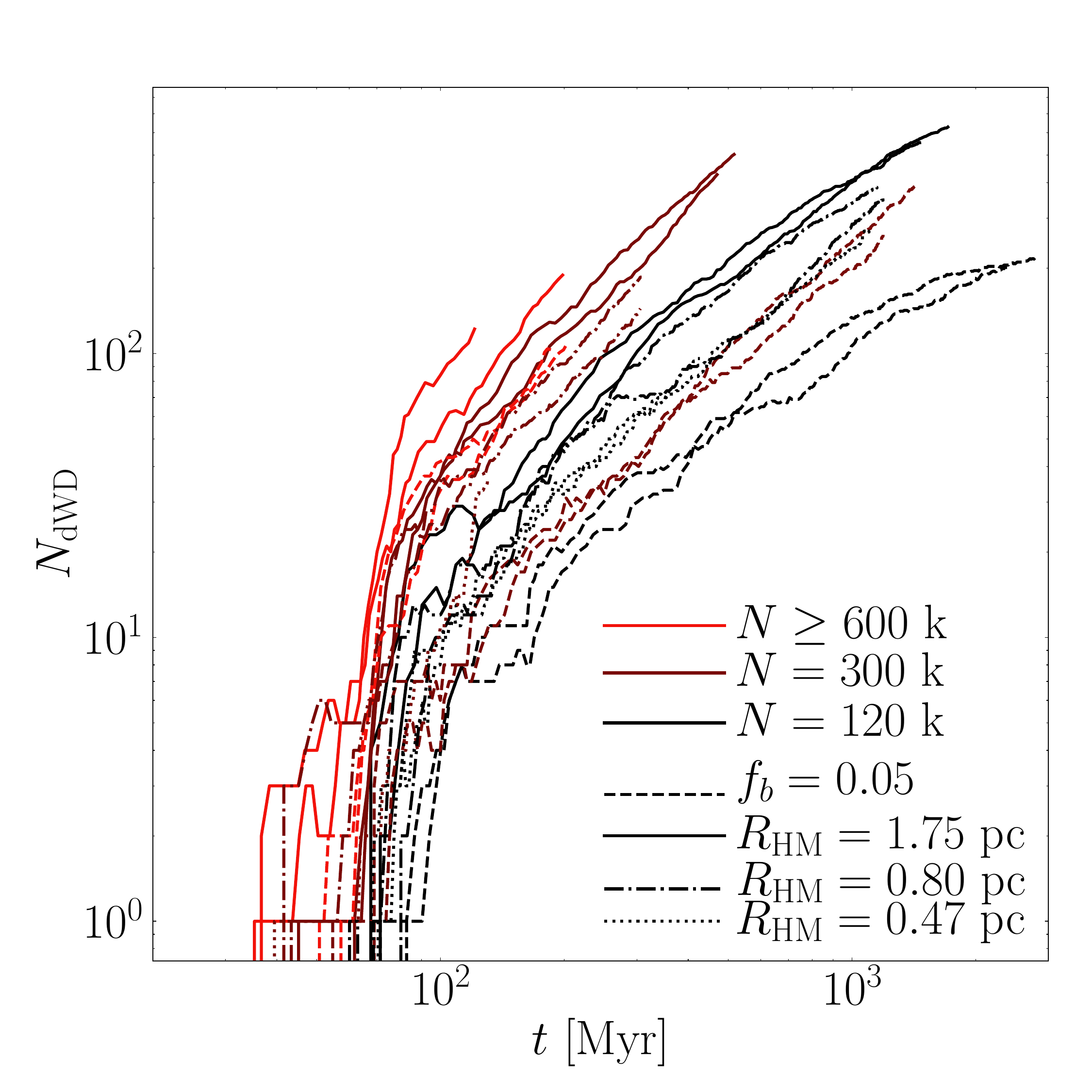}
    \includegraphics[width=\columnwidth]{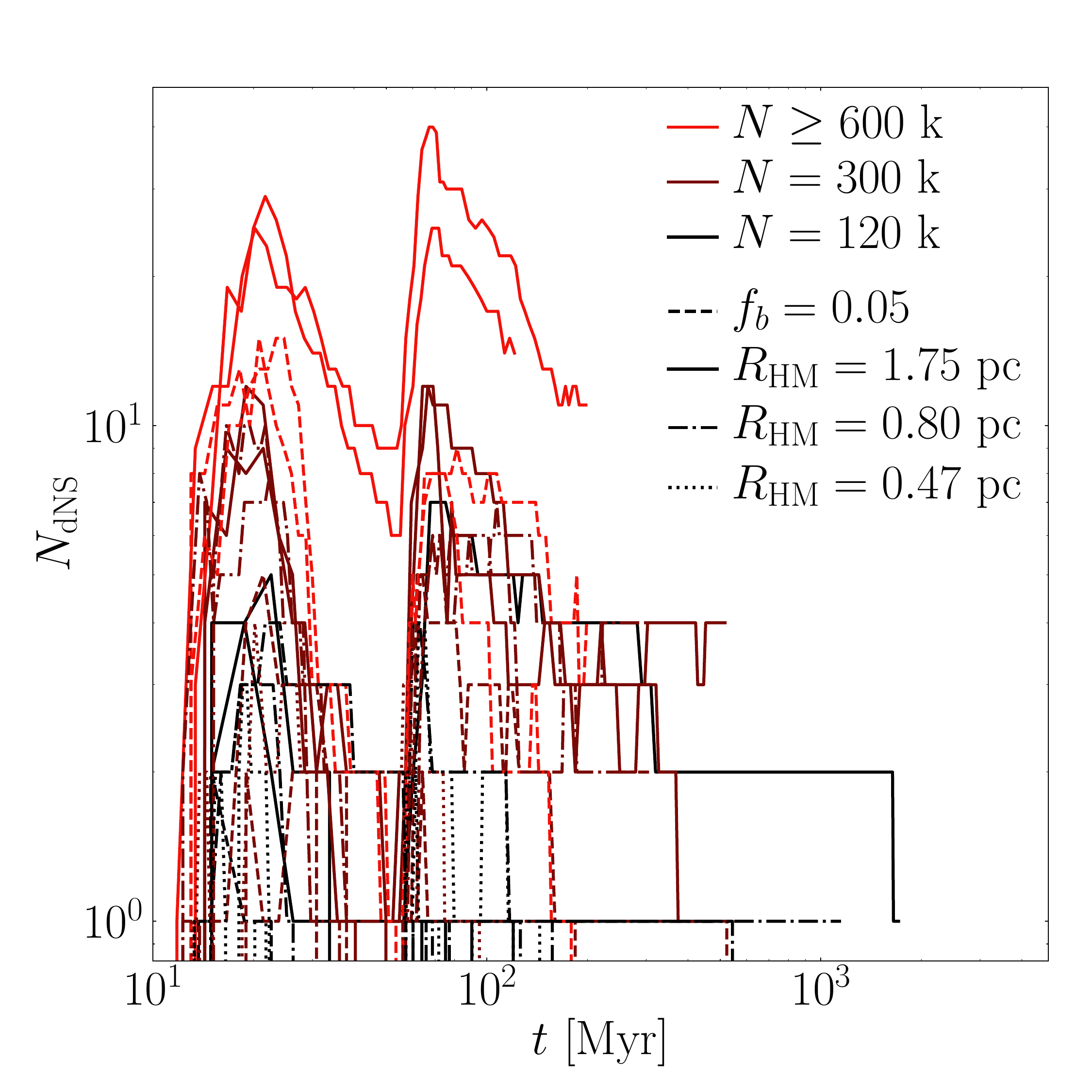}\\
    \includegraphics[width=\columnwidth]{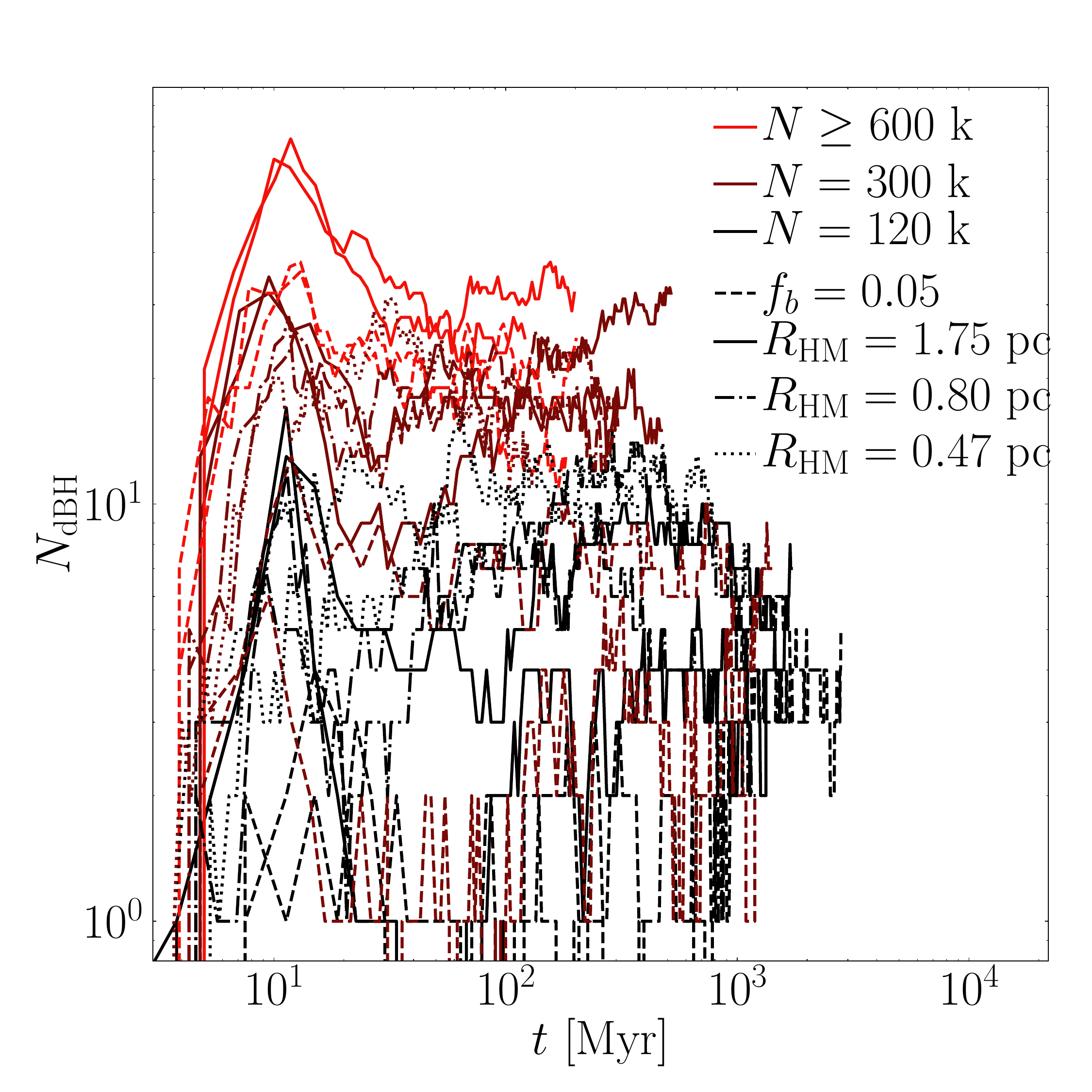}\\
    \caption{Number of binaries with at least one WD (upper panel), a NS (central panel), or a BH (lower panel) as a function of time for all \dragonii clusters. Different colours correspond to different values of the initial half-mass radius. There are two lines for each colour, corresponding to two realisations of the same cluster.}
    \label{fig:DCO}
\end{figure*}

Figure \ref{fig:DCO} shows the number of binaries with a WD, NS, or BH as a function of time for all models.

The population of binaries containing at least one WD (dWDs), $N_{\rm dWD}$, depends on the half-mass radius and binary fraction. At fixed half-mass radius, the number of binaries with a WD significantly decreases at decreasing $f_b$, because most of these binaries are of a primordial origin. In fact, at fixed $N$ stars and $R_\ham$, the ratio between the number of dWDs is $4-5$ times higher in models with $f_b=0.2$ compared to those with $f_b=0.05$, thus comparable to the ratio between the initial amount of primordial binaries in one case or the other. At fixed value of $f_b$, instead, the smaller the half-mass radius, the smaller is the number of dWDs. In general, by the end of the simulations we find $N_{\rm dWD}\simeq 200-700$ dWDs per cluster. The amount of binaries with a WD monotonically increases over the simulated time, highlighting the competition between WD depletion via dynamical encounters and the formation of new WDs, mostly via binary stellar evolution \citep[see also Figure 4 in][]{2022MNRAS.511.4060K}.

The evolution of the number of binaries with a NS (dNS) shows two clear peaks at $20$ and $\sim 100$ Myr. These peaks correspond to the formation of NSs from stars in the high-end (the first) and low-end (the second) of the NS progenitor mass range. The drop after each of the peaks is due to NS natal kicks, which cause the ejection of a substantial fraction of NSs from the parent cluster. The width of the peaks is  related to the time needed for the NS to leave the cluster, i.e. when their distance from the cluster centre exceeds twice the tidal radius. After the second peak, the number of binaries with a NS decreases in all simulations, regardless of the initial conditions. We find that the largest value of $N_{\rm dNS}$ is reached in the case of $R_\ham=1.75$ pc, $f_b=0.2$, and $N=600$k. At fixed value of $R_\ham$ and $N$ we find that a larger initial binary fraction leads to a more numerous population of binaries with a NS, around $50\%$ more for models with $f_b = 0.2$. At fixed value of $N$ and $f_b$ the number of binaries with a NS increases at increasing values of $R_\ham$ because in denser clusters it is more likely that massive stellar binaries either are ejected or merge before stellar evolution becomes dominant.

The population of binaries with a BH (dBH), similarly to those with a NS, are characterised by two peaks of formation, one at around $10$ Myr, driven by stellar evolution, and another at later times driven by dynamics. The number of binaries with a BH, $N_{\rm dBH}$, in the primary peak depends on the initial number of stars -- the larger $N_0$ the larger $N_{\rm bBH}$, whilst the number in the secondary peak depends on both the half-mass radius and binary fraction, although it is hard to discern the effects of different initial conditions in this case. 
the \dragonii clusters,

\subsection{Ejection of single and double compact objects}

\begin{figure}
    \centering
    \includegraphics[width=\columnwidth]{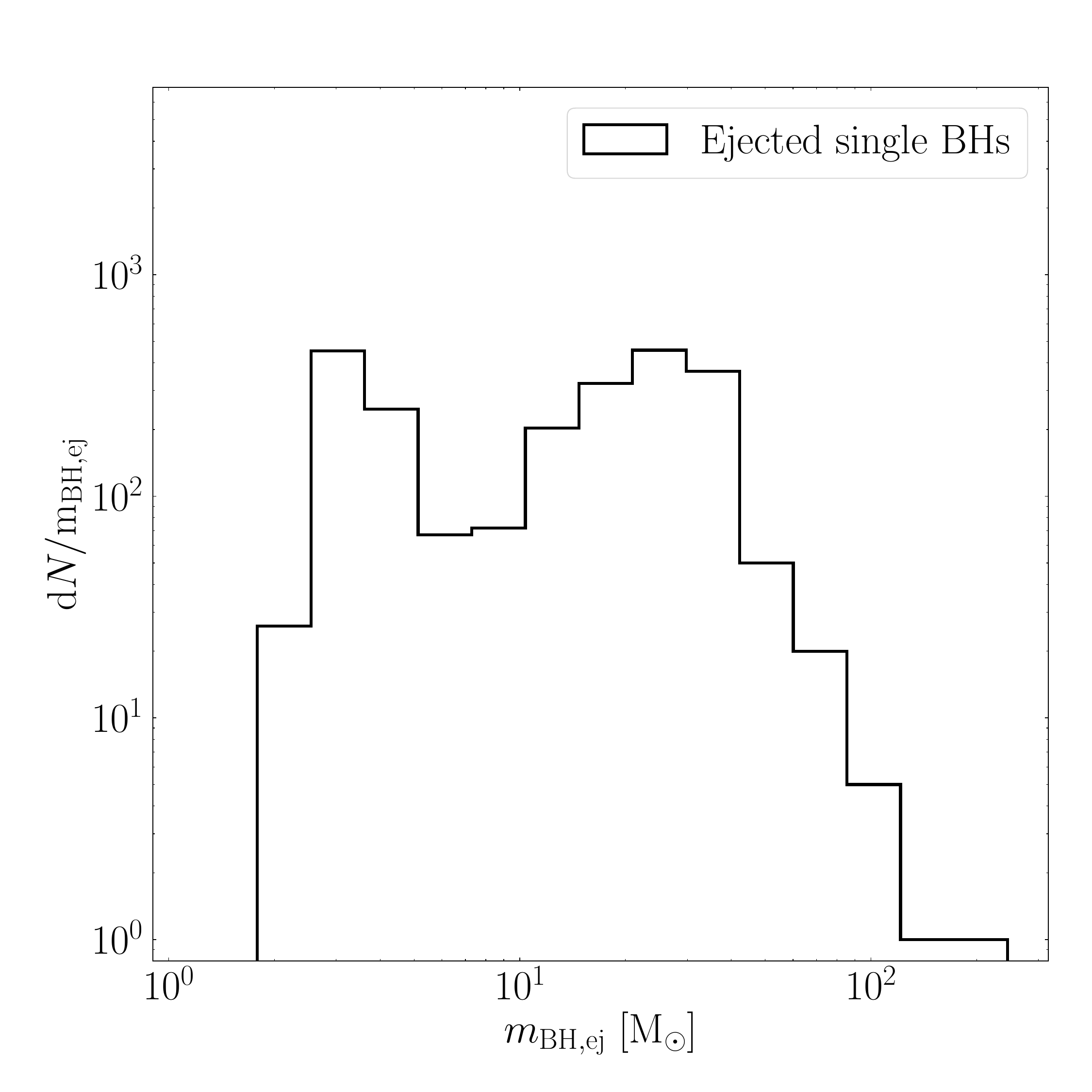}
    \caption{Mass distribution of ejected BHs in all \dragonii simulations.}
    \label{fig:sinBH}
\end{figure}

Over the simulated time, all clusters lose around 20--70 single BHs, depending on the cluster initial conditions, and 10--70 binaries containing either one or two compact objects. Figure \ref{fig:sinBH} shows the mass distribution of ejected single BHs, which is characterised by two  peaks, one at $m_{\rm BH} \sim 3\Ms$ and another at $m_{\rm BH} \sim 25\Ms$, and a tail that extends up to $m_{\rm BH} \sim 10^2\Ms$.
The first peak is due to the natal kick of NSs and low-mass BHs, with masses in the range $m_{\rm BH} = 2.5-6\Ms$, and develops in the first 10--50 Myr, whilst the secondary peak is due to dynamical interactions\footnote{In our simulations the minimum mass allowed for BHs is $m_{\rm BH,min} = 2.5\Ms$}.  

\begin{figure*}
    \centering
    \includegraphics[width=0.95\columnwidth]{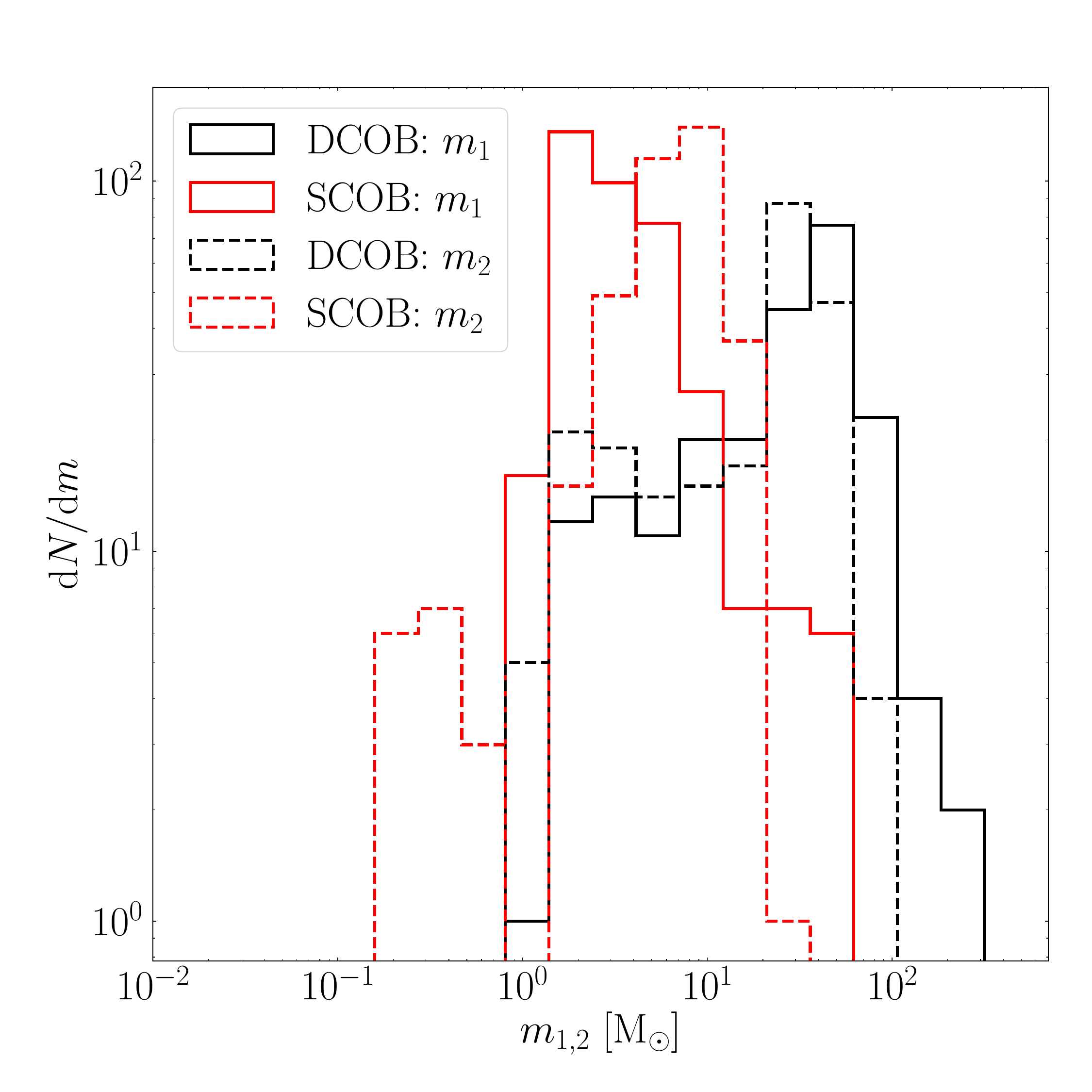}
    \includegraphics[width=0.95\columnwidth]{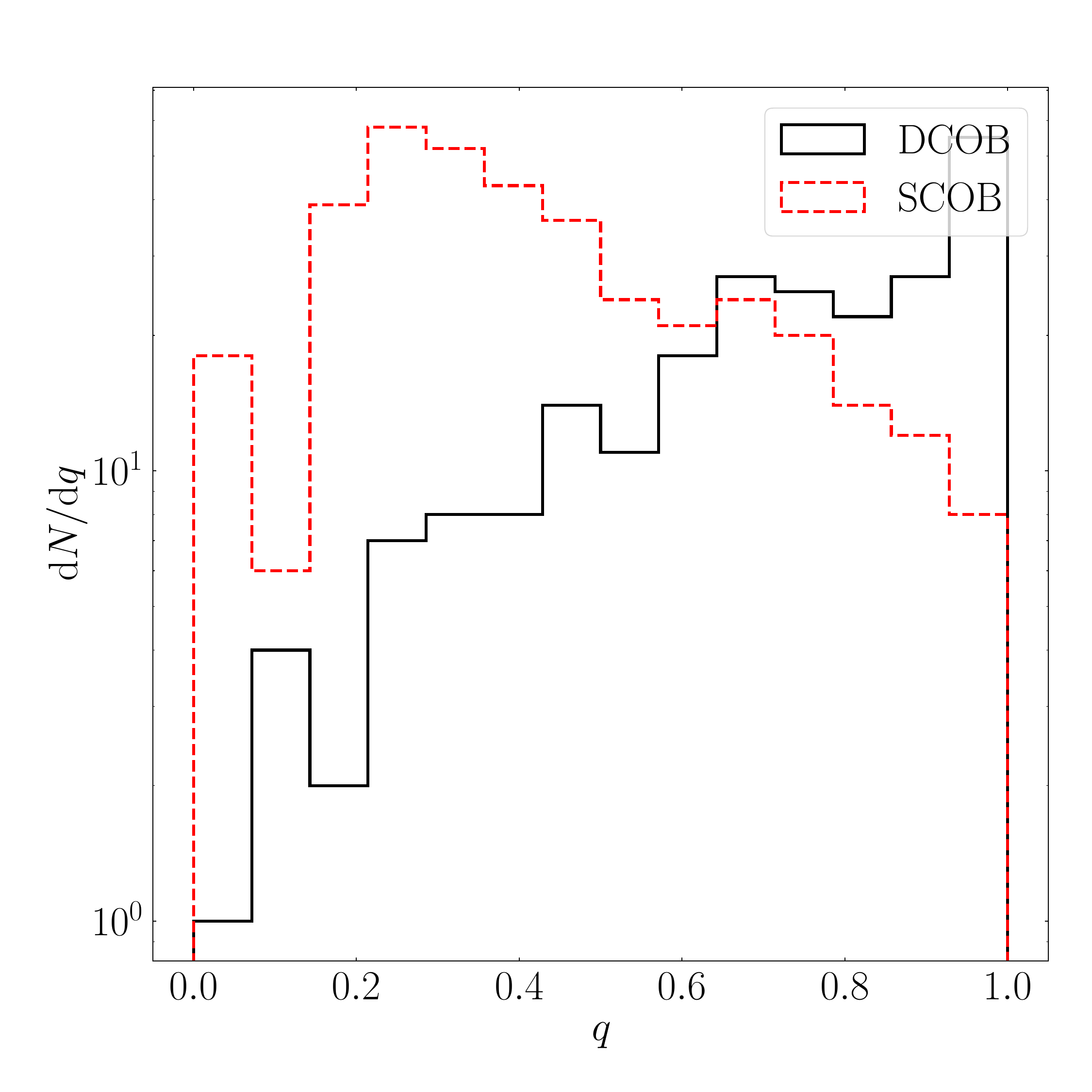}\\
    \includegraphics[width=0.95\columnwidth]{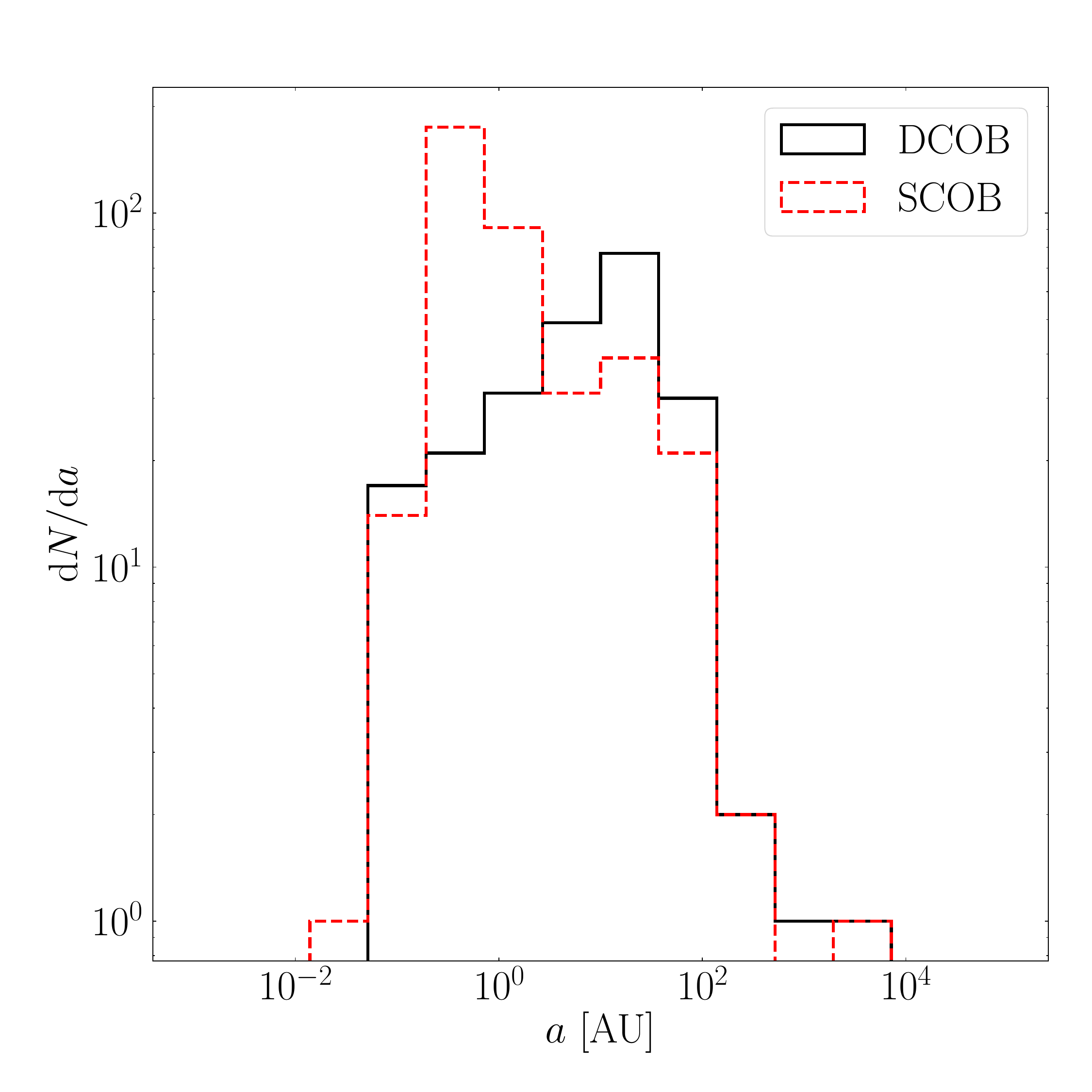}
    \includegraphics[width=0.95\columnwidth]{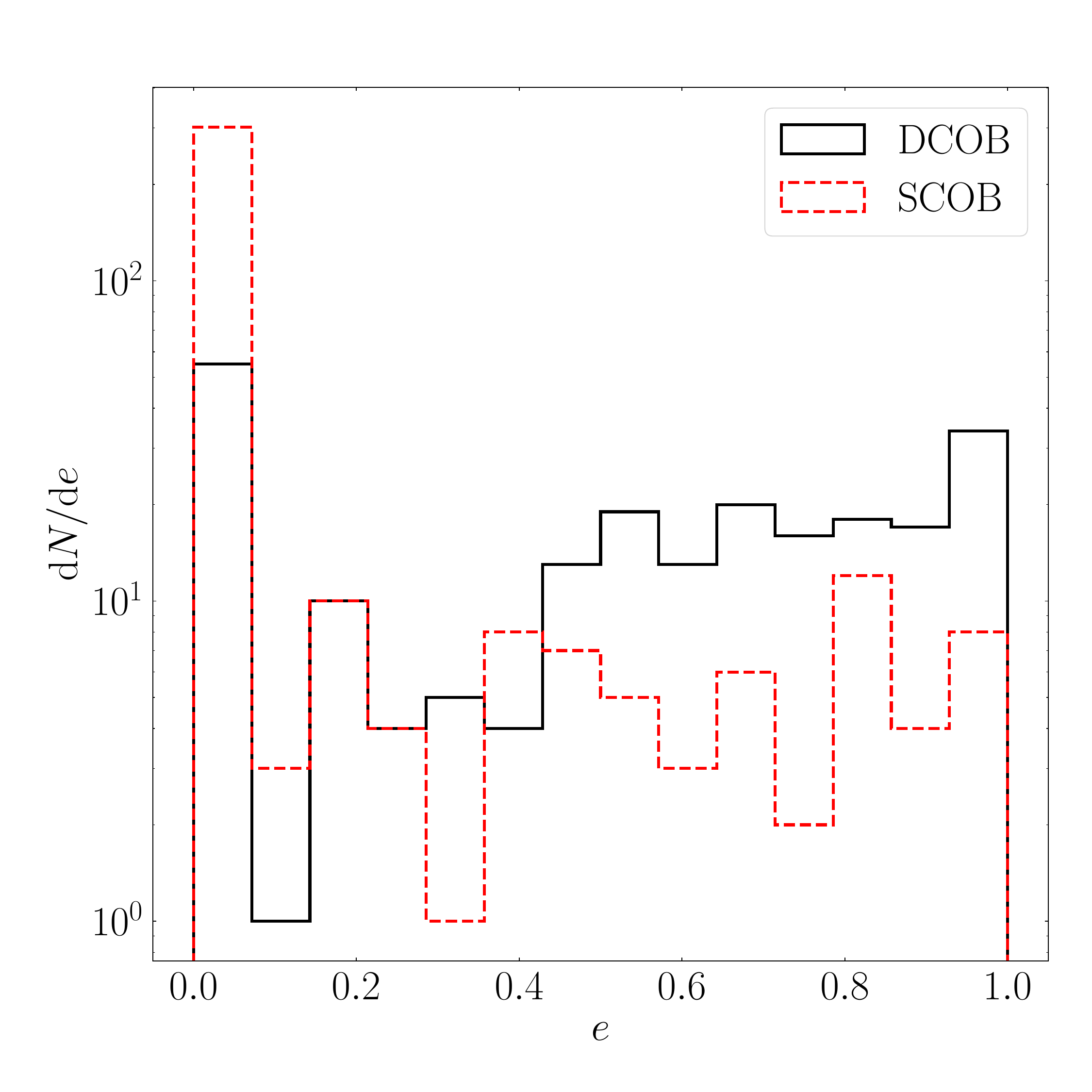}\\
    \caption{Upper panel: mass (left) and mass ratio (right) distribution of ejected binaries containing either two compact objects, i.e. BHs, NSs, or WDs. The mass distribution refers to the primary (black straight steps) and companion (black dashed steps) in the case of double compact object binaries, and to the compact object (red straight steps) and the star (red dashed steps) in the case of binaries with one compact object. Lower panels: semi-major axis (left) and eccentricity (right) distribution of ejected binaries containing two (black straight steps) or one (red dashed steps) compact objects.}
    \label{fig:binBH}
\end{figure*}

The population of ejected binaries hardly depends on the cluster initial conditions. Therefore, for the sake of simplicity, we gather the ejected binaries from all simulations to have a statistically larger sample.
In the following, we distinguish between binaries containing two compact objects, labelled as DCOB, and those containing one compact object and a star, labelled as SCOB.
Figure \ref{fig:binBH} shows the component mass, semi-major axis, and eccentricity distribution of the ejected binaries in all the \dragonii clusters.

Around $94\%$ of the ejected binaries are primordial.

A clear difference between double and single compact object binaries arises from these Figures. In total, we find 229 ejected DCOBs of both dynamical (144) and primordial (85) origin. The DCOBs exhibit a similar mass distribution for the primary and the companion, characterised by  a plateau in the $m_{1,2} = 2-20 \Ms$ and a clear peak at $m_1 \sim 45\Ms$ for the primary and $m_2 \sim 27\Ms$ for the companion. The resulting mass ratio distribution is quite peculiar, with a clear dominance of DCOB with a mass ratio $q>0.6$, owing to the tendency of dynamical interactions to pair objects of comparable mass. The eccentricity distribution is dominated by a peak around 0, caused by a sub-population of primordial binaries that underwent the common envelope phase ($64.7\%$), and a nearly flat distribution in the range $e=0.5-1$. 

Additionally, we find 375 ejected SCOBs, the vast majority of which coming from primordial binaries (353) with a small contribution from dynamically assembled systems (22). 
The mass distribution of the compact objects in SCOBs peaks at a value, $m_{\rm CO} \sim 2-4\Ms$, in the range of NSs and small BHs, definitely smaller compared to the mass distribution of the stellar companion, which peaks at $10\Ms$, but with a secondary peak at $\sim 0.3-0.5\Ms$. The binary mass-ratio distribution of SCOBs clearly differs from DCOBs, showing a peak at $q\sim 0.2$ and a decrease toward larger values. 
The compact object in the SCOBs is mostly a low-mass BH (200) -- typically with a mass $m_{\rm BH}<10\Ms$ (173) -- or a NS (173), and in only two cases a ONeWD (2). The stellar companion is a main-sequence star in the vast majority of the cases (353), followed by core He burning stars (20) (all with a primary weighing $<5\Ms$), and 2 naked He main-sequence (MS) star. 
Stellar companions in the MS phase are relatively massive: 18 of them have a mass $m_{\rm MS} < 1\Ms$, 245 have a mass in the range $1<m_{\rm MS}/\Ms<10$, 74 in the range $10<m_{\rm MS}\Ms<20$, and just one with a mass $m_{\rm MS} = 29\Ms$.
All stars in the CHeB phase have a mass in the $m_{\rm CHeB} = 5-16 \Ms$ range and are paired with an object lighter than $m_{\rm CO} < 5 \Ms$, all of them come from primordial binaries.

Focusing on DCOBs, we find a few peculiar and interesting systems. Among all ejected BBHs only 5 merge within a Hubble time, because most BBHs were ejected when the density and velocity dispersion of the cluster had already dropped due to its expansion and mass loss.

In two cases, the ejected BBH contains an IMBH with mass either $M_{\rm IMBH} = 120\Ms$ or $350\Ms$. In five cases, instead, we find an ejected BBH with a merging time smaller than a Hubble time. Table \ref{tab:t2} summarises the number of ejected single and binary BHs, and of BBHs and BH-IMBH binaries that merge within a Hubble time.

\begin{table}
    \centering
    \begin{tabular}{cccc|cccc}
    \hline
    \hline
        $N$    &$R_\ham$ & $f_b$ &  ID sim & \multicolumn{4}{c}{$N_{\rm ejec}$}\\
        $10^3$ &pc    &       & & ejected & BBH & GW & IMBH\\
    \hline
         $120$ &$1.75$ &$0.05$ & $0$ & $20$ & $16$ & $1$ & $0$ \\
         $120$ &$1.75$ &$0.05$ & $1$ & $11$ & $10$ & $0$ & $0$ \\
         $300$ &$1.75$ &$0.05$ & $0$ & $16$ & $11$ & $0$ & $0$ \\
         $300$ &$1.75$ &$0.05$ & $1$ & $31$ & $16$ & $0$ & $0$ \\
        $1000$ &$1.75$ &$0.05$ & $0$ & $25$ & $ 8$ & $0$ & $0$ \\
        $1000$ &$1.75$ &$0.05$ & $1$ & $29$ & $ 6$ & $1$ & $1$ \\
         $120$ &$1.75$ &$0.20$ & $0$ & $26$ & $ 7$ & $0$ & $0$ \\
         $120$ &$1.75$ &$0.20$ & $1$ & $39$ & $19$ & $0$ & $0$ \\
         $300$ &$1.75$ &$0.20$ & $0$ & $59$ & $21$ & $0$ & $0$ \\
         $300$ &$1.75$ &$0.20$ & $1$ & $47$ & $17$ & $0$ & $0$ \\
         $600$ &$1.75$ &$0.20$ & $0$ & $68$ & $13$ & $0$ & $0$ \\
         $600$ &$1.75$ &$0.20$ & $1$ & $68$ & $17$ & $0$ & $0$ \\
         $120$ &$0.80$ &$0.20$ & $0$ & $21$ & $14$ & $3$ & $1$ \\
         $120$ &$0.80$ &$0.20$ & $1$ & $14$ & $ 6$ & $0$ & $0$ \\
         $300$ &$0.80$ &$0.20$ & $0$ & $33$ & $10$ & $0$ & $0$ \\
         $300$ &$0.80$ &$0.20$ & $1$ & $37$ & $13$ & $0$ & $0$ \\
         $120$ &$0.47$ &$0.20$ & $0$ & $23$ & $10$ & $2$ & $1$ \\
         $120$ &$0.47$ &$0.20$ & $1$ & $14$ & $ 6$ & $0$ & $0$ \\
         $300$ &$0.47$ &$0.20$ & $0$ & $23$ & $ 9$ & $0$ & $0$ \\
    \hline
    \end{tabular}
    \caption{Col. 1-3: initial number of stars, half-mass radius, and binary fraction. Col. 4: simulation ID. Col. 5: number of ejected BHs, ejected BBHs, ejected BBHs that merge within 14 Gyr, and ejected BBHs that merger within 14 Gyr and involve one IMBH.}
    \label{tab:t2}
\end{table}

\subsection{Black hole -- main sequence star binaries }
The sample of known BH--MS star systems has significantly grown over the last few years \citep[e.g.][]{2018MNRAS.475L..15G,giesers19,2022MNRAS.511.2914S, elbadry22, shenar22, mahy2022}. Some of the BHs observed in a BH--MS binary appear to reside in star clusters both in the Milky Way \citep{2018MNRAS.475L..15G,giesers19} and the Large Magellanic Cloud \citep{2022MNRAS.511.2914S,shenar22}, whilst others appear to be in the Galactic disc \citep{elbadry22}. 
It is an open question whether these  BH--MS systems come from primordial or dynamically assembled binaries. In the case of a dynamical origin it is also unknown whether the stellar companion captured the BH or its progenitor. 

In these regards, the \dragonii models offer us a novel way to look for BH--MS binaries in simulated clusters and identify possible  properties of BH--MS binaries formed through different channels. Since the \dragonii cluster database is relatively small and limited to a single metallicity, we cannot perform a comprehensive comparison between observed and simulated BH--MS binaries. Nonetheless, it is illustrative to qualitatively compare the properties of BH--MS binaries formed in \dragonii models and the observed one. 

\begin{figure}
    \centering
    \includegraphics[width=0.9\columnwidth]{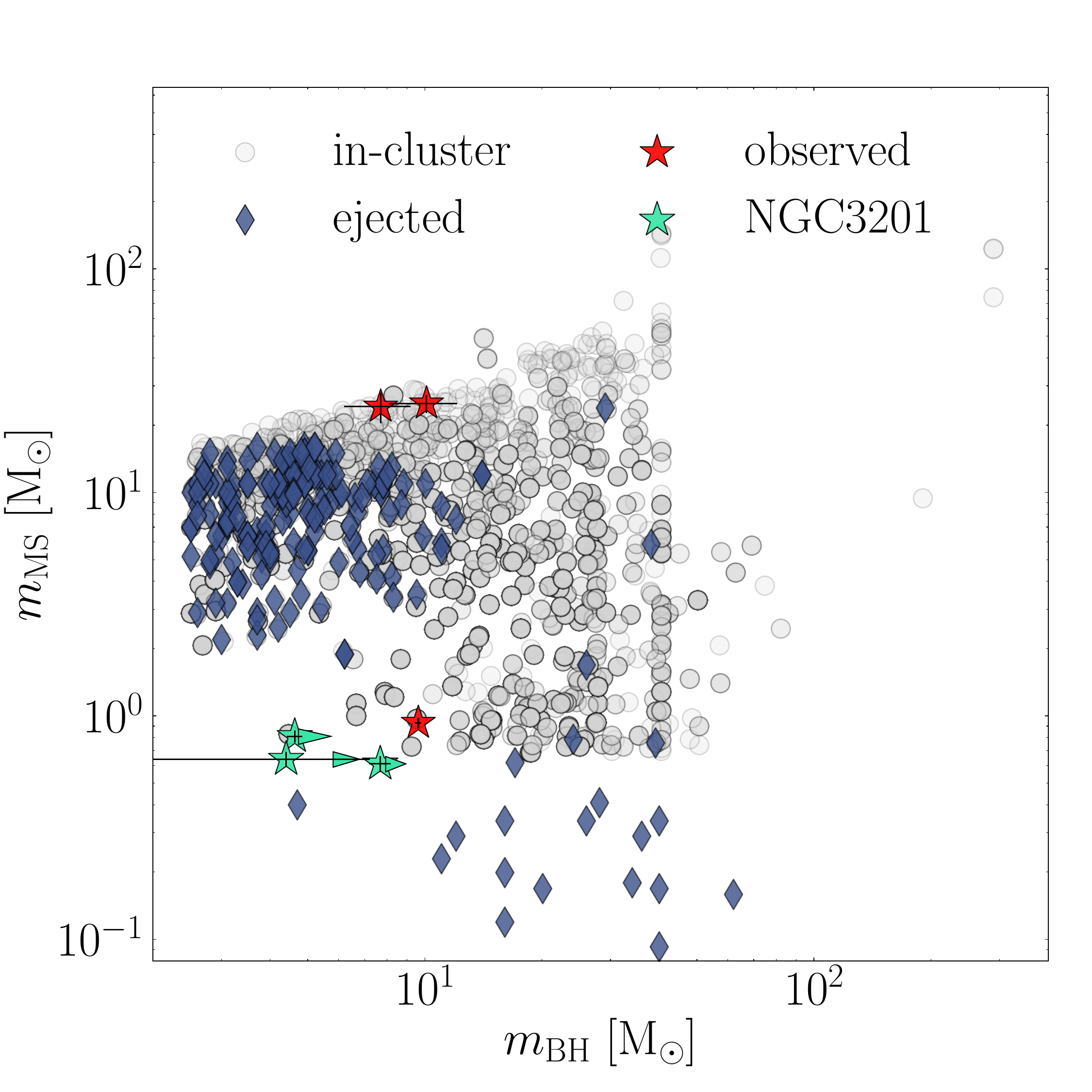}\\
    \includegraphics[width=0.9\columnwidth]{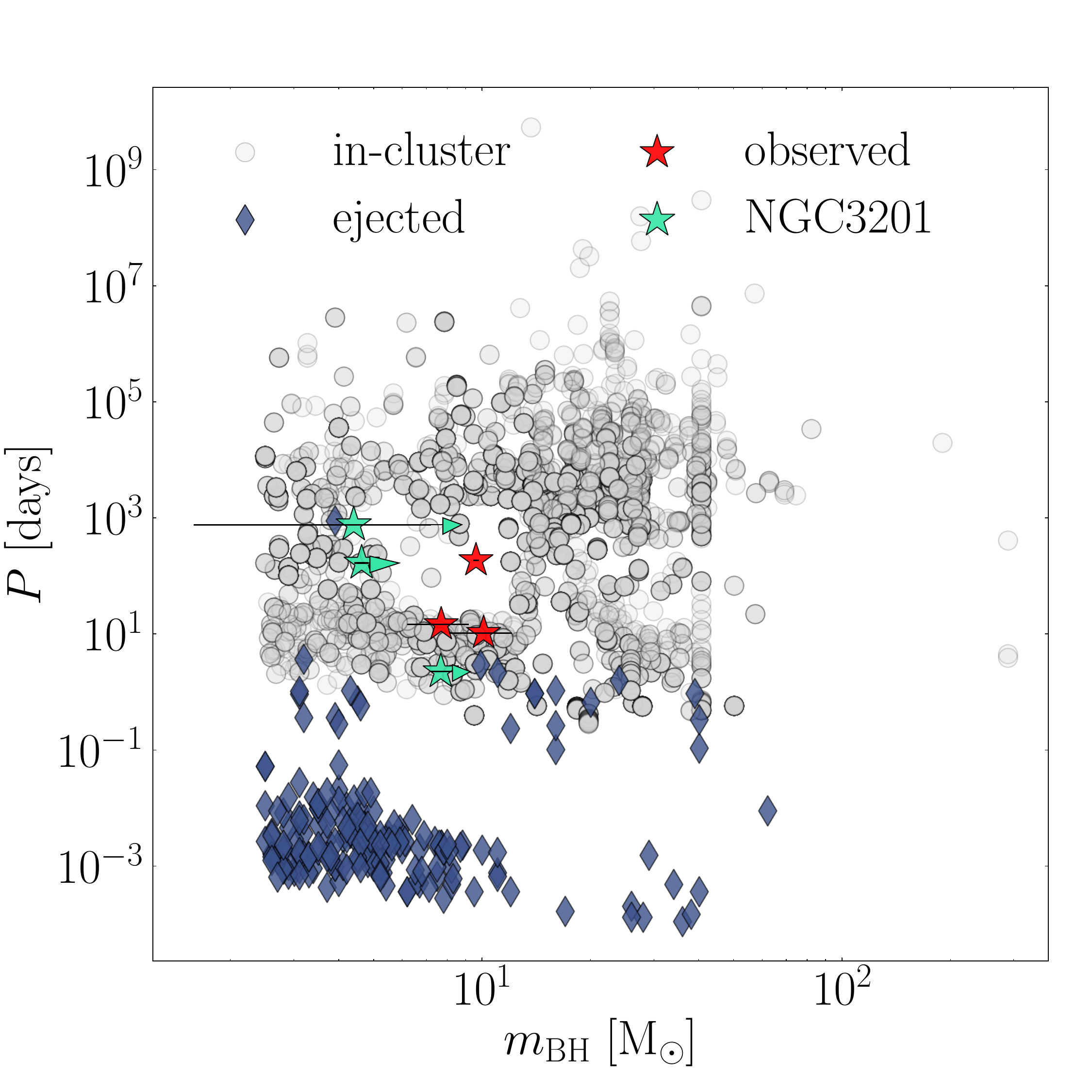}\\
    \includegraphics[width=0.9\columnwidth]{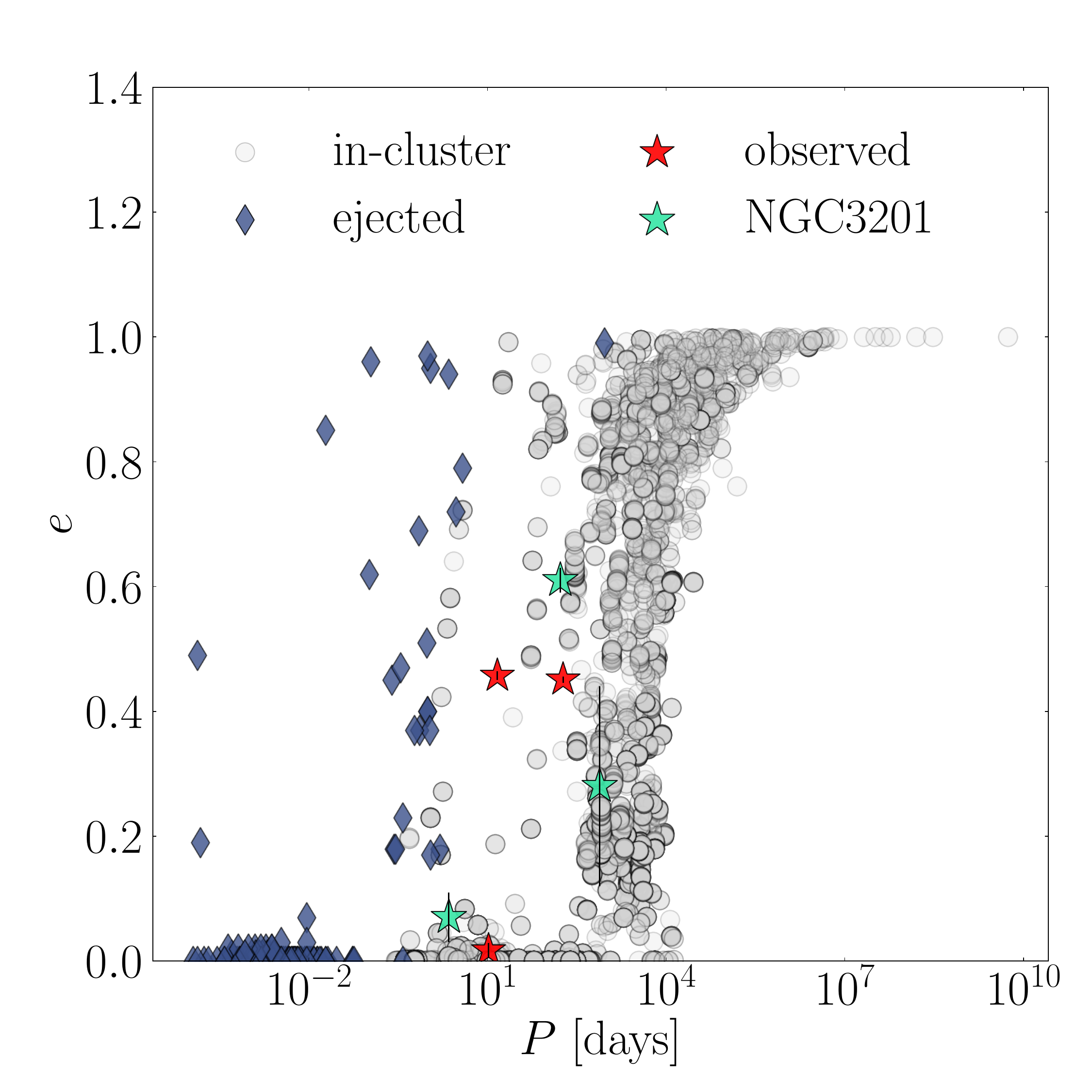}\\
    \caption{Masses, eccentricity, and period of in-cluster (grey dots) and ejected (blue diamond) BH--MS binaries. The red stars represent observed binaries taken from \citet{elbadry22},  \citet{shenar22}  and \citet{mahy2022}. The light grey stars represent binaries observed in the NGC3201 globular cluster \citet{giesers19}.}
    \label{fig:msbh}
\end{figure}

For example, \dragonii models permit us to dissect the population of BH--MS binaries into those forming inside the cluster, some of which have a lifetime much shorter than the cluster life and are disrupted via interactions with other cluster members, or that have been ejected from the cluster. Figure \ref{fig:msbh} shows the component masses, period, and eccentricity of {\it in-cluster} and {\it ejected} BH--MS binaries. 
We assume that in-cluster binaries are  those forming at any time over the simulated time, therefore the same binary or one or both components can appear multiple times in the plot. We see that in-cluster binaries are markedly different from ejected binaries. The latter can be divided in two sub-classes. The first sub-class exhibits a short period ($P<0.1$ day) and an almost null eccentricity, $e \sim 0$. Binaries in this sub-class are characterised by a BH with mass $m_{\rm BH} < 10\Ms$ and a MS star with a mass in the $2-10\Ms$ range. They originate from binary evolution, and, in particular, underwent a common envelope phase that shrank the semi-major axis and damped the eccentricity of the binary. The ejection engine of these binaries is a SN explosion. The second sub-class, instead, comprises  heavier BHs ($m_{\rm BH} = 10-100\Ms$) and lighter MS stars ($m_{\rm MS} < 1\Ms$), and is characterised by eccentricities in the range $e = 0.2-1$,  indicating that these binaries come from dynamical interactions sufficiently strong to eject the binary from the cluster.

In-cluster BH--MS binaries can contain BHs and MS stars significantly heavier than the ejected binaries and are characterised by longer periods ($P>10$ d) compared to ejected binaries. Most in-cluster binaries with a period $P\lesssim 10^3$ d have zero eccentricity, whilst practically all those with a longer period have eccentricity $>0.1$ and up to extreme values. 

From Figures \ref{fig:msbh}, it is evident that in-cluster binaries exhibit a peculiar distribution in the $m_{\rm BH}-m_{\rm MS}$, which suggests the existence of two sub-classes. We find that the first class is characterised by a companion with a mass $m_{\rm MS}/m_{\rm BH} = k\,{}(m_{\rm BH}/1\Ms)^{-1/2}$, with $k=2-10$. Most binaries falling in this class have a period shorter than $100$ d, whilst the second class involves binaries with $m_{\rm BH}>10 \Ms$ and $m_{\rm MS}<5 \Ms$. 

\begin{table*}
\begin{tabular}{ccccccccc}
\hline
\hline
name & ${\bm m_{\rm BH}}$ & ${\bm m_{\rm MS}}$ & ${\bm P}$ & ${\bm e}$ & \textbf{loc} & 
 \textbf{SYSTEM} & \textbf{Z} & \textbf{ref}\\ 
& $\Ms$ & $\Ms$ & days & & & & & \\
\hline
\textbf{BH1}  & $9.62 \pm 0.18  $         & $0.93 \pm 0.05$           & $185.59 \pm 0.05$          & $0.451 \pm 0.005$            & MW          & disc-field          & solar          & \citet{elbadry22}\\ 
\textbf{VFTS 243} & $10.1 \pm 2$            & $25.0 \pm 2.3$           & $10.4031 \pm 0.0004$          & $0.017 \pm 0.012 $           & LMC          & OC         & sub-solar          & \citet{shenar22}\\ 
\textbf{HD130298} & $7.7 \pm 1.5$            &$ 24.2\pm 3.8 $           & $14.62959 \pm 0.000854$          & $0.457 \pm 0.007 $           & MW          & runway          & solar         & \citet{mahy2022}\\
\textbf{ACS21859}$^*$ &$7.68\pm 0.50$ & $0.61\pm 0.05$ & $2.2422\pm 0.0001$ & $0.07\pm 0.04$ & NGC3201 & GC & sub-solar & \citet{giesers19} \\
\textbf{ACS21859}$^*$ &$4.53\pm 0.21$ & $0.81\pm 0.05$ & $167.01\pm 0.09$ & $0.61\pm 0.02$ & NGC3201 & GC & sub-solar & \citet{giesers19} \\
\textbf{ACS21859}$^*$ &$4.40\pm 2.82$ & $0.64\pm 0.05$ & $764\pm 11$ & $0.28\pm 0.16$ & NGC3201 & GC & sub-solar & \citet{giesers19}\\
\hline
\end{tabular}
\label{tab1}
\caption{Orbital properties of the observed BH--MS binaries. For the sources labelled with an $*$ only lower limits to the BH mass are available.}
\end{table*}

\begin{figure}
    \centering
    \includegraphics[width=\columnwidth]{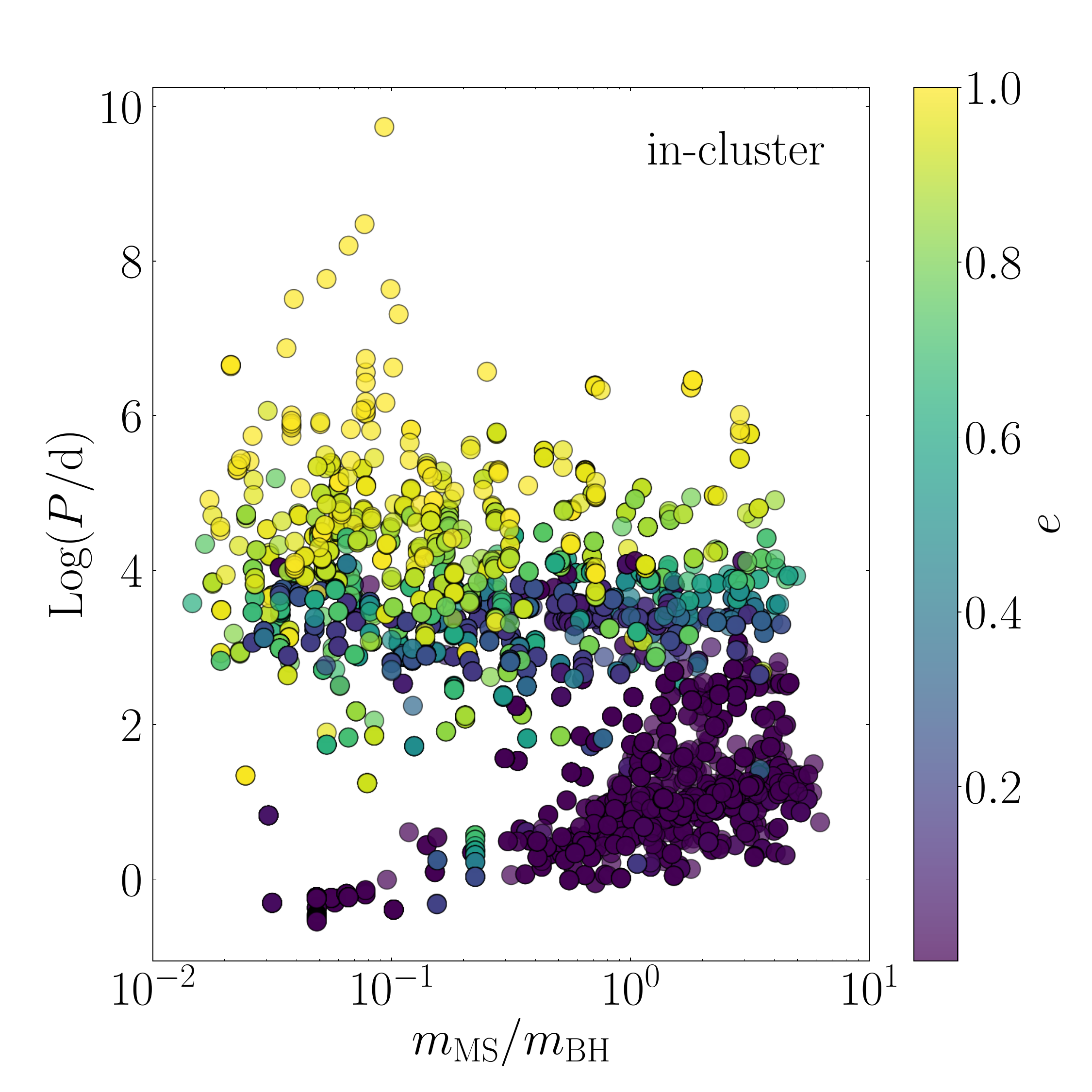}
    \caption{Orbital period as a function of the ratio between the MS and BH masses for all BH--MS binaries formed inside the clusters. The colour coding marks the binary eccentricity.}
    \label{fig:bhmsinc}
\end{figure}

An even more clear distinction is shown in Figure \ref{fig:bhmsinc}, where the MS-to-BH mass ratio is shown against the orbital period and eccentricity. This plot highlights four interesting peculiarities of in-cluster BH--MS binaries:
\begin{itemize}
    \item the vast majority of binaries with $e<0.1$ are primordial. Most of them are characterised by $m_{\rm MS}/m_{\rm BH} > 0.3$, heavy MS stars $m_{\rm MS} > 1$ M$_\odot$, and periods below $P < 100$ d;
    \item primordial binaries with $e > 0.1$ have  larger periods ($P = 10^2-10^6$ d), and similar mass ratio and MS mass as circular primordial binaries;
    \item the vast majority of dynamically formed binaries have $e>0.1$ and periods in the range ($P=10^2-10^9$ d). They are generally characterised by a mass ratio $m_{\rm MS}/m_{\rm BH} < 0.3$, MS stars with a mass $m_{\rm MS} < 10\Ms$ and a BH with mass $m_{\rm BH} = (10-100)\Ms$;
    \item only a handful dynamically formed binaries have $e < 0.1$, and  are all characterised by a period $P=1-10$ d.
\end{itemize}

As shown in Figure \ref{fig:bhmsinc}, we find that the longer is the orbital period the larger the binary eccentricity, and almost all binaries with eccentricity $e>0.1$ have a period $P>100$ d, with a handful exceptions. Most binaries with a period shorter than $P<100$ d, instead, are primordial and involve a MS star heavier than $m_{\rm MS} > 1\Ms$.

\begin{figure}
    \centering
    \includegraphics[width=0.9\columnwidth]{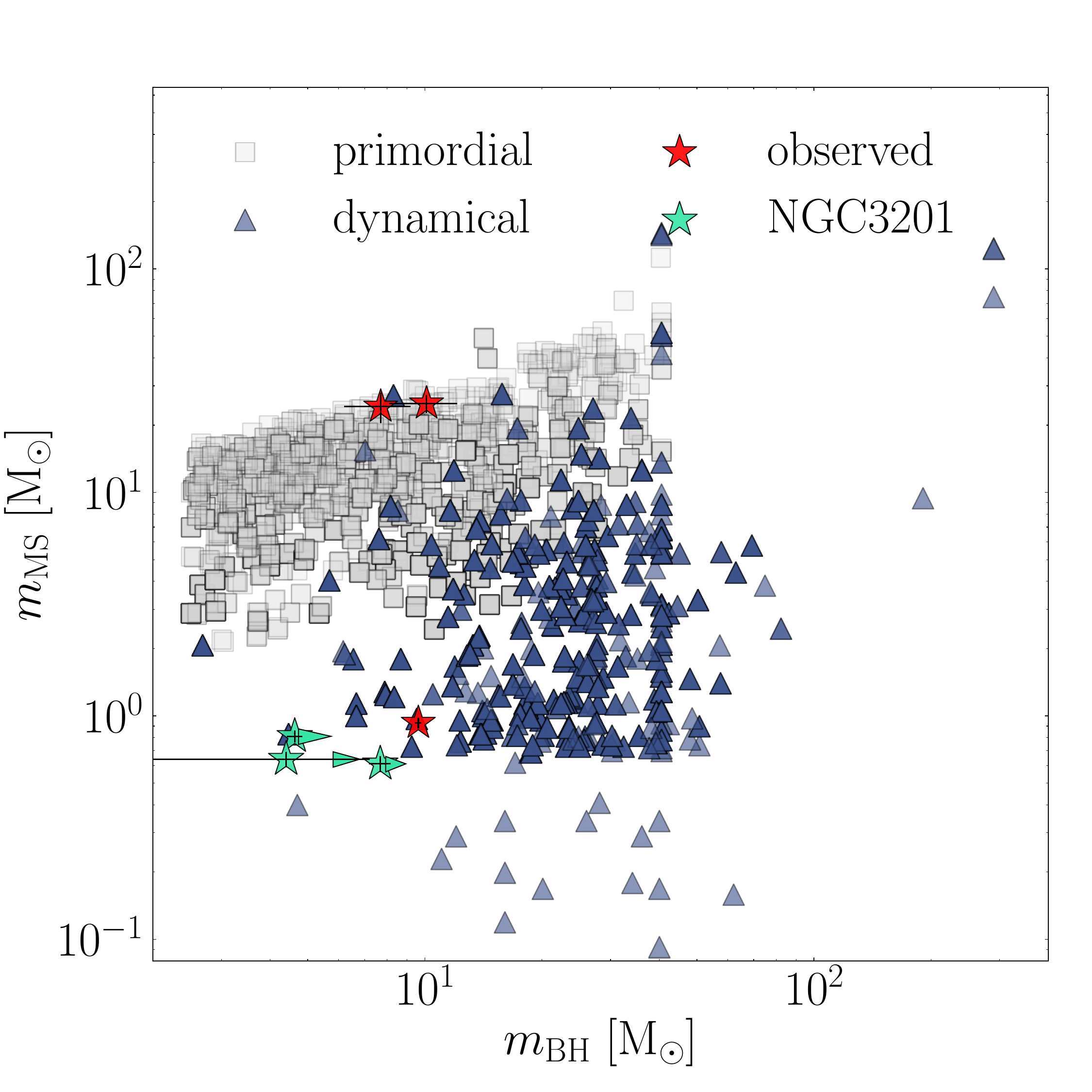}
    \caption{Masses of the MS star and BH in primordial (grey squares) and dynamical (blue triangles) BH--MS binaries. The red and green stars represent observed binaries as in Figure \ref{fig:msbh}. The plot refers to both in-cluster and ejected binaries.}
    \label{fig:msbh2}
\end{figure}

The difference between primordial and dynamical BH--MS binaries is further highlighted in Figure \ref{fig:msbh2}, which shows the component masses of these two classes of binaries. From the plot, it is apparent that dynamically assembled binaries dominate the region of the plane with $m_{\rm BH} > 10\Ms$ and $m_{\rm MS} < 10\Ms$. 

The observed BH--MS binaries have orbital properties quite different from our ejected binaries, especially if we consider the observed period and eccentricity. However, only the quiescent BH candidates in NGC3201 are still associated with a star cluster, whilst the origin of the other binaries is unknown. Two of the six observed binaries \citep{shenar22, mahy2022} have component masses compatible with our primordial binaries,  one of them \citep{elbadry22} falls in a range where only dynamically assembled binaries are present, and the three sources observed in the Galactic globular cluster NGC3201 have component masses compatible with both in-cluster and ejected binaries.

In our models, the vast majority of ejected binaries have a primordial origin and their small period ($P < 0.01$ d) owes to mass transfer episodes.
The few ejected binaries formed dynamically are characterised by a period $P<1$ d, still much shorter than observed values.
Wider, and more numerous, ejected binaries could form in substantially looser or lighter star clusters. On the one hand, decreasing the cluster mass or density would enlarge the hard-binary separation and possibly increase the semi-major axis of ejected binaries \citep{2015ApJ...800....9M}. On the other hand, a smaller cluster mass would correspond to a lower escape velocity and thus it is more likely for binaries to escape the parent cluster.

In principle, MS--MS binaries ejected in the earliest phase of the cluster life could further contribute to the population of BH--MS binaries, but these binaries are removed from our simulations before they can further evolve. Nonetheless, we find that only two ejected MS--MS binaries have at least one component with mass above the threshold for BH formation, i.e. $\sim 18\Ms$, thus ensuring that ejected MS--MS binaries do not contribute to the population of ejected BH--MS binaries. 

Among all observed data, the binaries observed in NGC3201 are probably the ones more suited for a comparison with our models, given the metallicity and mass of NGC3201. 

From the central and bottom panel of Figure \ref{fig:msbh}, it is apparent that our in-cluster binaries have periods, eccentricities, and BH masses compatible with those observed in NGC3201. The fact that our models do not match well the companion mass may be due to NGC3201's age. In fact, this cluster is relatively old \citep[$\sim 11.5\pm 0.4$ Gyr][]{2013ApJ...775..134V}, thus its population of binaries has likely been heavily processed over time, and most of its stellar population with super-solar mass already left the MS.
Figure \ref{fig:msbh2} favours this interpretation. Note that both the mass of BHs and MS stars in dynamically formed BH--MS binaries tend to be smaller compared to primordial binaries. As the BH-burning process proceeds, the average BH mass will keep decreasing, while stellar evolution processes will deplete the high-end tail of the MS mass distribution, possibly favouring the formation of BH--MS binaries in the region populated by NGC3201 sources.

\subsection{Black hole subsystem}
\label{sec:BHS}

In all \dragonii clusters, the segregation time is generally shorter  than the stellar evolution timescale of massive stars, therefore massive stars sink to the cluster centre before evolving to BHs. This implies a possible enhancement of the probability for stellar mergers and star--BH collisions.

\begin{figure}
    \centering
    \includegraphics[width=\columnwidth]{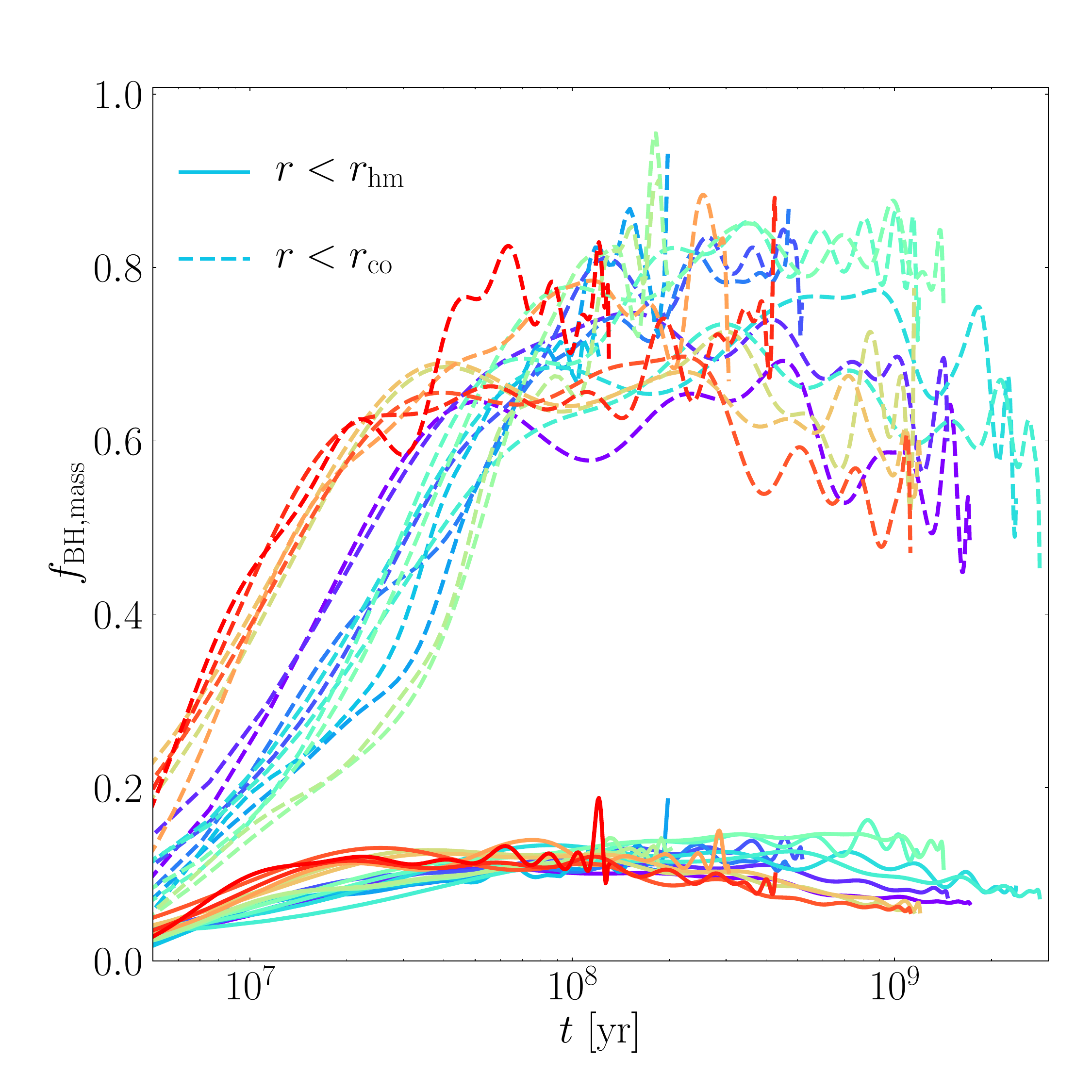}
    \caption{Fraction of mass in BHs within the cluster half-mass (straight lines) and core radius (dashed lines). Different colours correspond to different simulations.}
    \label{fig:fBH}
\end{figure}

Given the short segregation times, BHs dominate the dynamics in the cluster core already after a time $t=20-40$ Myr, making up the $50-80\%$ of the mass in the cluster core and around $10\%$ of the mass within the half-mass radius, as shown in Figure \ref{fig:fBH}. Given the amount of mass in BHs enclosed within the core radius, this length scale can be regarded as the BH sub-system scale radius \citep[see e.g.][]{2018MNRAS.479.4652A}.

A similar trend of the BH mass fraction inside $R_\ham$ has been found also in recent simulations performed with the Monte Carlo code \textsc{MOCCA} \citep{2019MNRAS.487.2412G,2022MNRAS.509.4713W} and the $N$-body code \textsc{PeTar} \citep{2022MNRAS.509.4713W}, which both exploit similar stellar evolution recipes. 

\begin{figure*}
    \centering
    \includegraphics[width=\columnwidth]{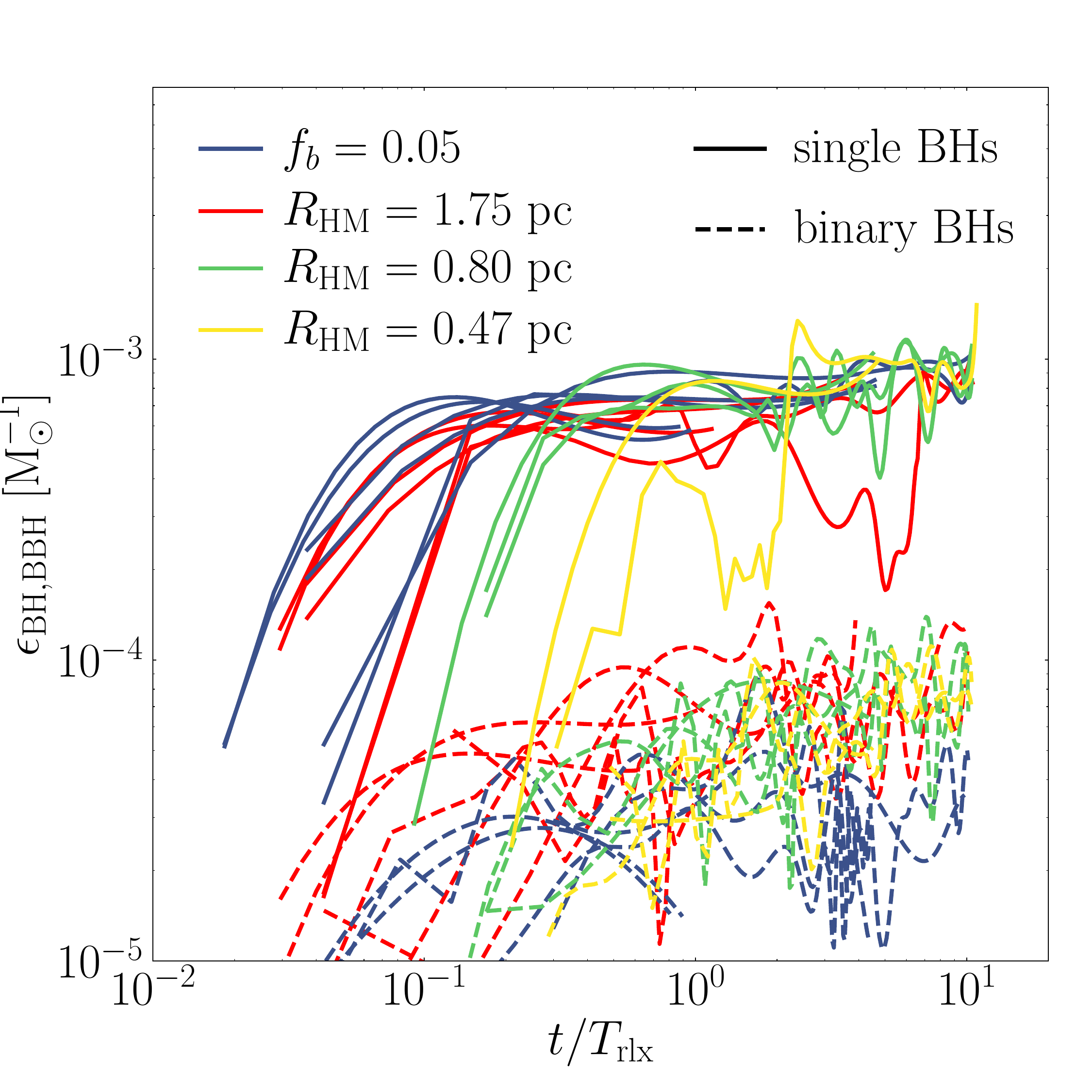}
    \includegraphics[width=\columnwidth]{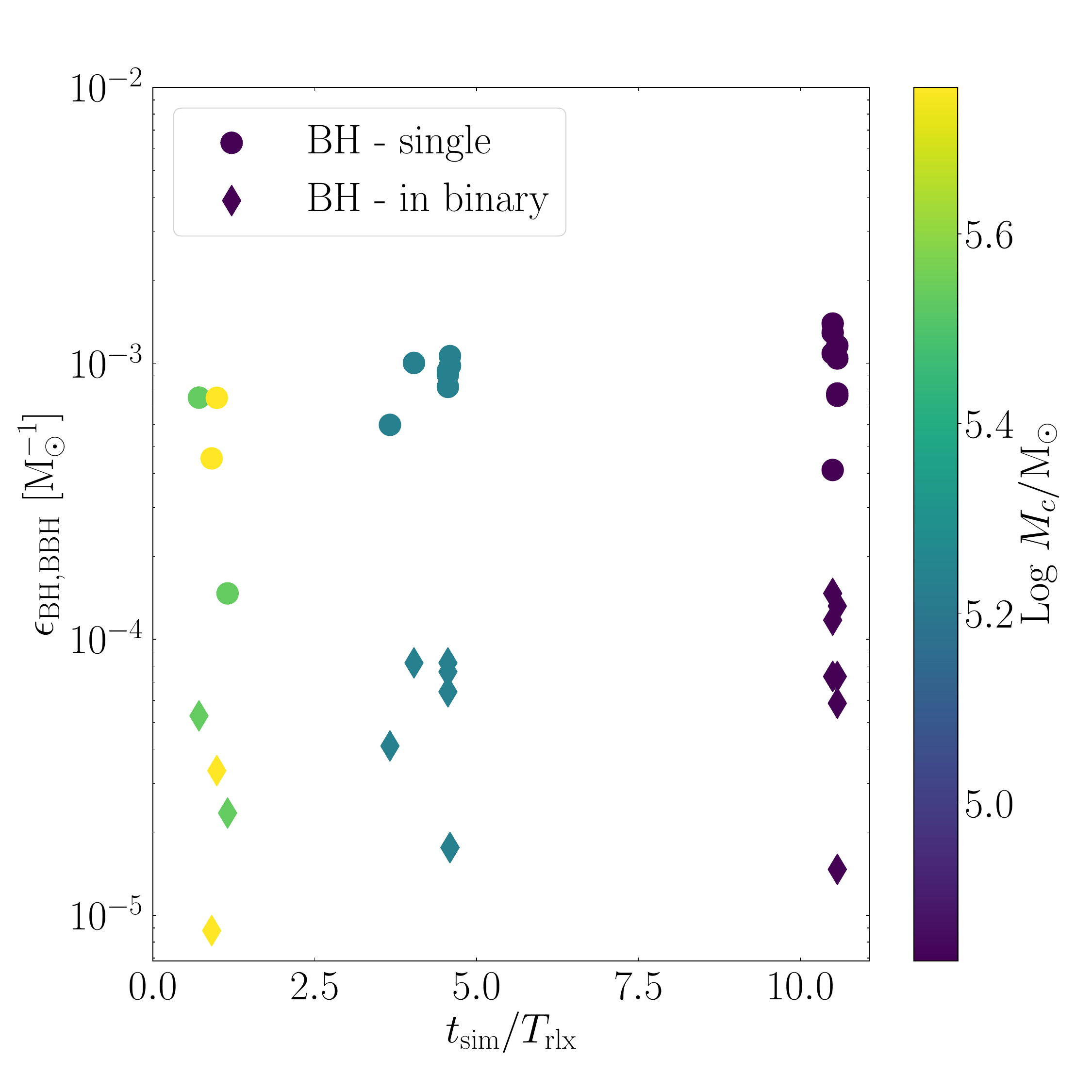}
    \caption{Left-hand panel: number of BHs (straight lines) and BBHs (dashed lines) inside the core radius normalised to the initial cluster mass (BH formation efficiency) as a function of time normalised to the initial cluster relaxation time. The large jumps follow the jumps observed in the core calculation. The colour-map identifies the cluster mass. Right-hand panel: BH formation efficiency $\epsilon$ for single (points) and binary BHs (diamonds) calculated at the end of each simulation as a function of the ratio between the simulated time and the cluster relaxation time.}
    \label{fig:BHeff}
\end{figure*}

Both the primordial binary evolution and the onset of three-body and multiple gravitational scattering favour the formation of binaries containing at least one BH.
Figure \ref{fig:BHeff} shows the {\it BH formation efficiency}, defined as the ratio between the number of BHs inside the cluster core radius and the initial cluster mass, i.e. $\epsilon_{\rm BH, BBH} = N_{\rm BH,BBH}(<R_c)/M_{\rm cl,0}$. We find that, regardless of the initial cluster mass, half-mass radius, or binary fraction, all models are characterised by $\epsilon_{\rm BH} \simeq (0.8-2)\times 10^{-3}\Ms^{-1}$ for single BHs and  $\epsilon_{\rm BBH} \simeq (0.8-2)\times 10^{-4}\Ms^{-1}$ for binary BHs. As shown in the right panel of Figure \ref{fig:BHeff}, the BH formation efficiency slightly increases with the simulation time, although it is unclear whether this quantity saturates already at $t_{\rm sim}/T_{\rm rlx} \gtrsim 10$. Note that our definition of $\epsilon_{\rm BBH}$ implies that a cluster with initial mass $7\times 10^4(6\times 10^5)\Ms$ contains around $7(60)$ BHs in a binary system after 10 relaxation times. 

It might seem trivial that $\epsilon$ is independent of the cluster initial conditions, as it suggests that it is just a consequence of the adopted mass function. However, the BH-burning mechanism \citep[e.g.][]{2010MNRAS.402..371B,2011MNRAS.416..133D,2010MNRAS.407.1946D,2013MNRAS.432.2779B,2018MNRAS.479.4652A, 2020IAUS..351..357K}, by which the most massive BHs pair in binaries that first eject the lighter BHs from the cluster and then get themselves ejected via super-elastic binary-single and binary-binary scatterings, could significantly affect the population of BHs. This does not seem the case in the \dragonii models. The small spread observed in the BH binary formation efficiency is related to the initial cluster half-mass radius and binary fraction, whilst the weak increase of $\epsilon_\bbh$ over time is the result of dynamically formed binaries.

\begin{figure}
    \centering
    \includegraphics[width=\columnwidth]{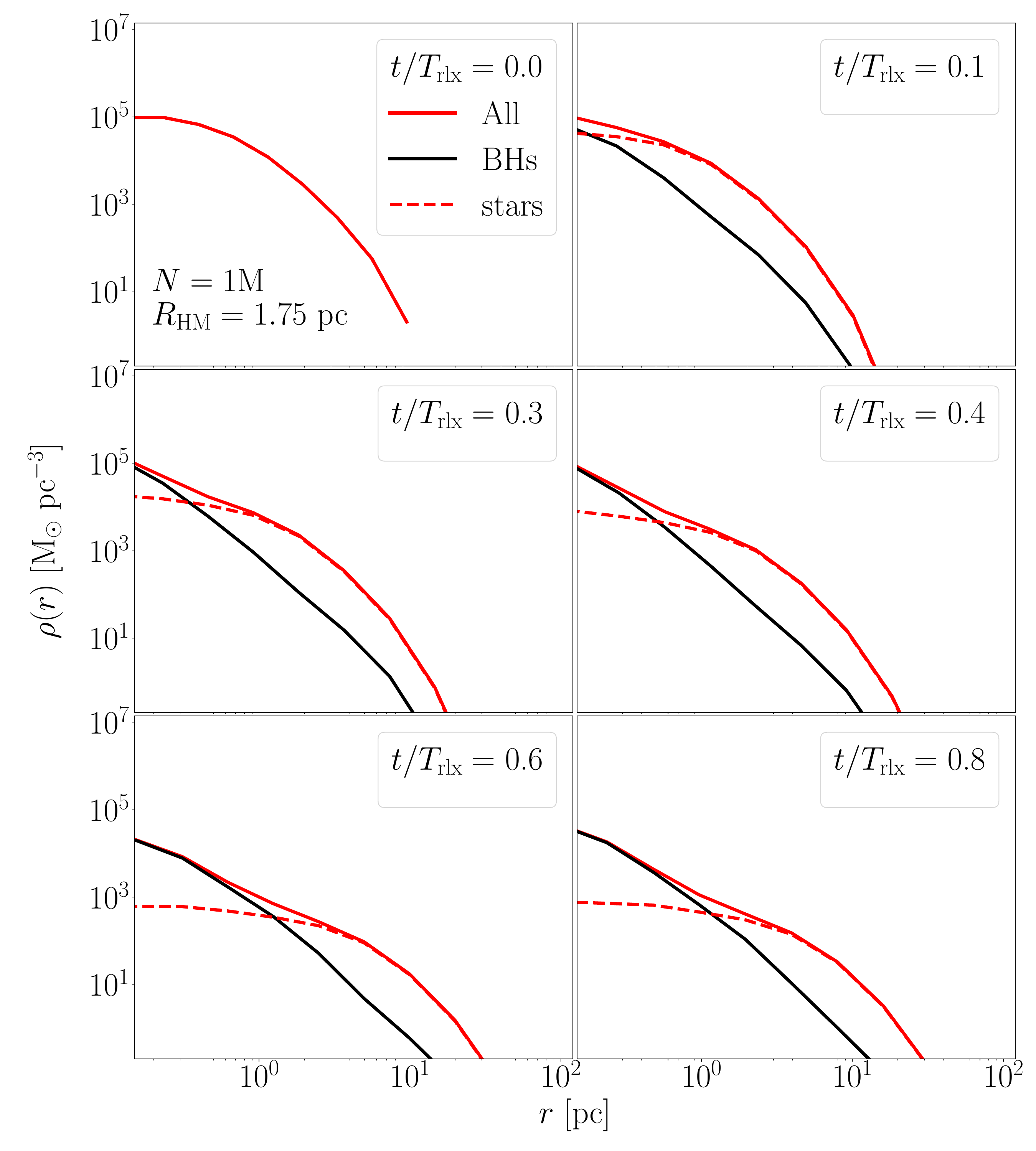}
    \caption{Density profile of the cluster (red straight line), stars (red dashed line), and BHs (black straight line) at different times for models with $N=10^6$ and $R_\ham=1.75$ pc.}
    \label{fig:BHS}
\end{figure}

\begin{figure}
    \centering
    \includegraphics[width=\columnwidth]{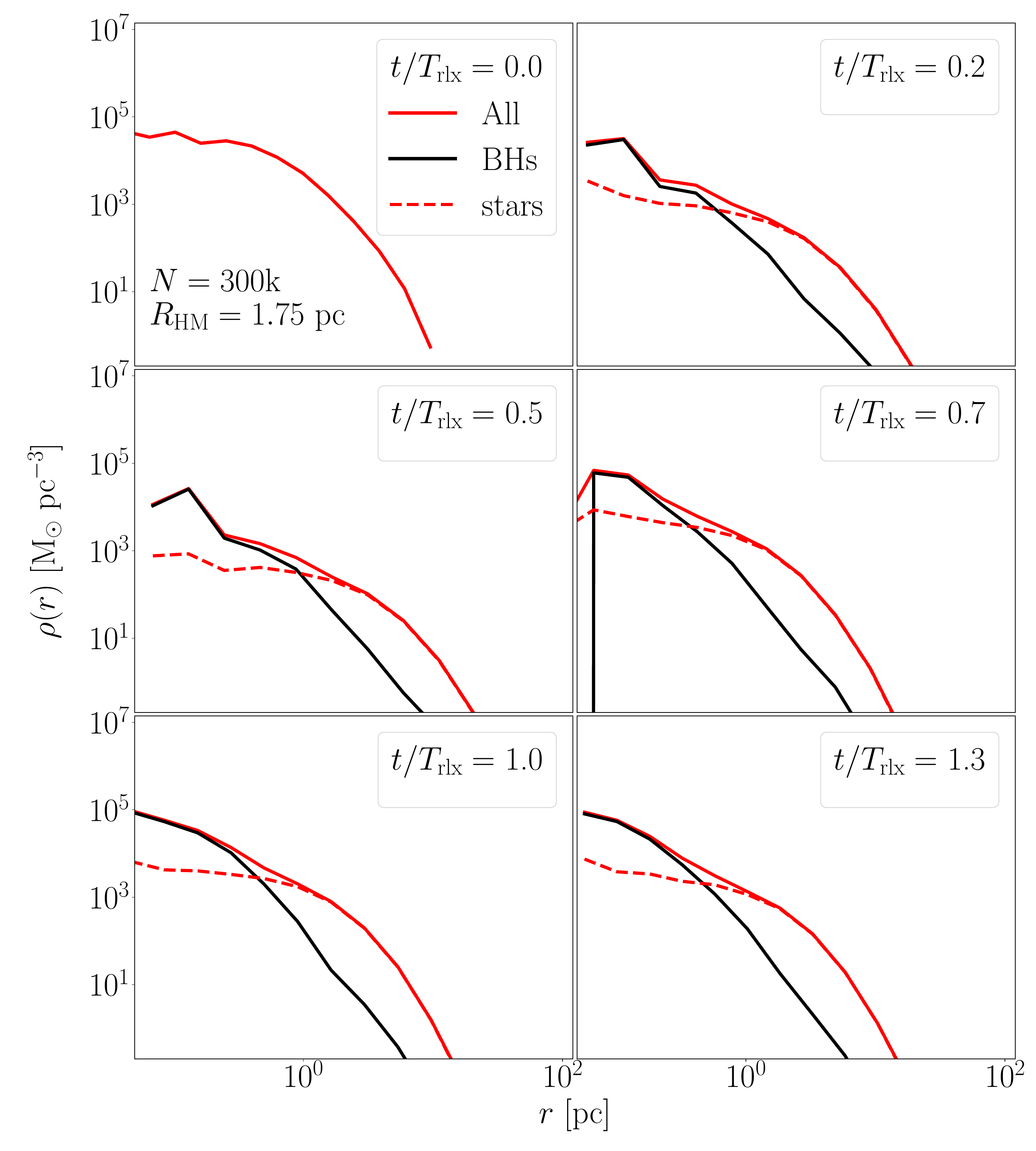}
    \caption{Same as in Figure \ref{fig:BHS}, but for one simulation with $N=3\times 10^5$ and $R_\ham=1.75$ pc.}
    \label{fig:BHS2}
\end{figure}

\begin{figure}
    \centering
    \includegraphics[width=\columnwidth]{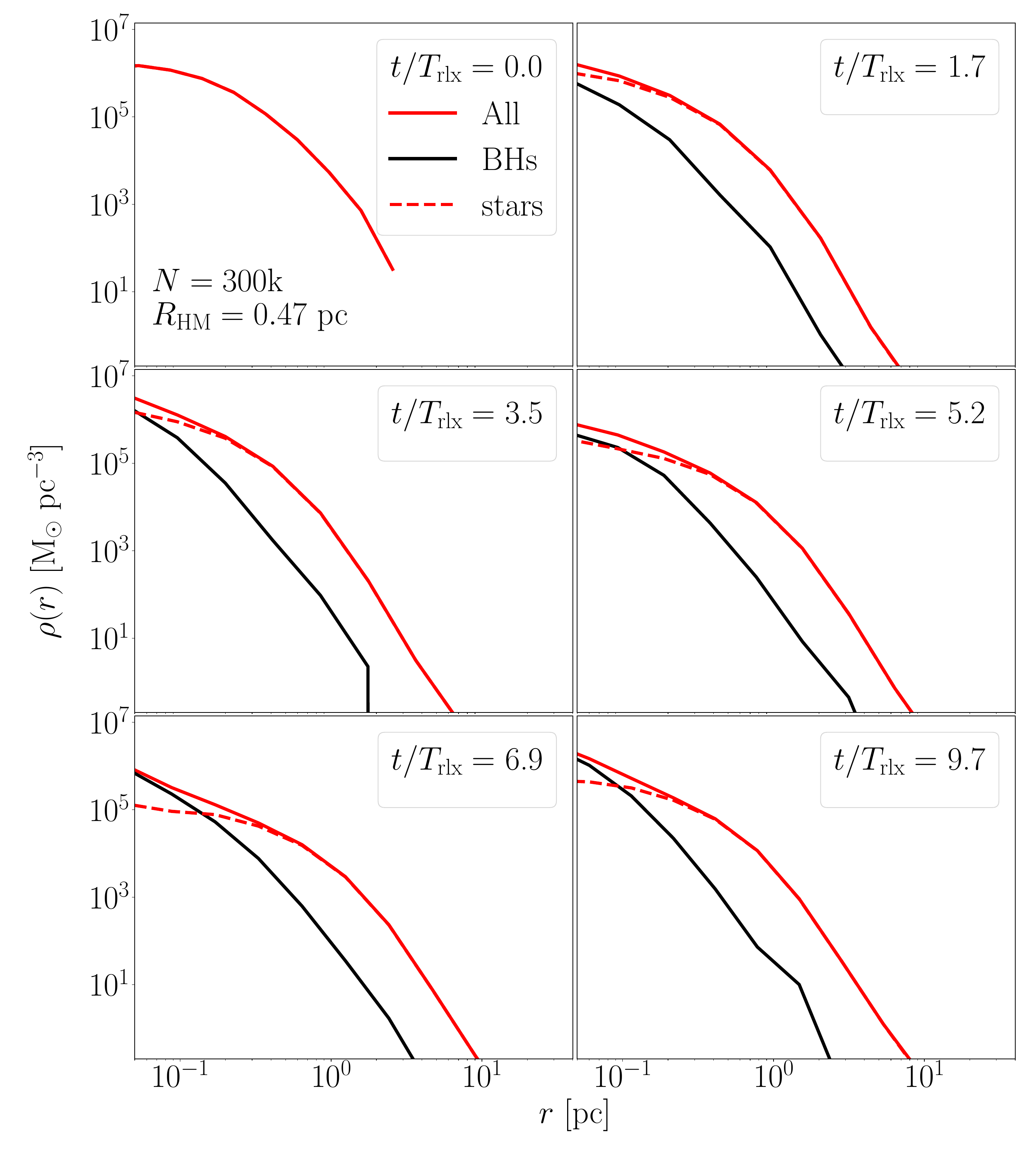}
    \caption{Same as in Figure \ref{fig:BHS}, but for one simulation with $N=3\times 10^5$ and $R_\ham=0.47$ pc.}
    \label{fig:BHS3}
\end{figure}

Figures \ref{fig:BHS}-\ref{fig:BHS3} show the cluster and BH subsystem density profiles at different times for three cluster models with $N = (0.3-1)\times 10^6$ and $R_\ham = 0.47-1.75$ pc. The central density of BH subsystems attains values around $\rho_{\rm BHS} \simeq (10^4-10^5)\Ms$~pc$^{-3}$, i.e. values 10--100 times larger than the  density of stars, whilst their scale radius is roughly $R_{\rm BHS} \simeq (0.5-1)$ pc in all models, corresponding to the radius at which the density contribution from the BHs and stars equal each other.

Looking at the different panels it is possible to identify the signatures of the whole BH burning process as described in \cite{2013MNRAS.432.2779B}. Firstly, BHs start forming and interacting, driving the formation of the BH subsystem and its subsequent expansion over a timescale $t\sim T_{\rm rlx}$. Secondly, dynamical BH interactions cause the steepening of the BHS density and the contraction of its structure, driven by BH ejections over a time $1<t/T_{\rm rlx}<5$. Thirdly, the BH subsystem rebounces and expands again, reaching a seemingly stable structure, at least within the simulated time.

\begin{figure*}
    \centering
    \includegraphics[width=\columnwidth]{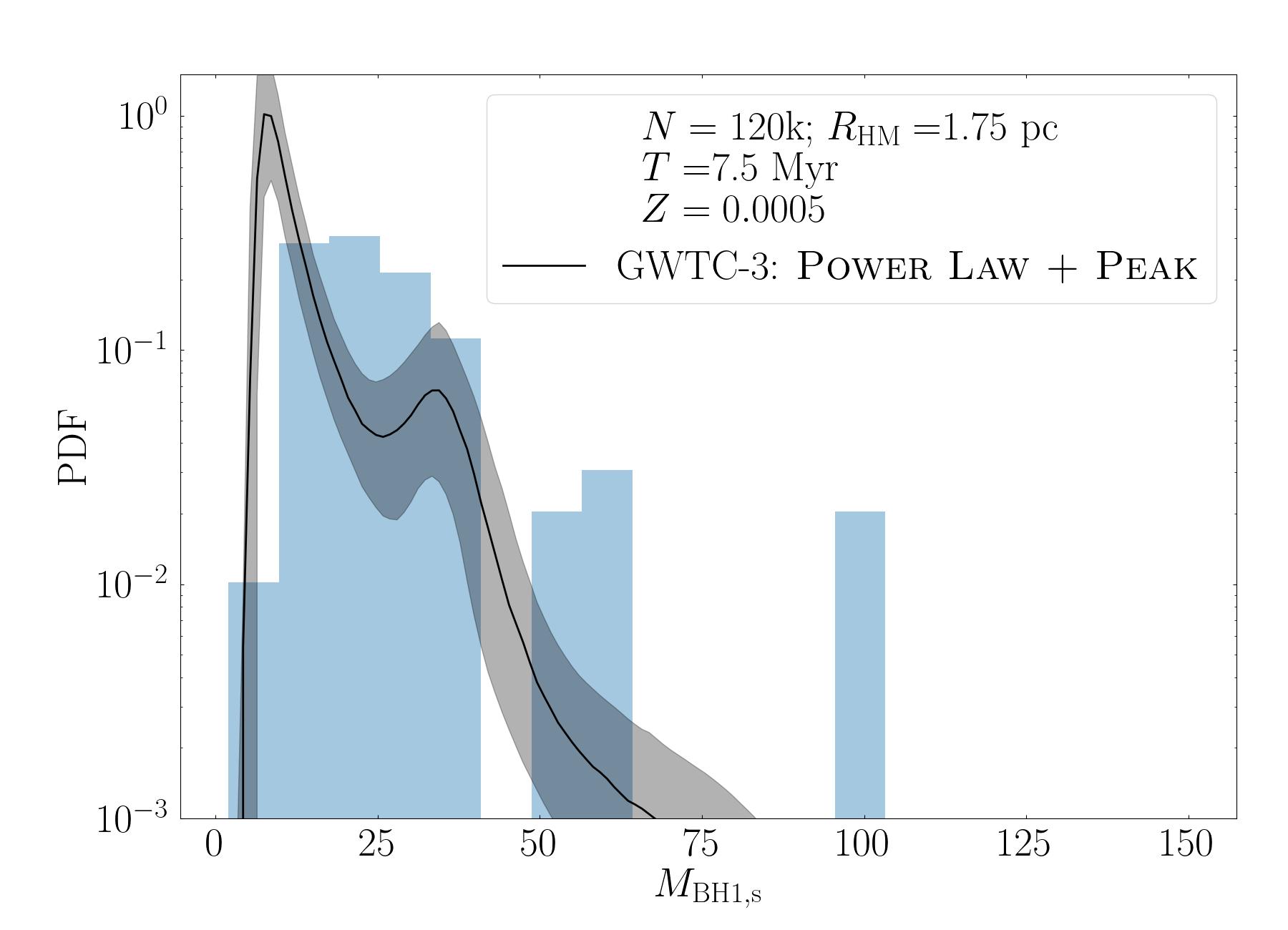}
    \includegraphics[width=\columnwidth]{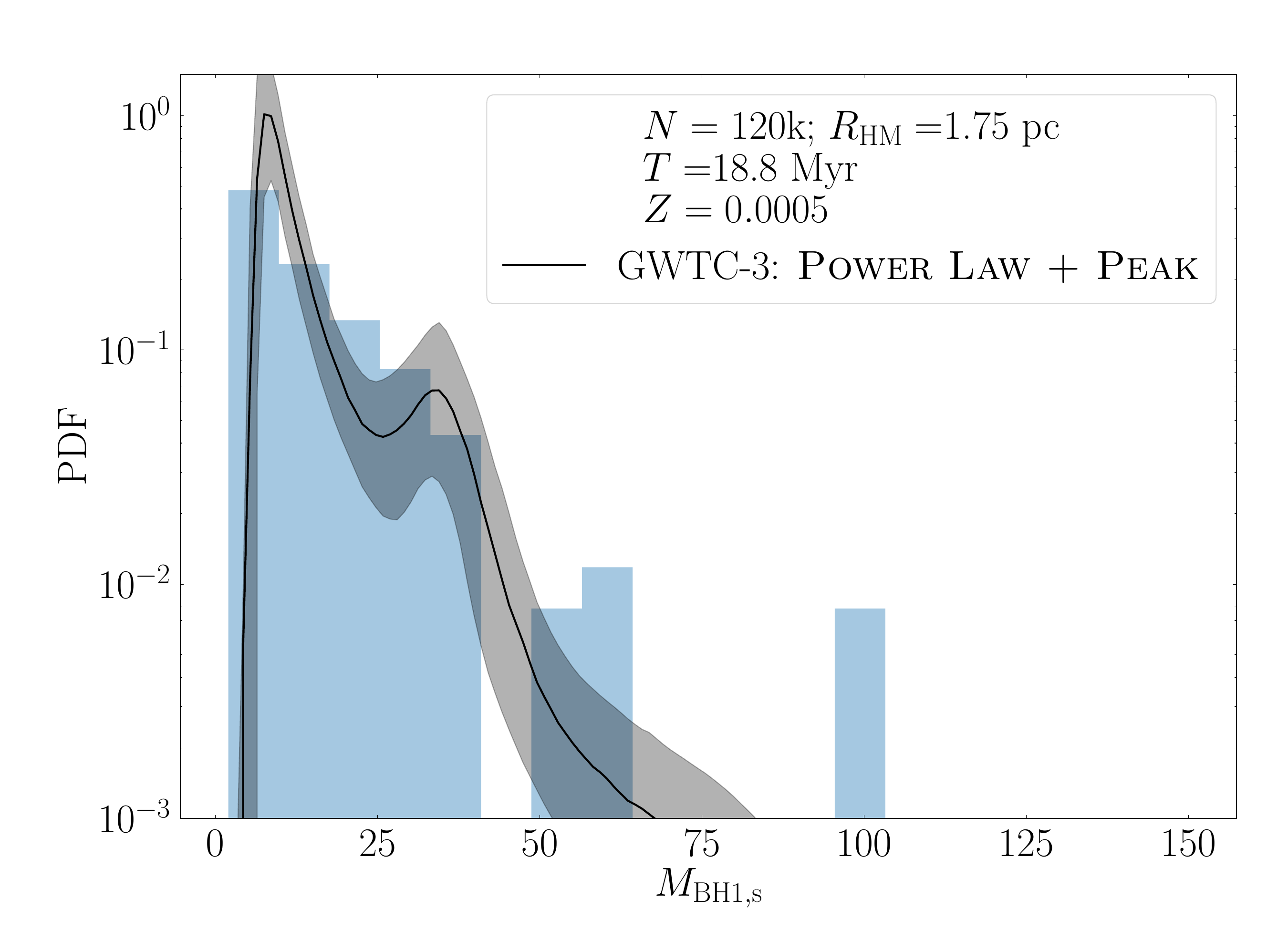}\\
    \includegraphics[width=\columnwidth]{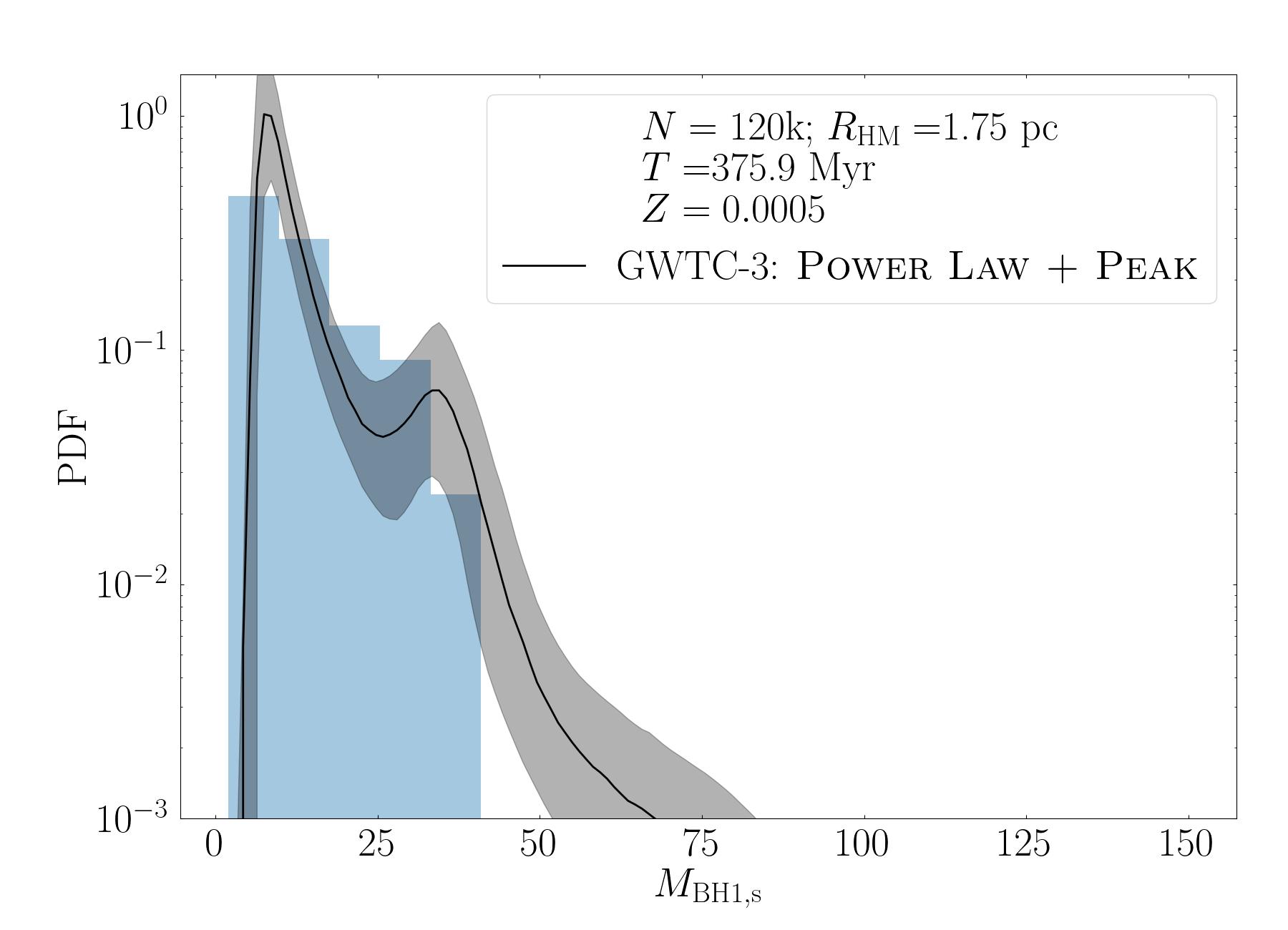}
    \includegraphics[width=\columnwidth]{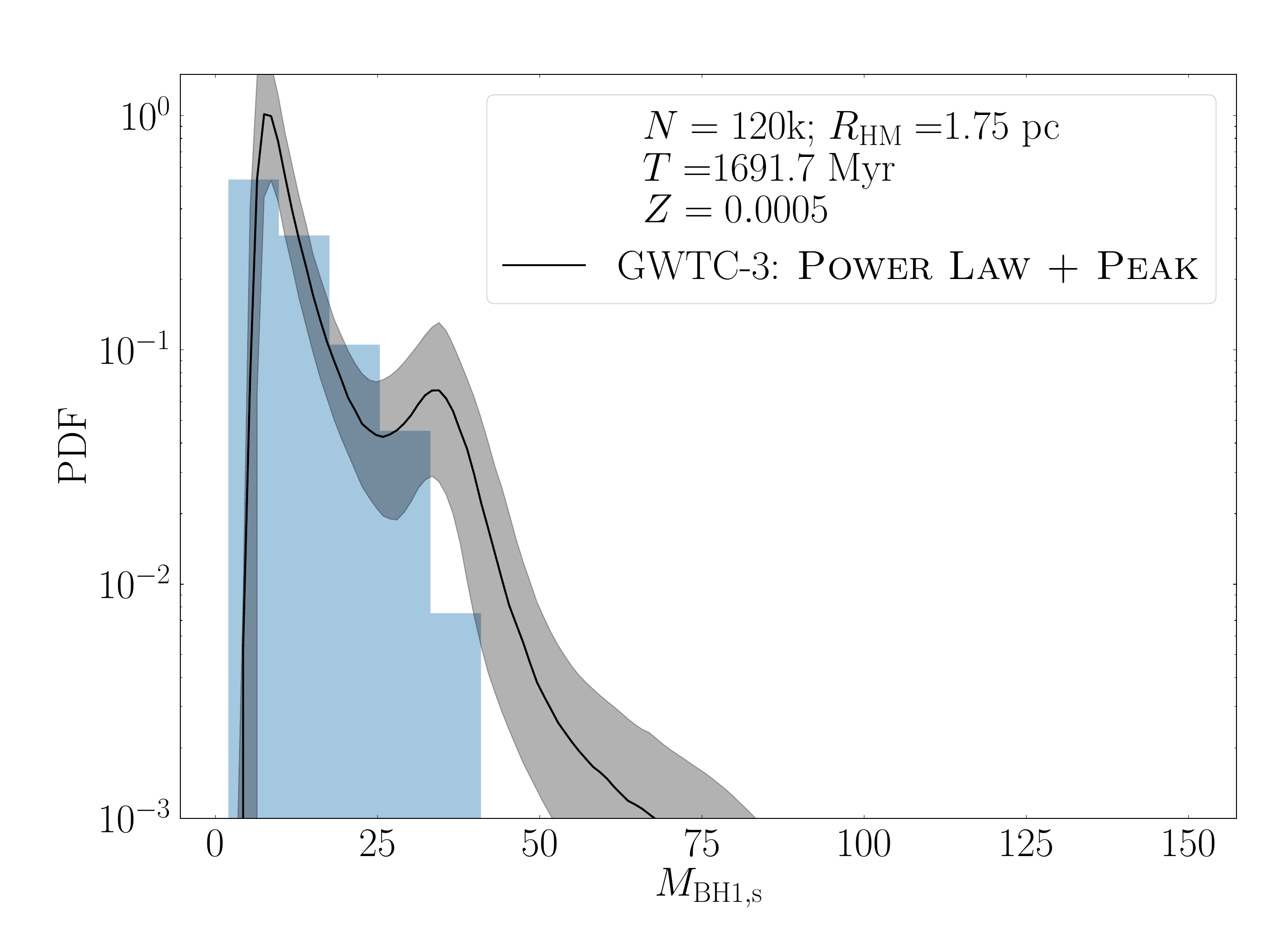}\\
    \caption{Mass distribution of BHs in a cluster model with $N=120$k stars, half-mass radius $R_\ham = 1.75$ pc, and initial binary fraction $f_b=0.2$. The black straight line and shaded area correspond to the mass distribution of primary BHs in merging binaries inferred from observed BBH mergers during the first, second, and third observing runs of the LIGO-Virgo-Kagra collaboration \citet{2021arXiv211103634T}.}
    \label{fig:BHbur}
\end{figure*}

Figure \ref{fig:BHbur} shows the BH mass distribution at different times for a model with $N=1.2\times 10^5$ stars, $R_\ham = 1.75$ pc, and $f_b=0.2$. This plot shows all BHs inside the cluster at a given time, regardless whether they are components of a binary system or single BHs. For the sake of comparison, we added in the plots the overall BH mass distribution inferred by the LVC \citep{2021arXiv211103634T}.
The plot highlights an initial phase in which the first BHs start to form, some of them falling in the upper-mass gap, but as the evolution proceeds new, lighter, BHs form while the most massive BHs are ejected via binary-single and binary-binary scatterings, as expected in the BH-burning scenario.

Interestingly, our simulations suggest that the evolution of the cluster can naturally lead to the peak around $10\Ms$ inferred from GW detections, mostly owing to stellar dynamics that crucially sculpts the BH population. Nonetheless, any comparison among our data, which show all BHs in the cluster, and LVC observations, which are representative of BH mergers, must be taken with a grain of salt. There are other potential explanations for the $10\Ms$ peak, like isolated binary stellar evolution \citep[e.g.][]{2022arXiv220903385V}, impact of primordial binary evolution in star clusters \citep[e.g.][]{2021arXiv210912119A}, metal rich star clusters \citep[e.g.][]{2020ApJ...894..133A,2021MNRAS.507.3612R,2022MNRAS.513.4527C}. Hopefully, the new data acquired during the forthcoming four LVC observation run could help pinning down the impact of different processes on the BH mass distribution.

We find that almost all BHs heavier than $>30\Ms$ are ejected from the simulated clusters reaching more than $\sim 15$ relaxation times. 

\begin{figure*}
    \centering
    \includegraphics[width=\columnwidth]{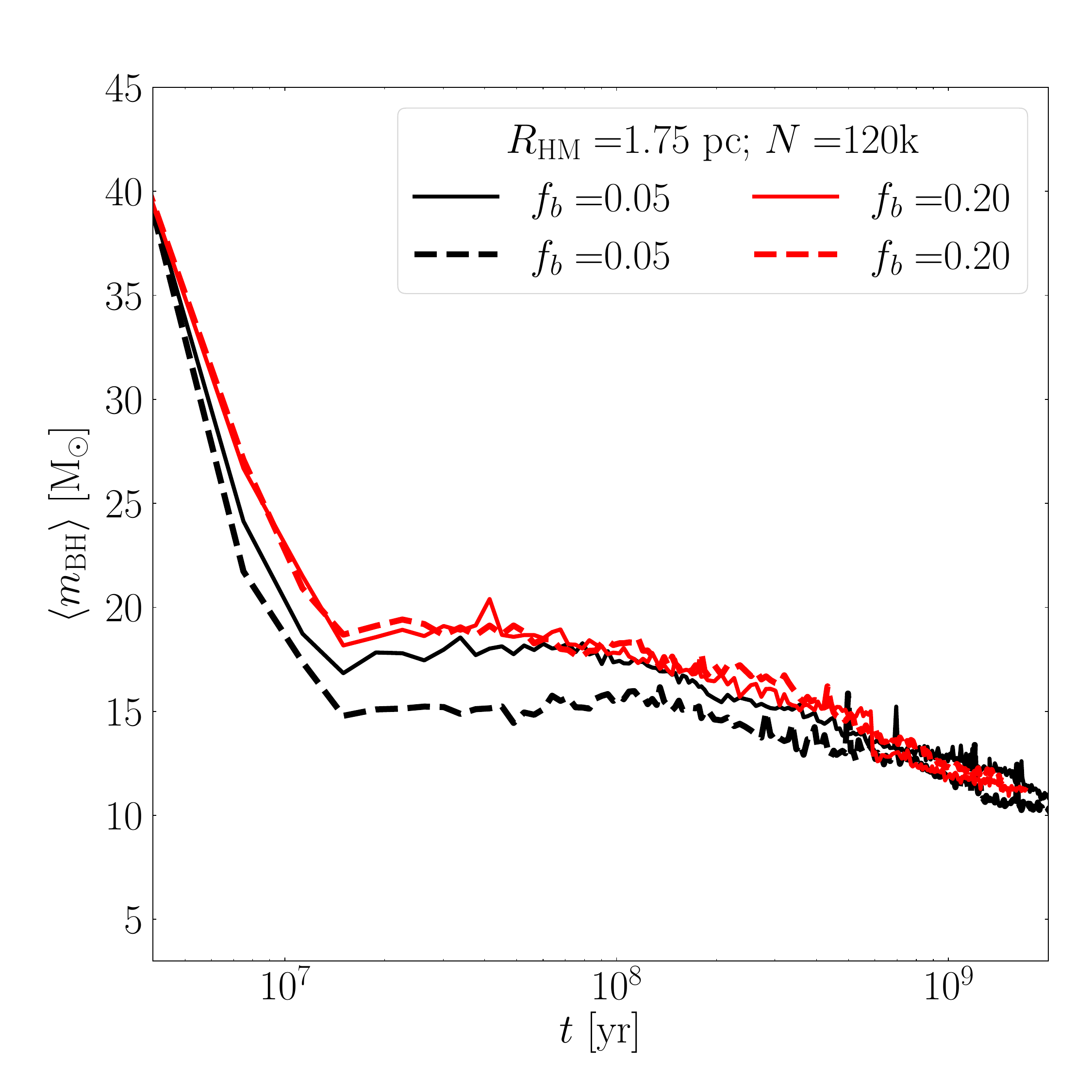}
    \includegraphics[width=\columnwidth]{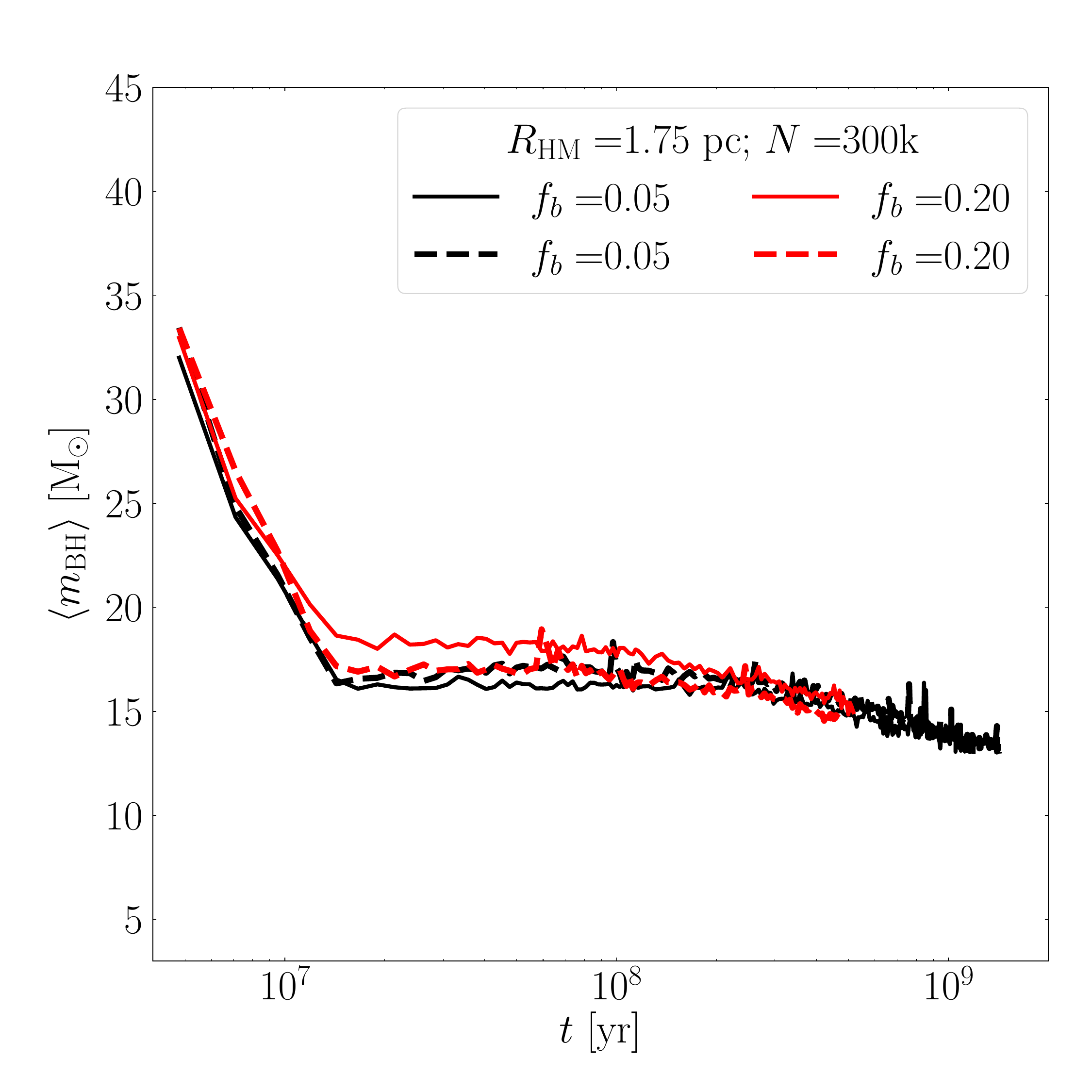}
    \includegraphics[width=\columnwidth]{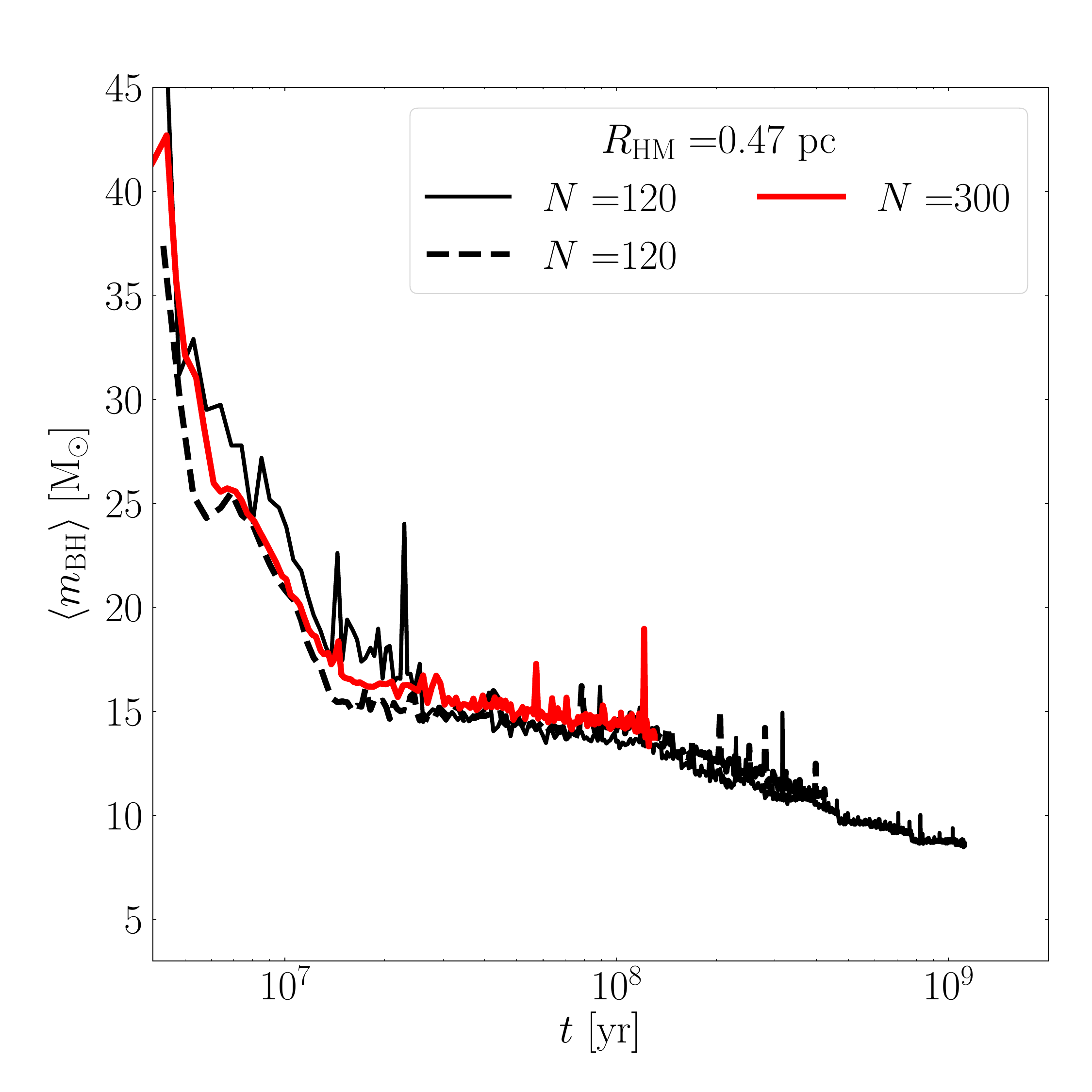}
    \includegraphics[width=\columnwidth]{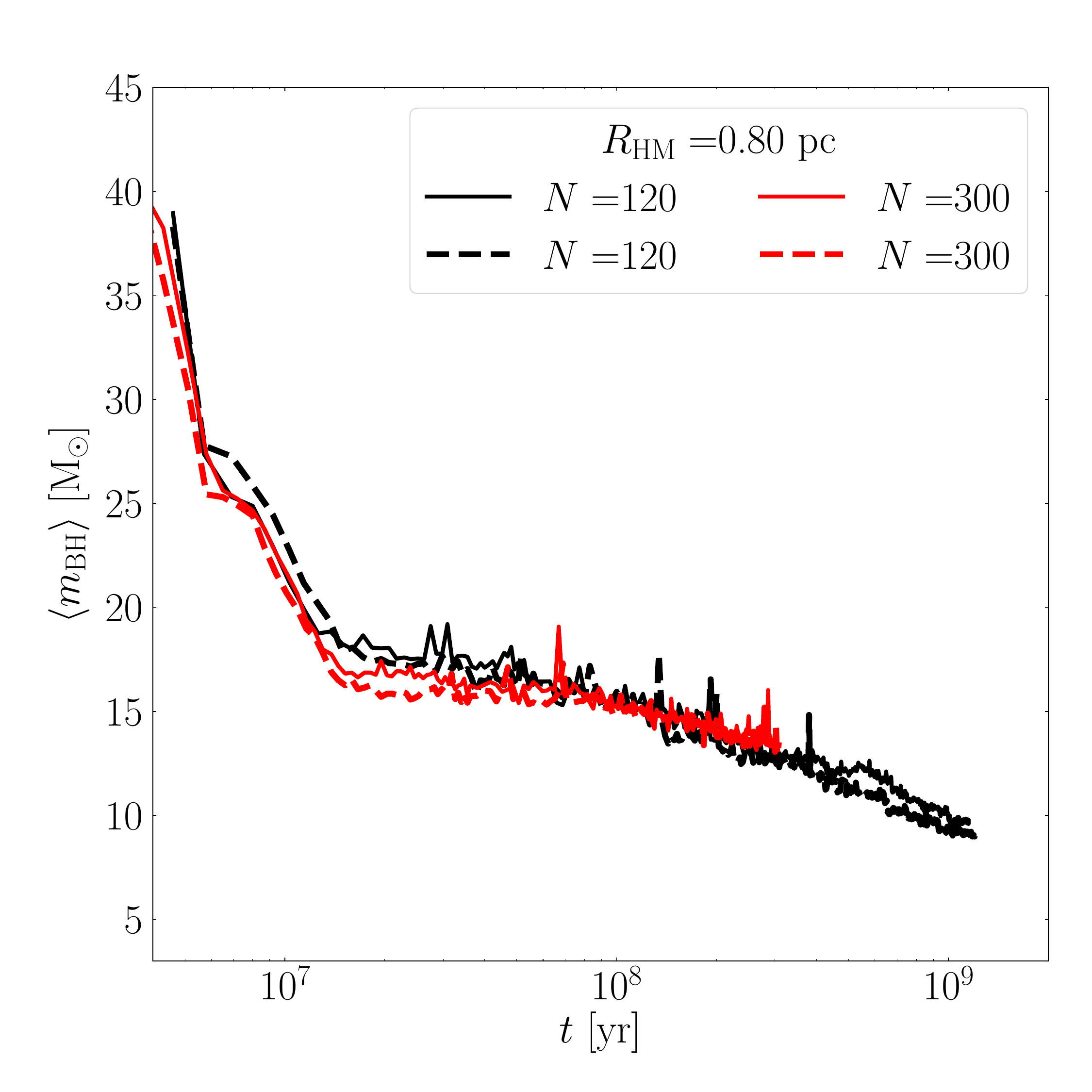}
    \caption{Average BH mass for BHs within the cluster half-mass radius. Note that the spikes can be due to heavy BHs orbiting on the edge of the half-mass radius, or on eccentric trajectories that occasionally enter the half-mass radius.}
    \label{fig:BHav}
\end{figure*}

To further highlight the BH burning process, we reconstruct the time evolution of the average BH mass, $\langle m_{\rm BH}\rangle$, for all BHs enclosed within the half-mass radius. As shown in Figure \ref{fig:BHav}, $\langle m_{\rm BH}\rangle$ follows the same trend regardless of the cluster initial condition, namely: i) the most massive BHs form first and the average mass sets close to the peak allowed by the adopted stellar evolution model ($35-40\Ms$); ii) more numerous, lighter BHs start to form causing a rapid decrease of the average mass down to $15-20\Ms$; iii) dynamical processes kick in and trigger BH ejection, leading to a secular decrease of the BH average mass down to $\sim 8-10\Ms$ \citep[see also][]{2019MNRAS.487.2412G}. 

\begin{figure}
    \centering
    \includegraphics[width=\columnwidth]{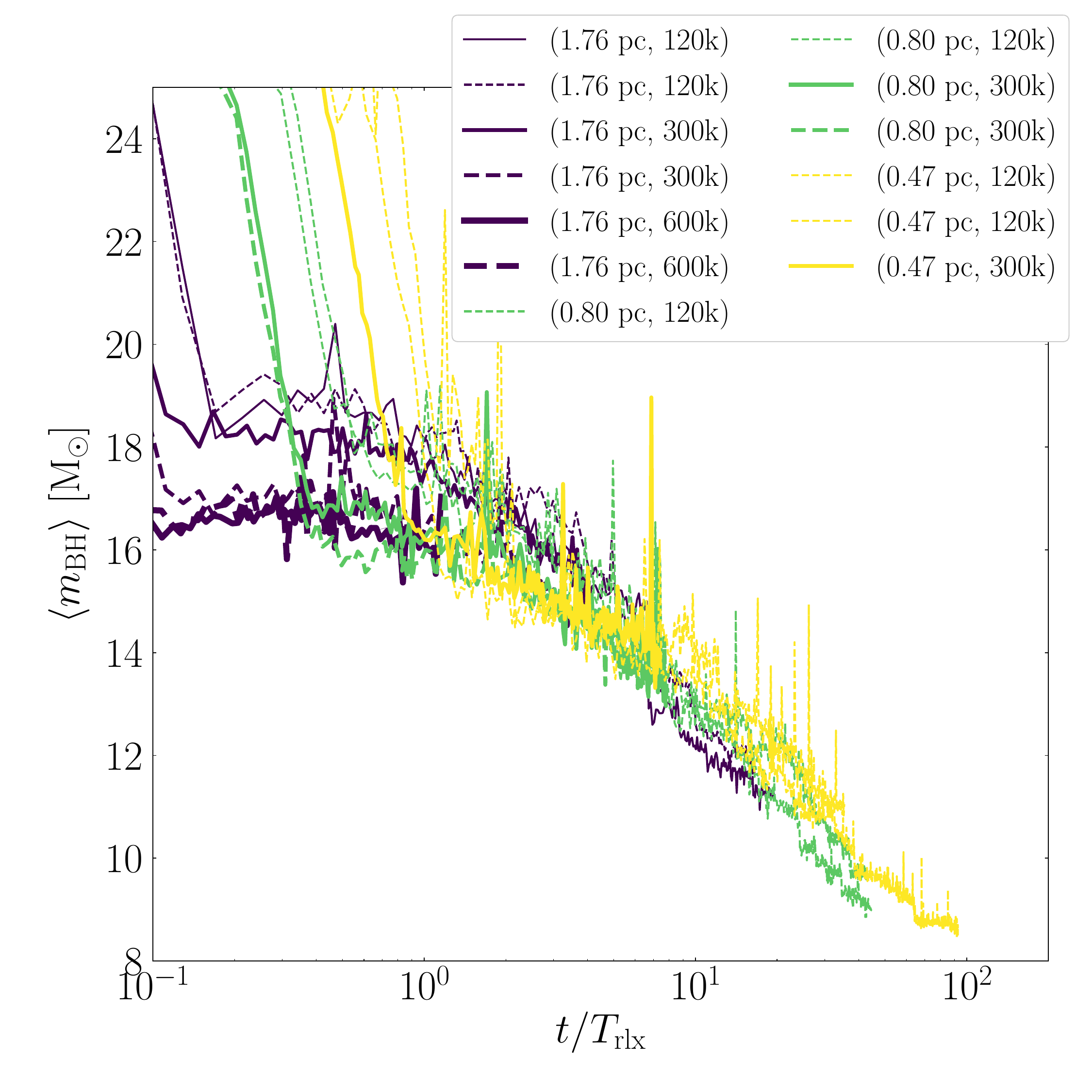}
    \caption{Average BH mass inside the half-mass radius as a function of the time normalised to the cluster initial relaxation time.}
    \label{fig:BHavNor}
\end{figure}

The similar $\langle m_{\rm BH}\rangle$ time evolution observed in different models supports the idea that the BH burning process is substantially due to dynamics. This is further highlighted in Figure \ref{fig:BHavNor}, which shows the BH average mass as a function of the time normalised to the cluster relaxation time. We find that at a time $t > T_{\rm rlx}$ the average BH mass is well described by a simple relation:
\begin{equation}
    \langle m_\bh(t) \rangle \simeq m_{\rm \bh,rlx} - 4{\rm Log}(t/T_{\rm rlx}),
\end{equation}
where $m_{\rm \bh, rlx} = 17.4\pm0.1\Ms$.

Although our models are not meant to be representative of any observed cluster, and although there are certainly many pathways leading to the same final cluster evolutionary stage, our results suggest that old Galactic globular clusters and massive clusters in the Small Magellanic Cloud could be harboring a population of relatively light BHs (see Figure \ref{fig:f2}). This would explain why observations of BHs in binary systems are generally characterised by masses $m_{\rm BH}<20\Ms$, relatively lighter than the typical value inferred for the population of merging BHs, i.e. $m_{\rm BH,GW}\simeq 30\Ms$.

\subsection{Using scaling relations as input for semi-analytic codes}
\label{sec:scal}

It is well known that $N$-body simulations of star clusters require generous computational resources to enable an exploration of the phase space and to reach an appreciably long simulated time. The \dragonii simulations make no exceptions, as they required in total approximately 2.2 million core hours. To overcome this problem, many works have proposed semi-analytic tools specifically devoted to study the evolution of compact objects in the last few years, and especially BH binary mergers \citep[e.g.][]{2020ApJ...894..133A,2021MNRAS.505..339M,2018PhRvL.121p1103F,2020MNRAS.492.2936A,2019MNRAS.486.5008A,2021arXiv210912119A,2022arXiv221010055K,2022MNRAS.511.5797M}.

One ingredient missing in some of these fast and accurate codes is a treatment of the co-evolution of the star cluster and the BH population, which may significantly affect the formation of merging compact objects \citep[see e.g.][]{2020MNRAS.492.2936A, 2021arXiv210912119A}. 

The \dragonii models could provide important fitting formulas to implement the evolution of under-filling cluster models in such semi-analytic tools. 

The overall evolution of \dragonii star clusters can be described by simple expressions (Equations \ref{eq:Mtime} and \ref{eq:rhalftime}). If the cluster initial mass and half-mass radius are known, the aforementioned relations enable an accurate description of its evolution, at least in the case of under-filling star cluster models. 

Moreover, our models offer also insights on the internal evolution of the cluster, providing, for example, details about the mass distribution of ejected single and double compact objects, and the properties of the central black hole subsystem. These ingredients can be easily implemented in semi-analytic tools to obtain a fast and accurate description of compact object dynamics in clusters too massive to be modelled with $N$-body models.

A simple implementation of the cluster evolution has been already developed by \cite{2021arXiv210912119A} in their \textsc{B-POP} code, showing that the inclusion of cluster mass loss and expansion causes a critical decrease of the probability of high-generation mergers in dense and massive star clusters \citep[see also][]{2020PhRvD.102l3016A}.

\section{Summary and conclusions}
\label{sec:sum}
In this work, we have presented the first results from the \dragonii star cluster simulations: a suite of 19 direct $N$-body simulations, performed with the \nbsix code, modelling the evolution of star clusters with up to 1 million stars and up to $33\%$ of stars initially in a binary, over a timescale of $\sim 0.5-2$ Gyr. These simulations contain up-to-date stellar evolution models, and for the first time a series of recipes to treat relativistic binaries in terms of merger remnant mass, spin, and post-merger recoil. Our models represent clusters initially under-filling their Roche lobe, and therefore their evolution can be considered quasi-isolated. The \dragonii models considerably expand the portion of parameter space covered with full $N$-body simulations, opening the possibility to compare with large-$N$ Monte Carlo models. Clearly, there is a vast number of parameters whose impact on the simulation results remains unclear. For example, adopting a sufficiently large value of the metallicity would imply the impossibility to form IMBHs from stellar collapse. However, we expect that our main conclusions about the properties of the BH population should not be severely affected by cluster metallicity, as they appear to be driven mostly by dynamics.

We find that the amount of primordial binaries seems to poorly affect the overall evolution of the cluster and the evolution of the BH population;  however the adopted initial orbital properties could become important when comparing our data with observations, like in the case of BH--MS binaries. For example, a different assumption on the initial mass-ratio distribution could lead to primordial binaries with final BH--MS component masses more similar to the observed one. However, discrepancies among observations and models could arise from a  combination of different assumptions, making it hard to pinpoint the main source of uncertainty.

Finally, our simulations model initially underfilling clusters, meaning that the impact of the Galactic field is almost negligible compared to clusters' internal dynamics. This choice enabled us to have a clean view at the impact of stellar interactions on the evolution of the whole cluster and its BH population, and incidentally lead to star cluster models that resemble observed clusters in term of mass and radius. Future simulations adopting filling or overfilling clusters may help understanding whether the evolution of BH subsystems is intrinsically linked to the overall evolution of the host cluster, for example in terms of mass-loss and expansion.

The main outcomes of the \dragonii models can be summarised as follows.
\begin{itemize}
    \item mass-loss and expansion of \dragonii clusters is mostly determined by internal dynamics and can be described by simple analytical expressions, with parameters that weakly depend on the initial conditions. The binary fraction varies mildly over the simulated time, within $10-15\%$ of its initial values. Nonetheless, stellar evolution and dynamics cause a progressive drop of the fraction of stars in binary systems for primary masses $m_1>2\Ms$ [Figures \ref{fig:f2}-\ref{fig:bmass}];
    \item over a Gyr timescale, \dragonii clusters contains around 200--700 binaries with at least one WD, whilst the number of binaries with a NS or a BH generally remains below 1--10 and 5--40, respectively. In general, binaries with at least one compact object are more numerous in clusters with a larger initial binary fraction, suggesting that most of these binaries have a primordial origin. Moreover, the denser the cluster is the smaller the number of binaries, owing to energetic dynamical interactions that disrupt binaries more efficiently [Figure \ref{fig:DCO}];
    \item ejected binaries with one (SCOB) or two (DCOB) compact objects have different properties. DCOBs exhibit masses following a nearly flat distribution around $2-20\Ms$ and a peak at $m_{\rm BH} = 45\Ms$, a peculiar mass-ratio distribution that peaks around $q\gtrsim 0.6$, and a flat eccentricity distribution in the range $e=0.5-1$. SCOBs, most of which formed from primordial binaries, typically involve low-mass BHs ($m_{\rm BH} = 3-10\Ms$) and fairly massive MS stars ($m_{\rm ST} = 1-10\Ms$) [Figure \ref{fig:binBH}];
    \item we find a substantial population of BH--MS binaries in \dragonii models. 
    Most BH--MS binaries forming inside the cluster have typical BH masses $m_{\rm BH}>10\Ms$, a companion star with mass $m_{\rm MS} = 0.7-100\Ms$, orbital periods $>10$ days, and span the entire eccentricity range. Ejected BH--MS binaries, instead, feature significantly smaller BH masses $m_{\rm BH} < 10\Ms$, shorter periods ($<10$ days), and are mostly contributed by primordial binaries. We find that the properties of the modelled binaries are compatible with some features of observed BH--MS binaries, especially those observed in the globular cluster NGC3201 [Figures \ref{fig:msbh}-\ref{fig:msbh2}];
    \item in all \dragonii models, BHs form a long-lived subsystem in the cluster centre already after 0.5 relaxation times, with a typical density $10-100$ times higher than that of stars. The cluster core radius represents a good proxy of the BH subsystem size, as BHs make up $50-80\%$ of the mass enclosed within this radius. We find that the ratio between the number of BHs inside the core radius and the bound cluster mass, which we refer to as formation efficiency, attains values of $\epsilon_{\rm BH,BBH}/$ M$_\odot=10^{-3}$($10^{-4}$), for single and binary BHs, respectively. This quantity is only mildly dependent on the initial conditions, suggesting that dynamical processes have a relatively minor effect on the overall BH population over the simulation time [Figures \ref{fig:fBH} - \ref{fig:BHeff}];
    \item dynamics in the BH subsystem critically affects the BH mass spectrum, owing to the BH-burning process. The peak of the mass distribution generally shifts from initial values $m_{\rm BH,pk} = 25\Ms$ down to $m_{\rm BH,pk} = 5-15$, and the average mass steadily decreases after one relaxation time, following an identical evolution regardless of cluster properties [Figures \ref{fig:BHbur}-\ref{fig:BHavNor}].
\end{itemize}
Our simulations suggest that dynamically old star clusters harbour in their centre a population of BHs whose amount scales linearly with the cluster bound mass. The older the cluster is, the smaller the peak of the BH mass spectrum and the average BH mass.

\section*{Acknowledgements}

The authors thank the referee for their insightful feedback, which helped us improving our analysis. The authors warmly thank Agostino Leveque for their help and assistance in using their implementation of the \mcl code, and Vincenzo Ripepi for useful discussions and comments. This work benefited of the support from the Volkswagen Foundation Trilateral Partnership through project No.~97778 ``Dynamical Mechanisms of Accretion in Galactic Nuclei'' and the Deutsche Forschungsgemeinschaft (DFG, German Research Foundation) -- Project-ID 138713538 -- SFB 881 ``The Milky Way System'' (in particular subproject A08), and by the COST Action CA16104 ``GWverse''. 

The authors gratefully acknowledge the Gauss Centre for Supercomputing e.V. for funding this project by providing computing time through the John von Neumann Institute for Computing (NIC) on the GCS Supercomputer JUWELS Booster at Jülich Supercomputing Centre (JSC).

MAS acknowledges funding from the European Union’s Horizon 2020 research and innovation programme under the Marie Skłodowska-Curie grant agreement No. 101025436 (project GRACE-BH, PI: Manuel Arca Sedda). 

AWHK is a fellow of the International Max Planck Research School for Astronomy and Cosmic Physics at the University of Heidelberg (IMPRS-HD).

The work of PB was supported by the Volkswagen Foundation under the special stipend No.~9B870. PB acknowledge the support within the grant No.~AP14869395 of the Science Committee of the Ministry of Science and Higher Education of Kazakhstan ("Triune model of Galactic center dynamical evolution on cosmological time scale"). 
The work of PB was supported under the special program of the NRF of Ukraine Leading and Young Scientists Research Support - "Astrophysical Relativistic Galactic Objects (ARGO): life cycle 
of active nucleus", No.~2020.02/0346.

RS thanks Max Planck Institute for Astrophysics (Thorsten Naab) for hospitality during many visits

MG was partially supported by the Polish National Science Center (NCN) through the grant UMO-2021/41/B/ST9/01191

GI, MM, and SR acknowledge financial support from the European Research Council for the ERC Consolidator grant DEMOBLACK, under contract no. 770017.

\section*{Data Availability}
The data from the runs of these simulations and their initial models
will be made available upon reasonable request by the corresponding author. 
The \textsc{Nbody6++GPU} code is publicly available\footnote{\url{https://github.com/nbody6ppgpu/Nbody6PPGPU-beijing}}. The \textsc{McLuster} version used in this work will soon be available. A similar version is described in \cite{2022MNRAS.514.5739L}.



\bibliographystyle{mnras}
\bibliography{example} 

\bsp	
\label{lastpage}
\end{document}